\documentclass[11pt]{article}
\usepackage[letterpaper,margin=1in]{geometry}
\usepackage{authblk}
\usepackage{placeins}
\usepackage{amsthm}
\usepackage{textcomp}
\usepackage{amssymb}
\usepackage{indentfirst}
\usepackage{bigints}
\usepackage{abstract}
\usepackage{bm}
\usepackage{natbib}
\usepackage{listings}
\usepackage{mathptmx}
\usepackage{CJKutf8}
\usepackage{amsmath}
\usepackage{cases}
\usepackage{amsfonts}
\usepackage{algorithm}
\usepackage{algorithmic}
\usepackage{subcaption}
\usepackage{titlesec}
\usepackage{appendix}
\usepackage{graphicx}
\usepackage{float}
\usepackage{verbatim}
\usepackage{enumitem}
\usepackage{lineno}
\usepackage[table]{xcolor}
\usepackage{array}
\usepackage{caption}
\usepackage{xfrac}
\usepackage{booktabs}
\usepackage{hyperref}
\usepackage{multibib}
\usepackage{url}
\usepackage{color}

\DeclareMathOperator{\arcosh}{arcosh}
\usepackage{mathrsfs}
\usepackage{xcolor}
\usepackage{tocloft}
\usepackage{tikz}
\usetikzlibrary{arrows.meta, positioning}

\hypersetup{
    colorlinks=true,
    linkcolor=black,
    filecolor=black,
    urlcolor=black,
    citecolor=black,
}

\titleformat{\part}[block]
{\centering\Huge\bfseries}
{}
{1em}
{}

\usepackage{ulem}
\usepackage{xcolor}
\definecolor{darkyellow}{RGB}{204, 153, 0}

\addtocontents{toc}{\protect\setcounter{tocdepth}{-1}}

\begin{document}

\begin{center}
\textbf{{\Large {Effective decoupling of mutations and the resulting loss \\ of biodiversity
caused by environmental change}}}

\bigskip

Ruixi Huang$^{a,b,1}$ and David Waxman$^{a,2}$

\smallskip

$^a$Centre for Computational Systems Biology, ISTBI,

Fudan University, 220 Handan Road, Shanghai 200433, PRC

\smallskip

$^b$Research Institute of Intelligent Complex Systems,

Fudan University, 220 Handan Road, Shanghai 200433, PRC
\end{center}

\noindent $^{1}$Email: rxhuang22@m.fudan.edu.cn \hfill ORCID ID: 0000-0002-3967-8818

\noindent $^{2}$Email: davidwaxman@fudan.edu.cn \hfill ORCID ID: 0000-0001-9093-2108\newline 
$^{2}$Corresponding author.

\begin{center}
{\large \textbf{Abstract}}
\end{center}

Many biological populations exhibit diversity in their strategy for survival and reproduction in a given environment, 
and microbes are an example. We explore the fate of different strategies under sustained environmental change by 
considering a mathematical model for a large population of asexual organisms. Fitness is a bimodal function of a quantitative trait, with two local optima, separated by a local minimum, i.e., a mixture of stabilising and disruptive selection. The optima represent two locally `best' trait values. We consider regimes where, when the environment is unchanging, the equilibrium distribution of the trait is bimodal. 
A bimodal trait distribution generally requires, for its existence,
mutational coupling between the two peaks, and it indicates two coexisting clones with distinct survival and reproduction strategies.
When subject to persistent environmental change, the population adapts by utilising mutations that allow it to track the changing environment. The faster the rate of change of the environment, the larger the effect of the mutations that are utilised. Under persistent environmental change, the distribution of trait values takes two different forms. At low rates of change, the distribution remains bimodal. At higher rates, the distribution becomes unimodal. This loss of a clone/biodiversity is driven by a novel mechanism where environmental change decouples a class of mutations.


\begingroup
  \renewcommand\thefootnote{}%
  \footnotetext{Please cite the formal publication: \href{https://doi.org/10.1016/j.jtbi.2025.112277}{https://doi.org/10.1016/j.jtbi.2025.112277}.}%
  \addtocounter{footnote}{-1}%
\endgroup

\newpage

\section{Introduction}

Anthropogenic activities have caused numerous changes in the environment over time, 
as exemplified by changes in ambient temperatures, salinity levels, and the eutrophication of
aquatic ecosystems \citep{lean2008natural, stern2014anthropogenic, williams2001anthropogenic, paerl2006anthropogenic, reid2019emerging}. 
Environmental change that persists over time poses a significant challenge that biological 
populations must overcome in order to survive \citep{domnauer2021proteome, wersebe2023resurrection}.
Indeed, many populations/ecosystems fail this challenge \citep{dudgeon2006freshwater, zhang2007global, zhao2017temperature}.

Existing genetic variation, and mechanisms such as genotype-environment interactions 
are likely to play a role in the adaptation of a population to environmental change, 
but chiefly only over the short term~\citep{lande1996role, gienapp2008climate,  chevin2010adaptation}. 
When environmental change persists over an extended period of time, adaptation primarily/ultimately depends on new mutations.

In the present work we explore the interplay between mutations and environmental change. At first sight, the only relations between environment 
and mutations are direct. For example, increased temperatures or mutagenic chemicals may 
elevate mutation rates \citep{bennett1990rapid, berger2021elevated, long2016antibiotic}. 
 However, we  work in a framework where mutational processes remain essentially
\textit{unchanged} and shall primarily explore what occurs when the environment changes at a 
\textit{constant rate} over time (we do, however, give some consideration to small cumulative environmental shifts).

We proceed assuming fitness is a bimodal function of a quantitative trait, with two local optima, separated by a 
local minimum, i.e., a mixture of stabilising and disruptive selection. The optima represent two locally `best’
trait values. Environmental change that occurs at a constant rate causes the fitness function 
to `move’ over time, with the trait 
values of the two optima (and all other features of the fitness function) occurring at progressively larger values as time increases. The population responds by incorporating new mutations, which we assume allow 
it to persistently adapt to the changing environment.

We determine properties of the distribution of the trait over time,
assuming the population size is very large, so the population does not exhibit stochastic effects. 
By focussing solely on the distribution (a probability density which always integrates to unity), 
this work does not touch on issues of population extinction, but rather on what occurs \textit{within} 
an adapting population. A natural limit to the applicability of this work is  when the mean absolute
fitness drops below unity, but to deal with such an issue requires a realistic measure of absolute fitness, 
which would take us beyond the current work.

The scope of this work is an investigation into how phenomena, occurring \textit{within} a population, depend on the rate of change of the environment.
For example, if we compare two different non zero rates of environmental change, then relative to the slower rate, 
the faster rate will repeatedly require larger effect mutations for 
the population to adapt, i.e., follow the environment, and the population will exhibit 
some effects of this, for example a higher genetic load \citep{whitlock2011genetic}. This indicates that 
consequences of adaptation of a population are a joint outcome of (i) the behaviour of the environment 
and (ii) the mutations that become utilised in adaptation. Indeed, the way a changing environment 
induces a population to adapt could be viewed as the environment probing the properties of the mutations
that are accessible to the population.

For a very large population, where extinction is not an issue, the response 
to a changing environment plausibly has a continuous dependence on the rate of change of the environment. 
We shall shortly present results where low and high rates of environmental change
lead to the population behaving in \textit{qualitatively} different ways. 
We demonstrate this phenomenon within the context of an asexual population of very large size. 
Such a description can, for example, apply to a variety of microorganisms \citep{schon2009lost}.

\section{Model}
We assume individuals are characterised by the value of a single quantitative trait, 
which may be associated, for example, with growth rate or metabolic efficiency, 
but there are many other possibilities.
We shall work at the level of the \textit{genotypic value} of the trait,
assuming phenotype to be additively determined from genotype and environmental
effect, with no genotype environment covariance \citep{Lynch}.
We assume the genotypic value of the trait, $g$, is controlled by the small effects 
of genes at a large number of loci, and so may be treated as a continuous variable,
and we take $-\infty<g<\infty$.
For more details of the model, see Section 4 of the Supplementary Material.

We obtain results within the framework of a basic lifecycle associated with
a very large population, which may be treated as behaving deterministically. 
Generations are discrete and labelled by $t=0,1,2,\ldots$ . A generation
starts with offspring, who are subject to viability selection.
The individuals that survive are adults. These reproduce asexually, 
and then die, leaving offspring that generally contain new mutations.

Selection occurs, according to the (genotypic) trait value of individuals, 
and mutations have the effect of changing the trait value.

\subsection{Static environment}

To set the context for later results, we first consider static environmental conditions.
In this case, the fitness of individuals with a given genotypic trait value, $g$, depends solely on $g$, and not on the time, $t$. We write the fitness of such individuals as $W(g)$. This fitness function arises from an average of fitness, as a function of phenotypic value, over the environmental effect of the phenotypic value.

The fitness function $W(g)$ generally has a lowest value of zero and a finite maximum value. As we shall shortly see (in Eq. \eqref{static differencems},
and its generalisation to time dependent fitness functions), the precise value of the maximum of $W(g)$ has no influence 
on the evolution of trait values (thus we could replace $W(g)$ by 
$k \times W(g)$, where $k$ is a positive constant, with no change in any feature or the behaviour of the distribution). This is an exact property 
of the discrete time equation that the distribution obeys. Because of this 
we could, for example, replace $W(g)$ by the \textit{relative fitness function} $w(g)=W(g)/\max_{g} W(g)$, which has 
a maximum value of unity, but it is not necessary to do so. Accordingly, all fitness functions specified in this work have mostly been chosen for convenience of form, rather than having a particular maximum value, and we shall often indicate their specification by a proportionality, rather than via an equation, as a reminder that the maximum value has no influence on the behaviour of the distribution of the trait.

Due to mutation, offspring trait values are distributed \textit{around}
the parental trait value. A parent with trait value  $h$ 
is taken to have offspring with trait values in the range $g$ to $g+dg$ with probability
$f(g-h)dg$ and we term $f(g)$ the \textit{distribution of mutant effects}. For simplicity and ease of exposition, throughout this work we have taken $f(g)$ to be a normal distribution with mean zero and variance $m^{2}$. 
Such a distribution can be derived (from the central limit theorem) if we assume that each individual carries an appreciable number of mutations at different loci, i.e., when the genome-wide mutation rate is large, $U=L\mu\gg 1$, where $L$ is the number of loci and $\mu$ is the mutation rate per locus. While small values of $U$ lead to other forms of $f(g)$ (see Section 4.2 of the Supplementary Material), the qualitative conclusions we shall come to, most notably the transition from a bimodal to a unimodal trait distribution as $v$ increases, persist for a broad class of mutational distributions. The precise numerical values we determine (e.g., for the critical velocity - see later) may change with the shape of $f$, but the qualitative phenomena themselves exhibit a level of robustness.

Explicitly, the form of distribution of mutant effects, $f(g)$, that we take is
\begin{equation}
    f(g)=\frac{\exp\left(-\dfrac{g^2}{2m^2}\right)}{\sqrt{2 \pi m^2}}        \label{f(g)ms}
\end{equation}
and the variance, $m^2$, is one of the parameters of the model, although it may have
more fundamental origins (see Section 4.2 of the Supplementary Material).

With $\Psi(g,t)$ the distribution (probability density) of genotypic values at the beginning of generation $t$, the distribution in the next generation results from 
selection, followed by mutation, and is thus given by
\begin{equation}
\Psi(g,t+1) = \frac{\displaystyle\int_{-\infty}^{\infty}{f(g-h)W(h)\Psi(h,t)dh}} {\displaystyle\int_{-\infty
}^{\infty}W(h)\Psi(h,t)dh}
.\label{static differencems}
\end{equation}

\begin{figure}[H]
\centering
\includegraphics[width=0.7\textwidth]{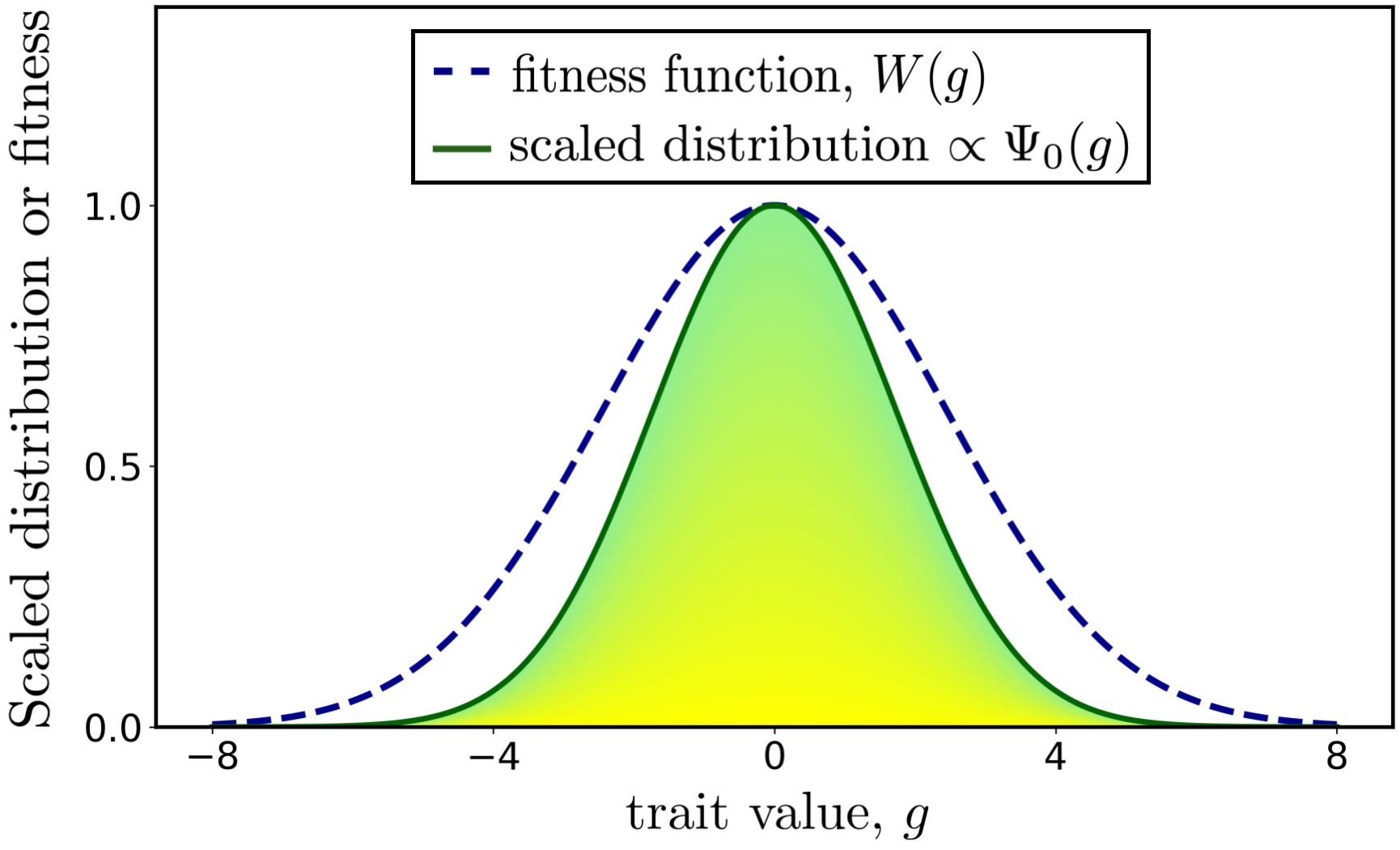}
\caption{\textbf{Unimodal fitness function, and the corresponding equilibrium distribution of genotypic trait values in a static environment.} In this figure we plot the particular \textit{unimodal} fitness function $W(g) = \exp\left(-g^{2}/(2s^2)\right)$  against
genotypic trait value $g$ (dashed blue line). The fitness function has a peak at $g=0$. The corresponding equilibrium genotypic distribution, 
which has been scaled so it has a maximum height of unity, is also plotted  against $g$ (green line).  
To calculate the figure we used the mutational distribution of Eq. \eqref{f(g)ms}. The parameter values adopted were $m=1$ and $s=\sqrt{6}$.}
\label{f1}
\end{figure}

\begin{figure}[H]
\centering
\includegraphics[width=0.7\textwidth]{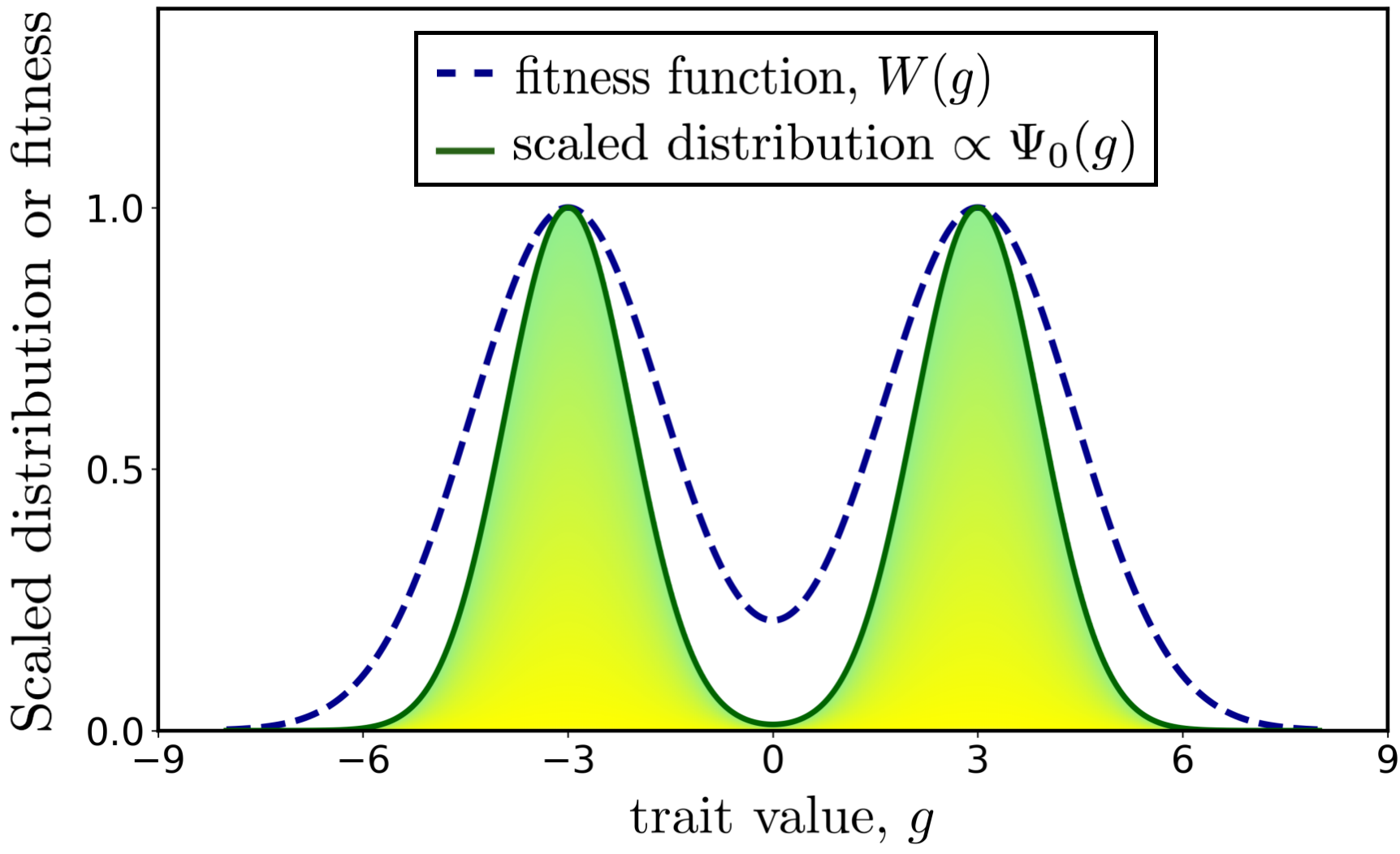}
\caption{\textbf{Fitness function, and the corresponding equilibrium distribution of genotypic trait values in a static environment.} 
In this figure we plot the particular \textit{bimodal} fitness function $W(g)=\exp\left(-\left(g+a\right)^{2}/(2s^2)\right)
+\exp\left(-\left(g-a\right)^{2}/(2s^2)\right)$ against
genotypic trait value $g$ (dashed blue line). The fitness function has peaks in the vicinity of $g=-a$ and $g=a$. 
The corresponding equilibrium genotypic distribution, which has been scaled so it has a maximum value of unity, 
is also plotted against $g$ (green line).  
To calculate the figure we used the mutational distribution of Eq. \eqref{f(g)ms}. The parameter values adopted were $m=1/2$, $s=\sqrt{2}$, and $a=3$.}
\label{f2ms}
\end{figure}

The origins of fitness can be complex, since it generally emerges from a multitude of underlying processes - see e.g., the work of \cite{peck2000mutation}. A common situation is where there is a single `best' or optimal value of the trait, and any deviations from this optimal value (i.e. trait values that are larger \textit{or} smaller than the optimal value) lead to reduced fitness. This amounts to \textit{stabilising selection}, of which there are many examples \citep{bell2008selection}. A commonly used unimodal fitness function has the Gaussian form
\begin{equation}
W(g) \hspace{0.2cm}\propto\hspace{0.2cm} \exp\left(-\dfrac{g^{2}}{2s^2}\right)  \qquad\text{unimodal
fitness function}\label{uni Gauss}
\end{equation}
where $s$ is a positive constant that is a direct measure of the width of the peak. In Figure \ref{f1} we illustrate such a fitness function, along with the equilibrium distribution of trait values, where at $g=0$ the fitness function $W(g)$ has its maximum value, while for $g=\pm \sqrt{2\ln(2)} s\simeq \pm 1.2s$ the value of $W(g)$ is half its maximum value. 
The other form of fitness we consider is \textit{locally disruptive} \citep{smith1993disruptive, hendry2009disruptive, strugariu2023factor}. In particular, we shall consider fitness that is locally disruptive  
in the vicinity of $g=0$, but we plausibly assume that fitness vanishes at large absolute values
of $g$ (i.e., at large $|g|$). Then the very simplest fitness function with such properties has a \textit{bimodal form}. While here we consider large (effectively infinite) populations,
some implications of a bimodal fitness function for a finite population have previously been considered \citep{whitlock1995variance}.
A possible example of a bimodal fitness function is
\begin{equation}
W(g) \hspace{0.2cm}\propto\hspace{0.2cm}\exp\left(-\dfrac{\left(  g+a\right)^{2}}{2s^2}\right)
+\exp\left(  -\dfrac{\left(  g-a\right)  ^{2}}{2s^2}\right)  \qquad
\text{bimodal fitness function}\label{bi Gaussms}
\end{equation}
in which we assume that the parameters satisfy $a/s>1$, so the fitness function has a trough at $g=0$ and two peaks \citep{eisenberger1964genesis}.
For $a/s\gtrsim 1.5$, the peaks are in the vicinity of $g=\pm a$, 
and each has a width of $\sim s$. Figure \ref{f2ms} contains an illustration of 
such a fitness function, which
for the chosen parameters, induces a bimodal distribution of the trait (green curve). 

\subsection{Changing environment}

We now ask what happens when constant environmental change occurs?

We consider a particular model of constant environmental change,  
based on the fitness functions
of Eqs. \eqref{uni Gauss} and \eqref{bi Gaussms} - see Figures \ref{f1} and  \ref{f2ms}.
Now, however, we assume the fitness functions are time dependent, and influenced by the environment, so that 
all features of a fitness function, for example any peaks it has, 
increase the value of $g$ at which they occur by the same amount each generation. 
Thus for  generation $t$, this amounts to replacing the fitness function $W(g)$ of Eqs. \eqref{uni Gauss} and \eqref{bi Gaussms} 
by $W(g-vt)$. 
Here $v$ is a constant that is arbitrarily taken to be positive; it 
corresponds to the amount by which features of the fitness function change their `position' each generation.

For a fitness function with a single optimum, problems related to the above have been considered elsewhere \citep{waxman1999sex, kopp2014rapid, matuszewski2015catch, garnier2023adaptation}, and for a fitness
function with two optima, the above is a natural generalisation of this previous work.
Indeed, if fitness is at all connected with an environmental chemical, or temperature, that e.g., increases 
over time, then a fitness function with features that increase their `position' every generation is 
a plausible assumption.

We shall call $v$ the `velocity' of environmental change even though it is not a physical velocity. 
For example, with generations labelled by an integer, the units of 
$v$ coincide with the units in which genotypic values are measured, and can
be associated with quantities that are very far from notions of physical space,
such as vitamin D storage and pigmentation \citep{haldane1990causes}.

The required generalisation of the dynamical equation (that governs the distribution $\Psi(g,t)$),
to include such a changing environment, is to replace $W(h)$ in Eq. \eqref{static differencems} 
by $W(h-vt)$. The equation then becomes
\begin{equation}
\Psi(g,t+1)=\frac{\displaystyle\int_{-\infty}^{\infty}{f(g-h)W(h-vt)\Psi
(h,t)dh}}{\displaystyle\int_{-\infty}^{\infty}W(h-vt)\Psi(h,t)dh}%
.\label{runningdistms}%
\end{equation}

\section{Methods}
In this work, a relatively small number of analytical methods are employed to obtain the main results. For clarity and reusability, we summarise these methods here in somewhat general terms, that are largely abstracted from the specific cases in which they are applied in the main text. This structure allows the same techniques to be readily applied to different scenarios, and facilitates cross-referencing between methods and the corresponding results. Detailed calculations and derivations are provided in the Supplementary Material.

\subsection{Static environment equilibrium}

Under biologically plausible conditions—specifically, when the fitness function $W(g)$ asymptotically vanishes  at large $|g|$ and the mutational distribution $f(g)$ has a finite variance—Eq.~\eqref{static differencems} admits a unique equilibrium solution, denoted $\Psi_{0}(g)$. This is independent of the initial distribution $\Psi(g,0)$. For a unimodal Gaussian fitness function, the explicit form of $\Psi_{0}(g)$ can be obtained analytically when the initial distribution is any Gaussian profile. For a bimodal fitness function with equivalent peaks, numerical investigation shows that there are parameter regimes where the equilibrium distribution of $g$, namely $\Psi_{0}(g)$, has a bimodal form. When the fitness function has two inequivalent peaks, the equilibrium distribution $\Psi_{0}(g)$ may still have a bimodal shape, as illustrated in the numerical example given in Section 6.3 of the Supplementary Material.

\subsection{Moving frame description}

Under constant environmental change at velocity $v$, Eq.~\eqref{runningdistms} describes the time evolution of the trait distribution. In this setting, the fitness function retains a fixed shape but shifts by an amount $v$ in trait space each generation, so that at time $t$ it takes the form $W(g - vt)$. To facilitate analysis, we introduce a `moving frame' that translates at the same velocity as the environment. The \textit{moving frame distribution}, written $\Phi(g,t)$,
is related to the original distribution, $\Psi(g,t)$, by the equation
\begin{equation}
    \Phi(g-vt,t)=\Psi(g,t).
\end{equation}
On substituting this form into Eq.~\eqref{static differencems} and making the substitutions $g \rightarrow g+v(t+1)$ and $h \rightarrow h+vt$ yields
\begin{equation}
    \Phi(g,t+1) = \frac{\displaystyle\int_{-\infty}^{\infty}f(g-h+v)W(h) \Phi(h,t)\,dh}
    {\displaystyle\int_{-\infty}^{\infty} W(h)\Phi(h,t) \, dh}.
    \label{Phi eq}
\end{equation}
This equation has a similar form to that of a static environment. In particular, the equation is now time homogeneous (apart from the time dependence of $\Phi(g,t)$, time does not enter elsewhere in the equation). The environmental velocity, $v$, enters the problem purely via the $v$ shift in the argument of the mutational distribution.

\subsection{Equilibration in the moving frame: leading to a rigidly moving distribution} \label{rigid}

Under the same biologically plausible conditions that guarantee the existence of a unique equilibrium in the static environment case (Eq. \eqref{static differencems}), the moving frame equation (Eq. \eqref{Phi eq}) leads to a distribution
that converges at long times to a time-independent distribution that we write as $\Phi_{\text{rigid}}(g)$. In terms of the original distribution $\Psi(g,t)$, we have that at long times $\Psi(g,t)$ approaches the distribution $\Phi_{\text{rigid}}(g-vt)$, i.e., a distribution of fixed shape that translates through trait space at the environmental velocity $v$. We refer to this as a \textit{rigidly moving solution}, and we 
shall refer to $\Phi_{\text{rigid}}(g)$ as its\textit{ profile}.

\subsection{Delta function limiting form}

We shall make use of the fact that
\begin{equation}
\lim_{s\rightarrow0_{+}}\frac{1}{\sqrt{2\pi s^{2}}}\exp\left(  -\frac{(g+b)^{2}   \label{limit delta}
}{2s^{2}}\right)  =\delta(g+b)
\end{equation}
where $\delta(x)$ denotes a Dirac delta function of argument $x$
\citep{hassani2009dirac}. Informally, $\delta(x)$ is a zero width, infinite
height spike, with unit area that is all concentrated where its argument vanishes, at $x=0$. 

This allows us to obtain an illustrative and exactly soluble case of the distribution of the trait for a bimodal fitness function. To see an example of this, we note that in Eq. \eqref{static differencems}, we can divide both occurrences of $W(g)$ (i.e., in numerator and denominator) by the same factor, without changing the distribution and its dynamics.
In Eq. \eqref{static differencems} we can thus use the fitness function 
\begin{equation}
    W(g)=\dfrac{1}{\sqrt{2\pi s^{2}}}\exp\left(-\dfrac{\left(g+a\right)^{2}}{2s^{2}}\right)+\dfrac{1}{\sqrt{2\pi s^{2}}}\exp\left(-\dfrac{\left(g-a\right)^{2}}{2s^{2}}\right).
    \label{bimodalfitnesssi}
\end{equation} 
In the limit $s\rightarrow0$ of the above form of $W(g)$, it  reduces to
\begin{equation}
W(g)=\delta(g+a)+\delta(g-a)\label{double delta}.%
\end{equation}
The fitness function in $W(g)$ in Eq. \eqref{double delta} consists of two arbitrarily sharp spikes that are located at $g=-a$ and $g=a$. In Section 6 of the Supplementary Material, we show how such a fitness function captures many essential properties of bimodal fitness functions and allows analytical derivations of numerous properties. While not fully realistic, biologically, this special case of the bimodal fitness function is illuminating because it leads to many exact results.

\subsection{Delta function approximation}
\label{delta function approx}
We shall also use an approximation that is related to the limiting form of a Gaussian distribution in Eq. (\ref{limit delta}). 
In particular, when we have the term $\exp\left(-\frac{g^{2}}{2s^{2}}\right)$ occurring within a $g$ integral, and we wish to determine the integral to leading order in the assumed small quantity $s$, we write $\exp\left(-\frac{g^{2}}{2s^{2}}\right)  =\sqrt{2\pi}s\times\left[\frac{\exp\left(-g^{2}/(2s^{2})\right)}{\sqrt{2\pi s^{2}}}\right]$ and approximate this by $\sqrt{2\pi}s\times\delta(g)$.

\section{Results}

\subsection{Static environment}

\subsubsection{Unimodal fitness function}

For the Gaussian fitness function plotted in Figure \ref{f1}, namely $W(g) = \exp\left(-g^{2}/(2s^2)\right)$, some exact results for dynamical and equilibrium quantities can be found. Specifically, in Section 5.1.5 of the Supplementary Material we show that the equilibrium distribution takes the form
\begin{equation}
    \Psi_{0}(g) = \sqrt{\frac{1}{2\pi\sigma_{0}^{2}}} \exp\left(-\frac{g^{2}}{2\sigma_{0}^{2}}\right).
\end{equation}
The equilibrium distribution of the trait has a mean,  written $\mu_{0}$, and a variance, written $\sigma_{0}^{2}$, that are given by
\begin{equation}
\mu_{0}=0\quad\text{ and }\quad\sigma_{0}^{2}=\dfrac{m^{2}+m\sqrt{m^{2}+4s^2}}{2},\label{sigma0 sq}
\end{equation}
respectively.
Furthermore,  the \textit{mean equilibrium fitness}, written $\overline
{W}_{0}$, is given by
\begin{equation}
\overline{W}_{0}=\displaystyle\int_{-\infty}^{\infty}W(g)\Psi_{0}(g)dg=\left(  1+\frac{\sigma_{0}^{2}}{s^2}\right)^{-1/2}.
\label{wbar0}
\end{equation}

\subsubsection{Bimodal fitness function}
Under static environmental conditions, a bimodal fitness function—such as the form plotted in Figure \ref{f2ms}—can lead to a bimodal equilibrium distribution of trait values, $\Psi_{0}(g)$. This can be the case even if the two fitness peaks are not precisely equivalent (see Section 6.3 of the Supplementary Material). While a sufficient separation of the two fitness peaks is necessary, this condition alone is not sufficient. A key additional requirement is \textit{mutational coupling}: mutations must occur at a frequency and scale such that some offspring, produced by a parent with trait value in the vicinity of one peak, will 
have a trait value in the vicinity of the other peak. We also note that bimodality can arise through mechanisms other than above. For example, frequency-dependent disruptive selection has been shown to generate bimodal trait distributions in certain settings \citep{kopp2006evolution, rettelbach2011effects}.

An equilibrium distribution of trait values that is bimodal can be interpreted as 
the \textit{coexistence, in a common environment}, of two distinct clones or morphs of 
the asexual organisms. In particular, the two clones will have
trait values that are centered around two different mean values (in Figure \ref{f2ms}
these mean values are in the vicinity of $\pm3$). We can interpret the two clones as representing
alternative strategies for surviving and reproducing in the same environment.
There are examples of different asexual clones of an organism coexisting in nature in a common environment (see, e.g., \citep{loaring1981ecological, carvalho1987clonal, zakrys1996genetic}), and mutational switching between different morphs of an asexual organism has been reported \citep{halme2004genetic}. 


As a special case, we consider the mean equilibrium fitness arising under a bimodal trait distribution when the width parameter $s$ of the fitness function is small $(s\ll m)$, such that the two fitness peaks are narrow and located at approximately $\pm a$.
Small $s$ may not be biologically accurate, but it serves a useful purpose of illustrating underlying theoretical results. 
In particular, we show in Section 6.1.4 of the Supplementary Material that the `double spike' fitness function in
Eq. \eqref{double delta} leads to an equilibrium distribution of the trait in a static environment that is given by
\begin{equation}
\Psi_{0}(g)=\dfrac{1}{2}\sqrt{\dfrac{1}{2\pi m^{2}}}\cdot\left[\exp\left(-\frac{(g+a)^{2}}{2m^{2}}\right)  +\exp\left(-\frac{(g-a)^{2}}{2m^{2}  }\right)\right].\label{Psi 0 s0}
\end{equation}
In this case, an approximation of the mean equilibrium fitness, that is valid to leading order in $s$, is given by
\begin{equation}
    \overline{W}_{0}=\int_{-\infty}^{\infty}W(g)\Psi_{0}(g)dg \approx \frac{s}{m}\left[1+\exp\left(-\dfrac{(2a)^{2}}{2m^{2}}\right)\right]\label{w0bimodal}
\end{equation}
as follows by application of the approximation in Section \ref{delta function approx}.

The behaviour of the form of the mean equilibrium fitness in Eq. (\ref{w0bimodal}) illustrates mutational coupling. 
When the fitness peaks are very far apart ($a/m$ large) most
mutant offspring of high fitness parents will have lower fitness.  However, when the peaks are not 
far apart ($a/m$ not large), some of the mutant offspring of such parents will lie near the other
fitness peak, and will not suffer such a fitness reduction. Thus we expect mean equilibrium fitness to
\textit{increase} when $a$ is \textit{decreased}, and this is a feature of the result in Eq. (\ref{w0bimodal}).

Generally, in a quantitative trait model of the type described, \textit{mutational coupling} is 
necessary for the continued coexistence of two or more clones, since a fitness asymmetry of the 
two peaks will, in the absence of mutations between the clones, 
lead to the extinction of the clone associated with a peak of lower fitness.

\subsection{Changing environment}



\subsubsection{Unimodal fitness function}

\begin{figure}[H]
\centering
\includegraphics[width=0.7\textwidth]{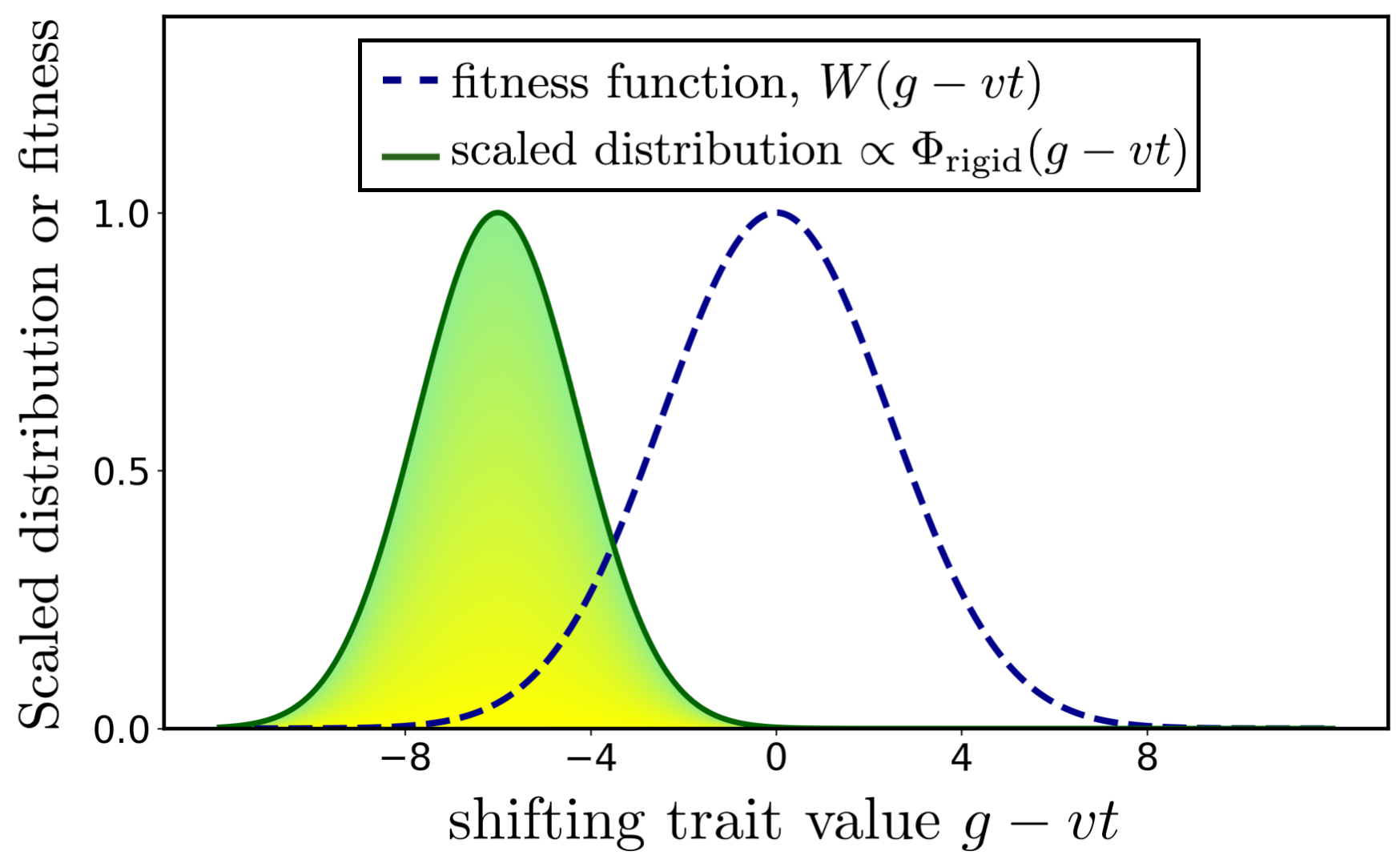}
\caption{ \textbf{Unimodal fitness function, and the corresponding rigidly moving genotypic distribution
in a uniformly changing environment.} In this figure we plot a unimodal fitness function, $W(g-vt)$,
against the shifting genotypic trait value $g-vt$ (dashed blue line). At time $t$, this fitness function 
has a peak  at $g=vt$, where $v$ is a constant - the `velocity' of environmental change.
The rigidly moving distribution, which depends on $g$ and $t$ only in the combination $g-vt$, is written as $\Phi_{\text{rigid}}(g-vt)$.
A scaled version of  $\Phi_{\text{rigid}}(g-vt)$, with a maximum height of unity, is also plotted against $g-vt$ (green line). 
To calculate the figure we used the mutational distribution of Eq. \eqref{f(g)ms} and the particular form 
of $W(g)$ plotted in Figure \ref{f1}. The parameter values adopted were $m=1$, $s=\sqrt{6}$
and $v=2$.}
\label{f3}
\end{figure}

Under stabilising selection, the fitness function is unimodal. Figure \ref{f3} illustrates 
a Gaussian form of such a fitness function, $W(g-vt)$, with an optimal genotypic value that uniformly moves with time. We also plot the \textit{rigidly moving distribution} (see Section \ref{rigid}), namely  
$\Phi_{\text{rigid}}(g-vt)$, against $g-vt$. 
A notable feature of Figure \ref{f3} is that for positive $v$
the rigidly moving  distribution has its peak to the left of the peak of the fitness function, $W(g-vt)$, 
so the distribution trails (lies `behind') the fitness function.

As in the static ($v=0$) case, when $W(g) = \exp\left(-g^{2}/(2s^2)\right)$, we can determine some exact results for various dynamical quantities. 
The rigidly moving solution has the form
\begin{equation}
\Phi_{\text{rigid}}(g-vt)=\sqrt{\dfrac{1}{2\pi\sigma_{0}^{2}}}\exp\left(-\dfrac{\left[(g-vt)+v\left(1+\frac{s^{2}}{\sigma_{0}^{2}}\right)\right]^{2}}{2\sigma_{0}^{2}}\right)
\end{equation}
where $\sigma_{0}^{2}$ is given in Eq. \eqref{sigma0 sq} (and is derived in Section 5.1.4 of the Supplementary Material).
In particular, the rigidly moving distribution, $\Phi_{\text{rigid}}(g-vt)$, has a mean at time $t$, 
written $\mu_{v}(t)$, and variance, written $\sigma_{v}^{2}$, that are given by
\begin{equation}
\mu_{v}(t)=vt-v\left(  1+\frac{s^2}{\sigma_{0}^{2}}\right)\quad\text{ and }\quad\sigma
_{v}^{2}\equiv\sigma_{0}^{2}=\frac{m^{2}+m\sqrt{m^{2}+4s^2}}{2}, \label{TW resultsms}
\end{equation}
respectively.

The mean fitness of the rigidly moving solution, written
$\overline{W}_{v}$, is defined by 
\begin{equation}
    \overline{W}_{v}=\displaystyle\int
_{-\infty}^{\infty}W(g-vt)\Phi_{\text{rigid}}(g-vt)dg,
\end{equation}
and the exact result is
\begin{equation}
\overline{W}_{v}=\left(  1+\frac{\sigma_{0}^{2}}{s^2}\right)  ^{-1/2}
\times\exp\left(  -\frac{v^{2}(s^2+\sigma_{0}^{2})}{2\sigma_{0}^{4}}\right)
\equiv\overline{W}_{0}\times\exp\left(  -\frac{v^{2}(s^2+\sigma_{0}^{2}
)}{2\sigma_{0}^{4}}\right). \label{wbarv}
\end{equation}
This expression reveals that the mean fitness, $\overline{W}_{v}$, decreases exponentially with the square of the environmental velocity, $v$. Such exponential decay underscores the increasingly severe fitness cost associated with faster environmental velocities, reflecting the growing mismatch between the population's trait distribution and the moving fitness optimum—this fitness cost is commonly referred to as the \textit{lag load} \citep{smith1976determines, chevin2013genetic}.

Providing the velocity of environmental change, $v$, does not produce excessively
large genetic loads \citep{whitlock2011genetic}, 
the above results indicate that a large asexual population, subject to a unimodal 
fitness function can persistently adapt to the changing environment.

\subsubsection{Bimodal fitness function}

\begin{figure}[H]
    \centering
    \begin{minipage}{0.5\textwidth}
        \centering
        \includegraphics[height=0.25\textheight]{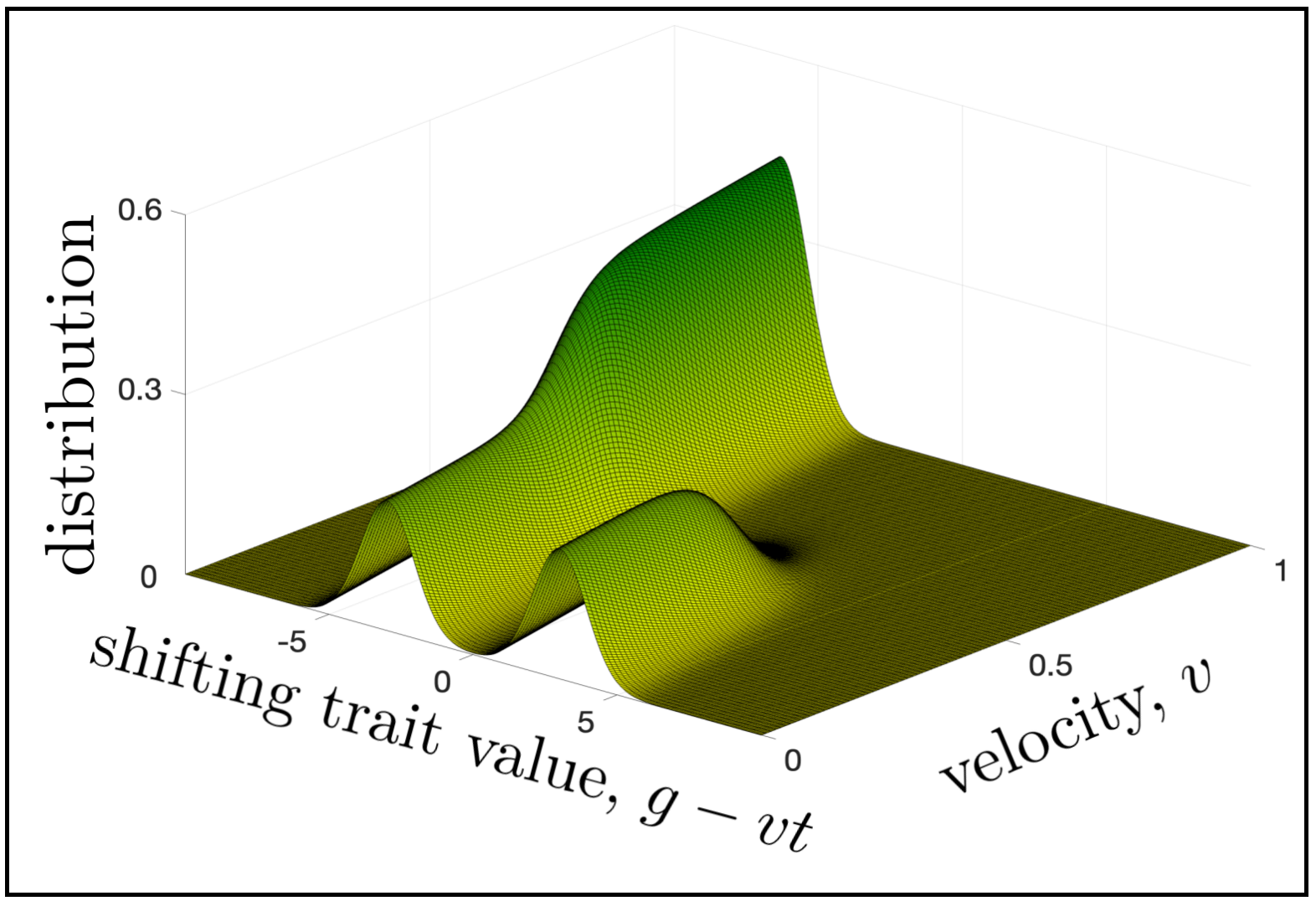}
        \caption*{A. Genotypic distribution}
    \end{minipage}\hfill
    \begin{minipage}{0.5\textwidth}
        \centering
        \includegraphics[height=0.25\textheight]{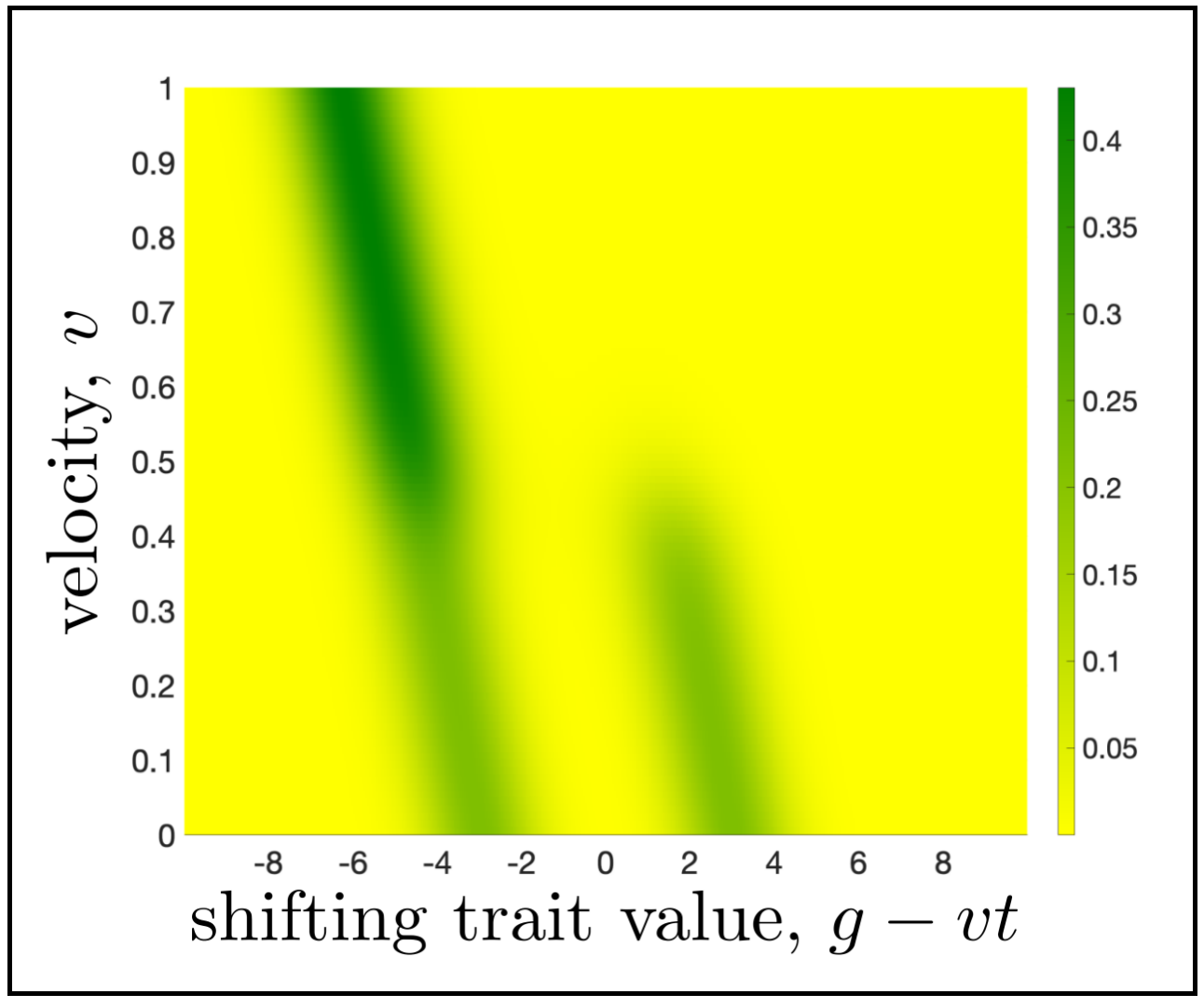}
        \caption*{B. Heat map}
        \label{fig:figure2}
    \end{minipage}
    \caption{{\textbf{Environmental velocity dependence of the rigidly moving distribution.} 
    \newline In Figure \ref{f2ms} we previously plotted a bimodal fitness function, $W(g)$, which has
peaks in the vicinity of $\pm 3$. For Figure \ref{f4ms} we used a fitness function with an identical shape
and parameters as this fitness function, with the exception that it is `moving' at 
constant velocity $v$, and which hence has the form $W(g-vt)$. 
For this figure the mutation distribution used was identical to that used for Figure \ref{f2ms}. \newline
\noindent \textbf{Panel A} contains a plot of the rigidly moving distribution of genotypic values, $\Phi_{\text{rigid}}(g-vt)$ against: (i) the shifting genotypic trait value $g-vt$, and (ii) the velocity $v$.  For small $v$ ($v \lesssim 0.4$) the distribution is bimodal, but for larger $v$ ($v \gtrsim 0.5$) the distribution is unimodal.\newline
\textbf{Panel B} shows a \textit{heat map}—effectively a top-down view of Panel A—depicting the distribution $\Phi_{\text{rigid}}(g - vt)$ as a function of the shifting trait value $g - vt$ and the environmental velocity $v$. 
The color intensity encodes the value of $\Phi_{\text{rigid}}(g-vt)$: darker (greener) regions correspond to higher probability density, thus allowing direct visualisation of the heights of the peaks.
For small $v$ ($v\lesssim 0.4$) two parallel green bars are visible, indicating the two peaks of a bimodal distribution. 
The two green bars are not vertical, indicating that the velocity has some effect on the location of the two peaks. 
For larger $v$ ($v \gtrsim 0.5$) only a single green bar is visible, indicating 
the single peak of a unimodal distribution. Generally, the distribution of trait values requires a topographical representation,
such as that in Panel B, since its behaviour cannot be captured by just its mean and variance.}
}
\label{f4ms}
\end{figure}

For a bimodal fitness function arising from locally disruptive selection, we shall work in a parameter regime where the $v=0$ equilibrium distribution, $\Psi_{0}(g)$, like the fitness 
function itself,  is bimodal in form. We then have numerical and analytical results that
indicate that the rigidly moving distribution that develops when $v \ne 0$ exhibits two qualitatively different behaviours (see Section 6.1 of the Supplementary Material):
\begin{enumerate}[label=(\roman*)]
\item For a range of low environmental velocities,
starting with $v=0$, the distribution has a \textit{bimodal} form. 
\item At larger environmental velocities, the rigidly moving  distribution is not bimodal in form 
but \textit{unimodal}. 
\end{enumerate}
These behaviours are illustrated in Figure \ref{f4ms}.

From Figure \ref{f4ms}, it is evident that there are a range of values of the environmental velocity, $v$, where a 
rigidly moving \textit{bimodal distribution} of genotypic values can persist over time ($v \lesssim 0.4$). 
However, at sufficiently large values of $v$  ($v \gtrsim 0.5$) only one peak in the distribution 
can be sustained over time, and the other peak is absent. 
From the form of the rigidly moving distribution in Figure \ref{f4ms}, it is very natural to say the single peak, 
that becomes established at higher velocities, develops from the left hand peak of the distribution and 
the right hand peak \textit{disappears}.

Given the interpretation of the two 
peaks in the distribution, when present, as corresponding to two distinct clones that can coexist in a common 
environment, it follows that the disappearance of one of these peaks signals the population 
\textit{losing biodiversity}, with only a single clone remaining.  This loss is evidently driven by the environment 
changing too rapidly.  

Finally, we theoretically investigate the dynamics under the bimodal fitness function $W(g-vt)$ where $W(g)$ is the particular `double spike' form in Eq. \eqref{double delta}. In Section 6.1.2 of the Supplementary Material, we prove that under this fitness function, \textit{any} initial distribution will ultimately reach a rigidly moving distribution $\Phi_{\text{rigid}}(g-vt)$, where $\Phi_{\text{rigid}}(g)$ can be written as
\begin{equation}
\Phi_{\text{rigid}}(g)=\dfrac{e^{av/m^{2}}f(g+v+a)+e^{-av/m^{2}}f(g+v-a)}{2\cosh\left(
\dfrac{av}{m^{2}}\right)}.
\end{equation}
We also provide exact expressions for the mean and variance of this distribution, written $\mu_{v}(t)$ and $\sigma _{v}^{2}$, respectively, which are given below, in Eq. \eqref{mu_v, sigma_vms}, when we discuss surrounding issues.

From numerical findings (see Figure \ref{f4ms}) we observe that the rigidly moving distribution of $g$ undergoes a crossover from bimodal to unimodal form, as the environmental rate of change, $v$, is increased. 
In the parameter regime $s\to 0$ and $a>m$, 
in a static environment, the equilibrium distribution is bimodal (see Section 6.1.4 of the Supplementary Material). 
When considering a constantly changing environment in the same parameter regime, the exact \textit{critical environmental velocity} at which one peak disappears is
\begin{equation}
v_{c}=\sqrt{a^{2}-m^{2}}-\dfrac{m^{2}}{a}\arcosh\left(  \dfrac{a}{m}\right).\label{vcsii}%
\end{equation}
For $v<v_{c}$ the rigidly moving distribution is bimodal, whereas for $v\geq v_{c}$, the right peak is absent and the distribution is unimodal.
The most interesting aspect of this is that such a critical velocity exists (rather than its specific form, which is given in Eq. \eqref{vcsii}), and that this specific example analytically captures the phenomenon of the vanishing peak shown in Figure \ref{f4ms}.

In Section 6.1.4 of the Supplementary Material we show that the position of this
peak, immediately prior to its disappearance, is given by $g-vt$ having the value
\begin{equation}
g-vt=\dfrac{m^{2}}{a}\arcosh\left( \dfrac{a}{m}\right).
\end{equation}
We can also compute the mean fitness $\overline{W}_{v}$ for a rigidly 
moving distribution. The mean fitness in this case is
\begin{equation}
    \overline{W}_{v}=\int_{-\infty}^{\infty}W(g-vt)\Phi_{\text{rigid}}(g-vt)dg
\end{equation}
and we now consider the special illustrative case where the `width parameter' $s$ of the bimodal fitness function is 
small ($s \ll m$) but not zero. Then both peaks of the fitness function have narrow widths, and are approximately $2a$ apart. Paralleling the derivation of Eq. (\ref{w0bimodal}), the mean fitness of 
an adapting population under a constantly moving bimodal fitness function is given by
\begin{equation}
    \overline{W}_{v}\approx \frac{s}{m}\left[1+\exp\left(-\dfrac{(2a)^{2}}{2m^{2}}\right)\right]\exp\left(-\dfrac{v^{2}}{2m^{2}}\right)=\overline{W}_{0}\cdot\exp\left(-\dfrac{v^{2}}{2m^{2}}\right)  \label{Wv bi approx}
\end{equation}
where $\overline{W}_{0}$ is given in Eq. \eqref{w0bimodal}.

The result in Eq. \eqref{Wv bi approx} highlights how mean fitness depends principally on the ratios $a/m$ and $v/m$. A larger value of $a/m$ reflects a wider separation between the fitness peaks relative to the mutation scale, reducing mutational coupling and thus lowering the mean fitness $\overline{W}_{v}$. Similarly, a larger value of $v/m$ implies a faster-moving environment relative to the mutational step size, and results in more rapid exponential decay of fitness. The prefactor $s/m$, which arises from our asymptotic analysis in the regime $s \to 0$, plays a secondary role and acts primarily as a scaling factor.


We emphasize that the $s\to 0$ limit we have used here is not a biologically motivated assumption,
but rather a theoretical device that allows us to expose limiting behaviours and obtain analytically tractable results. This limit is thus for illustrative purposes. Numerical results, such as those leading to Figure \ref{f4ms}, were conducted using \textit{finite values} of $s$, corresponding to a moderate level of selection. Thus the change in the form seen in the distribution of the trait, under a
moving bimodal fitness function, exhibits a level of robustness to the strength of selection, as measured by $1/s$.

\subsubsection{Mechanism of the transition}
It is natural to ask about the detailed mechanism that drives the loss of a peak/biodiversity that is seen in Figure \ref{f4ms}.

Previously, we have considered a bimodal fitness function of the form given in Eq. \eqref{bi Gaussms} and which
was plotted
in Figure \ref{f2ms}. To gain some insight into what occurs, we again consider such a fitness function in the 
limit of arbitrarily narrow peak widths, i.e., we again take the limit $s \rightarrow 0$.
This corresponds to a fitness function with two extremely narrow peaks that are located at $\pm a$. 

In the Section 6.1 of the Supplementary Material, we analytically demonstrate that the corresponding distribution of the trait will evolve over time towards 
a rigidly moving form that only depends on $g$ and $t$ in the combination $g-vt$, written $\Phi_{\text{rigid}}(g-vt)$. 
We find that the exact form of mean, $\mu_{v}(t)$, and the variance, $\sigma_{v}^{2}$, are given by
\begin{equation}
\mu_{v}(t)=vt-v-a\tanh\left(\frac{av}{m^{2}}\right)\quad\text{ and }\quad\sigma_{v}^{2}=m^{2}+a^{2}-a^{2}\tanh^{2}\left(\frac{av}{m^{2}}\right)  \label{mu_v, sigma_vms}
\end{equation}
respectively (see Section 6.1.4 of the Supplementary Material). The term $vt$ in $\mu_{v}(t)$ corresponds to the rigidly moving solution constantly 
`following' 
the constantly moving fitness function, and hence the mean value uniformly changes over time. It is also worth emphasizing that the functional form of $\mu_{v}(t)$ indicates that, as the environmental velocity $v$ increases, the population mean shifts toward the `left' (trailing) optimum, reflecting a reduction in the weight associated with the right peak.

Although the distribution converges to a rigidly moving form for large $t$, 
different rates of environmental change, $v$, result in distinct distribution characteristics — maintaining either a bimodal or unimodal shape. 
Thus even in the simplifying regime of vanishingly small $s$, 
this key feature, which is visible in Figure \ref{f4ms}, occurs. 
This is hinted at by the behaviour of the variance, $\sigma_{v}^{2}$, in Eq. \eqref{mu_v, sigma_vms}. For sufficiently
large $v$ ($v \gtrsim 3m^{2}/a$) it can be seen that the variance in Eq. \eqref{mu_v, sigma_vms}
is close to $m^2$, i.e., effectively independent of $a$ and hence independent of the positions of the two peaks 
of the fitness function, which lie at $\pm a$. Thus as far as this statistic is concerned, 
in this large $v$ regime, the distribution of the trait has no information about bimodality 
of the fitness function, and hence the distribution will not reflect such a property of the fitness function.


\subsubsection{Intuition about the transition}
Here, we give intuitive arguments for the occurrence of a transition in the form of the rigidly 
moving solution, when the environmental velocity, $v$, is increased. 

We argue that at sufficiently large $v$ an entire class of mutations become
significantly suppressed to the extent that they decouple from the problem. 
These mutations are in offspring whose trait values lie
in the vicinity of the \textit{right peak} of the distribution, but were
produced by parents with trait values in the vicinity of the \textit{left peak}
(mutations going in the opposite direction are less affected).

Verifying this involves demonstrating how environmental change alters the balance between mutation and selection, leading to an asymmetric outcome—specifically, the eventual disappearance of one of the peaks in the trait distribution.

In the arguments we give, we shall omit
inessential factors, integrals, and other mathematical niceties. Throughout, we shall imagine a 
fitness function, $W(g)$, that has
a bimodal form, such as that in Eq. \eqref{bi Gaussms}, and which has peaks at
$-a$ and $a$. 
In what follows, we shall see that the dominant contributions to the population distribution 
plausibly arise from trait values in the vicinity of these optimal trait values \citep{forien2022ancestral}.
Furthermore, with a $p$ or $o$ subscript labelling a quantity of a parent or offspring, respectively, we shall use the result that when a parent with trait value $g_{p}$  
produces an offspring with trait value $g_{o}$, 
the probability of this occurring is proportional to
$f(g_{o}-g_{p})$ where $f(g)\equiv f(-g)$ is the distribution of mutant effects of the trait.

First we consider the case of a \textit{static environment} (i.e., with $v=0$)

\begin{figure}[H]
\centering

\begin{tikzpicture}[scale=3, every node/.style={font=\large}]
\tikzset{>={Stealth[scale=1.5]}}

\node at (0.25, 1.65) {$g_{p} \sim -a$}; 
\node at (2.75, 1.65) {$g_{p} \sim a$}; 
\node at (0.25, 0) {$g_{p} \sim -a$};   
\node at (2.75, 0) {$g_{p} \sim a$};   

\node at (0.75, 1.05) {$(ii)$};   
\node at (2.25, 1.05) {$(iii)$};   

\draw[->, ultra thick] (0.2, 1.5) -- (0.2, 0.2) node[midway, left] {$(i)$}; 
\draw[->, ultra thick] (0.2, 1.5) -- (2.8, 0.2) node[midway, below right] {}; 
\draw[->, ultra thick] (2.8, 1.5) -- (2.8, 0.2) node[midway, right] {$(iv)$}; 
\draw[->, ultra thick] (2.8, 1.5) -- (0.2, 0.2) node[midway, above left] {}; 
\end{tikzpicture}
\caption{\textbf{Four mutation types in a static environment.} 
The probability of a parent with trait value $g_{p}$ producing a mutant offspring with trait value $g_{p}$ 
is proportional to $f(g_{p}-g_{p})$. For the four types of mutations shown, for a static environment where 
the fitness optima are at $\pm a$:\\
type (i) from $g_{p}=-a$ to $g_{p}=-a$, have a probability proportional to  $f(0)$;\\ 
type (ii) from $g_{p}=-a$ to  $g_{p}=a$, have a probability proportional to $f(2a)$; \\
type (iii) from $g_{p}=a$ to $ g_{p}=-a$, have a probability proportional to $f(-2a)$; \\
type (iv) from $ g_{p}=a$ to $g_{p}=a$, have a probability proportional to  $f(0)$.}
\label{f5ms}
\end{figure}
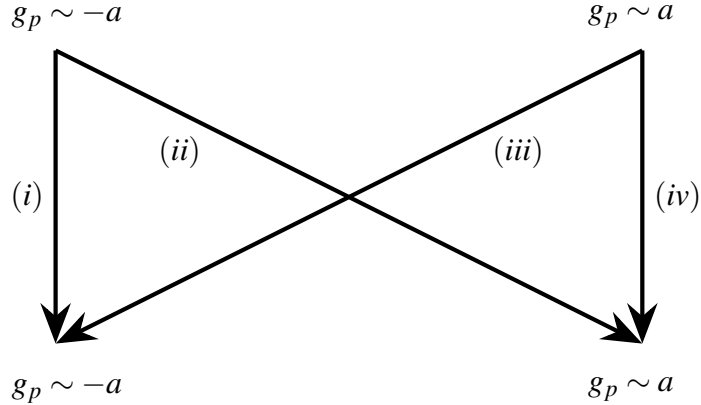

In this case, fitness maximally selects for trait values at $g=-a$ and $g=a$. We shall consider
the mutations that occur when the population is at equilibrium, assuming the distribution of
trait values is peaked in the vicinity of $-a$ and $a$. Then there are
four types of mutations that occur. These are summarised in Figure \ref{f5ms}.

\bigskip

Note in particular that type (ii) and type (iii) mutations
have the same probability, because  $f(g)=f(-g)$.

\bigskip

We now consider the case of a \textit{changing environment} (i.e., with $v>0$).

The form of the fitness in generation $t$ is given by $W(g-vt)$, 
and hence maximally selects for
both $g-vt=-a$ and $g-vt=a$. We shall consider the mutations that occur when the
population is described by a rigidly moving distribution that, in generation
$t$, is of the form $\Phi_{\text{rigid}}(g-vt)$. We assume this distribution is peaked in
the vicinity of both $g-vt=-a$ and $g-vt=a$. In
generation $t+1$, i.e., the next generation, the rigidly moving distribution will be
peaked in the vicinity of both $g-v(t+1)=-a$ and $g-v(t+1)=a$.  
Since it is the
occurrence of mutations that produces these new values we now have the mutation
types summarised in  Figure \ref{f6ms}.

\bigskip

\begin{figure}[H]
\centering
\begin{tikzpicture}[scale=3, every node/.style={font=\large}]
\tikzset{>={Stealth[scale=1.5]}}

\node at (0.25, 1.65) {$g_{p} \sim vt-a$}; 
\node at (2.75, 1.65) {$g_{p} \sim vt+a$}; 
\node at (0.25, 0) {$g_{p} \sim v(t+1)-a$};   
\node at (2.75, 0) {$g_{p} \sim v(t+1)+a$};   

\node at (0.75, 1.05) {$(ii)$};   
\node at (2.25, 1.05) {$(iii)$};   

\draw[->, ultra thick] (0.2, 1.5) -- (0.2, 0.2) node[midway, left] {$(i)$}; 
\draw[->, ultra thick] (0.2, 1.5) -- (2.8, 0.2) node[midway, below right] {}; 
\draw[->, ultra thick] (2.8, 1.5) -- (2.8, 0.2) node[midway, right] {$(iv)$}; 
\draw[->, ultra thick] (2.8, 1.5) -- (0.2, 0.2) node[midway, above left] {}; 
\end{tikzpicture}

\caption{\textbf{Four mutation types in a changing environment.} 
This figure is the version of Figure \ref{f5ms}, for a changing environment, 
where in generation $t$ the fitness optima lie at $vt \pm a$. 
For the four types of mutations shown: \\
type (i) from $g_{p}=vt-a$ to $g_{p}=v\left(t+1\right)-a$, have a probability proportional to  $f(v)$;\\
type (ii) from $g_{p}=vt-a$ to $g_{p}=v\left(t+1\right)+a$ , have a probability proportional to $f(v+2a)$; \\ 
type (iii) from $g_{p}=vt+a$ to $g_{p}=v\left(t+1\right)-a$, have a probability proportional to $f(v-2a)$;\\ 
type (iv) from $ g_{p}=vt+a$ to $g_{p}=v\left(t+1\right)+a$, have a probability proportional to  $f(v)$.}

\label{f6ms}
\end{figure}


Here we see that mutations of type (ii), from parents at the `left peak’ of the rigidly moving distribution to offspring contributing to the `right peak’, are significantly suppressed, with a probability proportional to $f(v+2a)$, which is typically much smaller than $f(v)$ or $f(v-2a)$. In contrast, type (iii) mutations, which move from the right to the left peak, have a higher probability proportional to $f(v-2a)$ and are therefore less suppressed. This asymmetry in transition probabilities arises because type (ii) mutations require a large trait shift in the same direction as the environmental change, leading to a larger mutational displacement, and hence a lower probability under the mutation distribution in this study. Consequently, the right peak receives fewer incoming mutations compared to the left, and this imbalance in replenishment becomes more pronounced as the environmental velocity $v$ increases. For sufficiently large $v$, this suppression can destabilize the right peak entirely, leading to its disappearance in the equilibrium distribution.

\section{Discussion}
In this work we have considered a population of asexual organisms that are primarily characterised by a single quantitative trait.  
This trait is subject to mutation, and the fitness of all individuals with trait value $g$ is controlled by the value of $g$. Environmental change manifests itself by causing the fitness function to change with time. We focussed on persistent environmental change, aiming to capture changes that 
continue to occur over time, such as the environmental changes that have origins in anthropogenic activities \citep{reilly2000irreversibility, hasselmann2003challenge, solomon2010persistence}.  


The novel fitness function we considered in this work is 
locally disruptive, i.e., in the vicinity of
$g=0$, but at large $|g|$ the fitness vanishes. The simplest fitness function with these properties
is double peaked (or bimodal) in form. 

We considered the fitness function to have all features (such as peaks) occurring at progressively larger trait values, each generation.
Such `moving' fitness functions are plausible if fitness is related to a chemical or temperature of the 
environment that, e.g., increases over time.

We restricted considerations to cases where, in the presence of a static bimodal 
fitness function, the equilibrium distribution is itself bimodal. It is then natural to interpret the two peaks 
in the distribution as representing two clones of the asexual organism, i.e., two distinct groups, where 
members of a group have similar genotypic trait values. The two clones represent two alternative ways, or strategies, 
to survive and reproduce in an environment, and practitioners of these two strategies can coexist. 
This coexistence generally relies on some of the mutant offspring of each clone being members
of the other clone. Under a constantly changing environment, the distribution 
describing the trait settles down to have a rigidly moving form, but exhibits some level of
complexity. The rigidly moving distribution has a bimodal form for low rates of change of the 
environment, and a unimodal form for higher values, with a somewhat narrow crossover region. 

While some of the `small $s$' illustrative results we have given above for a `moving' bimodal fitness function may be described as `strong selection', this is not a fully accurate statement. For a unimodal fitness function, strong selection means that after selection, there is very little variation in the population. With the bimodal fitness function of the present work, even with $s\rightarrow 0$, there is still appreciable variation in the population (because two distinct types of individual remain present after selection), which is a feature that makes it unlike strong selection.

We attribute the qualitative change in the shape of the distribution to result from the demands put on the 
population, by the changing environment, to adapt. When the rate at which the environment changes is relatively low
or zero, there is a balance between different types of mutations and selection, that leads to a bimodal 
distribution. When the rate of environmental change is sufficiently high, an asymmetry enters. A class of mutations, 
that are needed for both adaptation and coexistence of the two clones, have to have large effects,
but their probability of occurrence is low. We have described this as an effective decoupling of mutations, 
and this  upsets the balance in the population required for a bimodal distribution. The loss of one peak
is the outcome, and a unimodal distribution emerges. 


As an example, apart from persistent environmental change, as we have considered, the environment may also undergo discrete changes \citep{kopp2014rapid, kopp2007adaptation}—referred to here as sudden environmental jumps—manifested as extreme events that occur abruptly in a given generation \citep{otto2023attribution, stott2016climate}. In response to such a jump, our results suggest that the population will, given sufficient time, fully adapt to the new environmental conditions and ultimately attain a new equilibrium. In this case, the shape of the trait distribution, after a long time, will be a translated (moved) version of the equilibrium distribution that applied before occurrence of an environmental  jump. Importantly, if this equilibrium distribution is bimodal, then the distribution that is ultimately attained will also be bimodal. Thus, environmental change, occurring as a single extreme event, does not diminish the population's \textit{ultimate} biodiversity. We present analyses of this scenario in the Supplementary Material: for a unimodal fitness function, see Section 5.2 of the Supplementary Material, while for a bimodal fitness function see Section 6.2.  We note, however, that the mutational decoupling mechanism 
we have introduced in this work may also have a manifestation here. After an event, the distribution over the short term will depend on the magnitude of the event. A large event may disrupt the population sufficiently that the two peaks in the distribution become mutationally uncoupled. Then initially one clone will go extinct. It will then take time for the distribution to recover a bimodal form, and hence its original level of biodiversity. 

A natural next step is to examine how populations respond under multiple successive environmental jumps, rather than a single event. In particular, one may ask whether the population can preserve bimodality when the environment undergoes repeated abrupt shifts. Preliminary numerical experiments suggest that the ability to retain a bimodal trait distribution depends sensitively on both the frequency and magnitude of these jumps. If environmental shifts occur too frequently, or if each jump is too large, then the population may not have sufficient time to recover equilibrium between events, leading to a collapse into unimodality. These observations point to the existence of critical thresholds in jump frequency and size, beyond which long-term biodiversity cannot be sustained. A more systematic investigation of this regime, including quantification of these thresholds, is a possible direction for future work.

There remains the issue how the population actually adapts to a changing environment.
For an asexual population that is subject to a \textit{unimodal} fitness function, we have 
recursion relations for the mean and the variance of the trait value
(see Section 5.1.1 of the Supplementary Material) that give an idea of how these two primary 
statistics, and hence the distribution, adapt from a somewhat arbitrary starting point, 
but for this case, adaptation can also be considered at the 
\textit{lineage level} \citep{forien2022ancestral}. When, however, the fitness 
function is \textit{bimodal}, the present work has exposed a richer set of phenomena. 
We have seen mutational coupling (or `cross talk') between the peaks
in the equilibrium distribution in a static environment, and when the environment
changes, the balance between mutation and selection is modified. The result is that
adaptation is different under different circumstances. Persistent adaptation at low rates
of constant environmental change allows continued bimodality of the trait distribution, but
in a modified form. Persistent adaptation at higher rates of constant environmental change still
occurs, but it only allows the presence of a single peak in the trait distribution.
Similar behaviour is observed in models with frequency-dependent selection, such as that of \cite{johansson2008evolutionary}, where under slow environmental change two competitors can coexist, but under rapid change, one of them is lost.
With environmental `jumps' the different timescales determine the nature of the adaption
that can occur, for example, sufficiently rapid repeating jumps may mimic the effects of
a relatively rapid rate of constant environmental change, and will not lead back to an equilibrium
distribution between jumps.

The calculations we have presented in this work apply to large (effectively infinite) populations. 
A natural question to ask is whether the main findings hold in the case of \textit{finite populations}. 
At the time of writing we only have preliminary results on this, from simulations. For populations of finite size, 
such as $10^3$, we find, like an infinite population, that there are low rates of environmental change that lead to a bimodal distribution of trait values, and faster rates that lead to a unimodal distribution. We plan to report on this elsewhere.

To summarise, in the present work, we primarily consider a large asexual population that is
subject to selection on a trait, where there are `two best values'. Under static environmental
conditions we consider situations where both best values of the trait can be present in the population, in the form of a bimodal equilibrium distribution. Under constant environmental change, the rate of change determines whether the 
distribution of the trait can maintain two peaks over time, or just one peak. This has direct 
implications for the biodiversity of a population. The loss of a peak of the distribution, due to sufficiently rapid
environmental change, arises from a novel mechanism of mutational decoupling. 
We believe this mechanism may have manifestations in other evolutionary scenarios.

\bigskip

\noindent \textbf{\large CRediT authorship contribution statement} 

Both authors (Ruixi Huang and David Waxman) made equal contributions to conceptualization, formal analysis, investigation, methodology, visualization, writing-original draft and writing-review \& editing.

\medskip

\noindent \textbf{{\large Funding}} 

Not applicable.

\medskip

\noindent \textbf{{\large Declaration of competing interest}} 

We declare we have no competing interests.

\medskip

\noindent \textbf{{\large Acknowledgments}} 

We thank Mingbo Yin for very helpful conversations and comments on the manuscript.

\medskip

\noindent \textbf{\large Supplementary Material} 

The Supplementary Material for this paper is divided into two sections: Core background and Detailed calculations. It is accompanied by Supplementary References (1 - 4).

\medskip

\noindent \textbf{{\large Data availability}} 

This study is theoretical and does not use empirical data. The code for analysing evolutionary dynamics is available at:
\href{https://github.com/TRxHuang/Moving-Optimum-Asexual}{https://github.com/TRxHuang/Moving-Optimum-Asexual}.
This repository includes all scripts necessary to reproduce the results and create animations illustrating the evolutionary process over time.

\newpage
\bibliographystyle{apalike}
\bibliography{MSRef}
\end{document}


\newtheorem{notation}{Notation}[section]
\newtheorem{definition}{Definition}[section]
\newtheorem{remark}{Remark}[section]
\newtheorem{lemma}{Lemma}[section]
\newtheorem{claim}{Claim}[section]
\newtheorem{example}{Example}[section]
\newtheorem{proposition}{Proposition}[section]
\newtheorem{property}{Property}[section]
\newtheorem{problem}{Problem}[section]

\cleardoublepage
\setcounter{page}{1}
\renewcommand{\thepage}{S\arabic{page}}

\setcounter{figure}{0}
\renewcommand{\thefigure}{S\arabic{figure}}

\setcounter{section}{0}
\numberwithin{equation}{section}
\renewcommand{\theequation}{S\thesection.\arabic{equation}}

\addtocontents{toc}{\protect\setcounter{tocdepth}{3}}

\begin{center}
    \LARGE
    \textbf{SUPPLEMENTARY MATERIAL} \\
    \vspace{0.6cm}

    \Large
    \textbf{Effective decoupling of mutations and the resulting loss \\ of biodiversity caused by environmental change}
        
    \vspace{0.6cm}
    
    \large
    Ruixi Huang\textsuperscript{1,2} and  David Waxman\textsuperscript{1} \\
    {\textsuperscript{1}Centre for Computational Systems Biology, ISTBI, Fudan University, \\220 Handan Road, Shanghai 200433, PRC} \\
    {\textsuperscript{2}Research Institute of Intelligent Complex Systems, Fudan University, \\220 Handan Road, Shanghai 200433, PRC}
\end{center}

\vspace{0.6cm} In this work, we investigate the evolutionary dynamics of a large
population of asexual organisms subject to selection on a mutable quantitative
trait. This Supplementary Material has two parts. In Part \ref{part1} we give core background of the paper
while Part \ref{part2} contains detailed calculations that lead 
to the principal results of the main text.

\begin{center}
	\renewcommand{\contentsname}{\color{black}\Large\bfseries Contents} 
	\tableofcontents
\end{center}

\hypersetup{linkcolor=black}

\newpage

\part[Core background]{Part I: Core background}

\label{part1}

\section{Population}

\label{p1s1}

We consider the evolutionary dynamics of a very large population of asexual
organisms. The population is subject to selection on a quantitative trait that
is also susceptible to mutation. The population evolves in discrete
generations, labelled by $t=0,1,2,\cdots$. Individuals are taken to
be characterised by a single quantitative trait, where a possible genotypic value
is denoted by $g$, with $-\infty<g<\infty$. We consider the following life
cycle, with census made in offspring, immediately after birth.\bigskip

\noindent(i) offspring $\rightarrow$ juveniles.

\noindent Offspring undergo viability selection, and those that survive are termed juveniles.

\noindent In principle, there is a `number thinning' stage, that reduces the
population back to a regulated size, with the remaining individuals termed
adults. For a very large population, as we assume, thinning has no effect on
the distribution of trait values.

\noindent(ii) juveniles $\rightarrow$ adults.

\noindent The individuals that survive viability selection \textit{and} thinning are
termed adults. These reproduce asexually and die shortly afterwards, with
mutations taken to occur at the production of offspring.\bigskip

Selection acts on phenotype, which we assume is additively determined by the
sum of genotypic value and the environmental noise. The quantitative trait
under selection is controlled by $L$ haploid loci, with each allele
contributing in a purely additive manner to the trait.

With $z$ the phenotypic value, $g$ the genotypic value and $\varepsilon$
environmental effect, we have
\begin{equation}
z=g+\varepsilon=\sum_{i=1}^{L}x_{i}+\varepsilon.
\end{equation}
where $x_{i}$ is the effect of the allele at the $i^{\prime}$th locus
controlling the trait (loci are labelled in an arbitrary manner). Following
Kimura, each $x_{i}$ is taken to be continuous
\cite{kimura1964number, kimura1965stochastic}
\begin{equation}
-\infty<x_{i}<\infty,\qquad i=1,2,\cdots,L.
\end{equation}
and corresponds to an effectively infinite number of alleles at a locus.

Fertility is assumed to be independent of phenotype, and an individual's
viability (hence fitness) is determined by their phenotypic value, $z$. For
organisms with phenotypic value $z$ the fitness in generation $t$ is written
as $W_{p}(z,t)$. The fitness of organisms with genotypic value $g$ in
generation $t$, written $W(g,t)$, is obtained by averaging $W_{p}%
(g+\varepsilon,t)$ over all values of the environmental effect, $\varepsilon$.
We take the environmental effect to be a random variable that is independent
of $g$ and is normally distributed with
mean zero and variance $V_{E}$, thus
\begin{equation}
W(g,t)=\sqrt{\dfrac{1}{2\pi V_{E}}}{\bigintsss_{-\infty}^{\infty}}%
W_{p}(g+\varepsilon,t)\exp\left(  -\dfrac{\varepsilon^{2}}{2V_{E}}\right)
d\varepsilon.
\end{equation}

For a parent with genotypic value $h$, the probability that any
offspring has a genotypic value in the range $g_{1}$ to $g_{2}$ is taken to be
$\displaystyle\int_{g_{1}}^{g_{2}}f(g-h)dg$, where $f(g)$ is termed the
\textit{distribution of mutant effects}. We use
\begin{equation}
f(g)=\sqrt{\frac{1}{2\pi m^{2}}}\exp{\left(-\frac{g^{2}}{2m^{2}}\right)
}\label{fgsiiiii}%
\end{equation}
and in Section \ref{subsection13} of the Supplementary Material we provide justification for
selecting $f(g)$ as a normal distribution with mean zero.

Let $\Psi(g,t)$ denote the \textit{distribution of genotypic
values} in offspring in generation $t$. The distribution, $\Psi(g,t)$, is a probability density. 
After selection the distribution of
genotypic values becomes
\begin{equation}
\dfrac{W(g,t)\Psi(g,t)}{\displaystyle\int_{-\infty}^{\infty}W(h,t)\Psi(h,t)dh}.
\end{equation}
Because we deal with an arbitrarily large population, no change occurs when
the population is regulated (thinned), and hence the same distribution applies to adults.
On reproduction, offspring are subject to mutation and in this way the
distribution in the next generation is%
\begin{equation}
\Psi(g,t+1)=\dfrac{\displaystyle\int_{-\infty}^{\infty} f(g-h)W(h)\Psi(h,t)dh}{\displaystyle\int_{-\infty
}^{\infty}W(h)\Psi(h,t)dh}.\label{dyn}%
\end{equation}

For completeness, and to gain a broader perspective, this work explores
t\textit{wo distinct types of fitness function}. The first is stabilizing
selection, marked by a single optimal trait value, where fitness is
represented as a unimodal (single-peaked) function of the trait value. The
second type generalizes stabilizing selection to include two optimal trait
values, resulting in a bimodal (double-peaked) fitness function.

\section{Static environment}

\label{p1s2} We first consider a static environment, which corresponds to a
fitness function that is independent of time, $t$, and written as $W(g)$.

For two different forms of $W(g)$, and for an arbitrary initial distribution,
$\Psi(g,0)$, we find, through a combination of theoretical and numerical
analysis that the distribution $\Psi(g,t)$ converges to an equilibrium
distribution, written $\Psi_{0}(g)$. (see Section
\ref{subsection14}  of the Supplementary Material for details).

\subsection{Unimodal fitness function}

\label{p1s21} In Section \ref{subsection216}  of the Supplementary Material we determine the
exact form of the equilibrium distribution of genotypic values, for a Gaussian fitness function.
In particular, Section \ref{subsection216}  of the Supplementary Material establishes that for
$W(g)$ of the form
\begin{equation}
W(g)=\exp\left(  -\frac{g^{2}}{2s^{2}}\right) \label{Gauss w(g)}%
\end{equation}
the equilibrium distribution is given by
\begin{equation}
\Psi_{0}(g)=\sqrt{\frac{1}{2\pi\sigma_{0}^{2}}}\exp{\left(  -\frac{g^{2}%
}{2\sigma_{0}^{2}}\right)  }%
\end{equation}
which has a vanishing mean, and a variance of%
\begin{equation}
\sigma_{0}^{2}=\dfrac{m^{2}+m\sqrt{m^{2}+4s^{2}}}{2}. \label{sigma0 sq}%
\end{equation}
The mean equilibrium fitness in this case is given by
\begin{equation}
\overline{W}_{0} =\int_{-\infty}^{\infty}W(g)\Psi_{0}(g)dg=\left(
1+\dfrac{\sigma_{0}^{2}}{s^{2}}\right) ^{-1/2}.
\end{equation}

\subsection{Bimodal fitness function}

\label{p1s22} We consider a bimodal form of $W(g)$ that has two equivalent
peaks. Numerical investigation shows that there are parameter regimes where
the equilbrium distribution of $g$, namely $\Psi_{0}(g)$, has a bimodal form.

Additionally, when the fitness function has two \textit{inequivalent} peaks, the equilibrium
distribution $\Psi_{0}(g)$ may still have a bimodal shape, as illustrated in a
numerical example in Section \ref{subsection33}  of the Supplementary Material.

To obtain an illustrative and soluble case, we note that in Eq. \eqref{dyn},
we can divide both occurrences of $W(g)$ by the same factor, without changing
the dynamics. We can thus use the fitness function
\begin{equation}
W(g)=\frac{1}{\sqrt{2\pi s^{2}}}\exp\left(  -\frac{\left(  g+a\right)  ^{2}%
}{2s^{2}}\right)  +\frac{1}{\sqrt{2\pi s^{2}}}\exp\left(  -\frac{\left(
g-a\right)  ^{2}}{2s^{2}}\right)  \label{bimodalfitnesssi}%
\end{equation}
in Eq. \eqref{dyn}. In the limit of $s\rightarrow0$ the above form of $W(g)$
reduces to
\begin{equation}
W(g)=\delta(g+a)+\delta(g-a)\label{double delta}%
\end{equation}
where $\delta(x)$ denotes a Dirac delta function of argument $x$. Informally,
$\delta(x)$ is a zero width, infinite height spike, that is integrable,
with unit area that is all concentrated at $x=0$. Thus $W(g)$ 
in Eq. \eqref{double delta} consists of
two arbitrarily sharp spikes that are located at $g=-a$ and $g=a$. In
Section \ref{subsection3}  of the Supplementary Material we show how such a fitness function
captures many essential properties of bimodal fitness functions and allows
analytical derivations for numerous properties in this context. Biologically, this is not
fully realistic, { the variance of the environmental component of the phenotypic value 
will generally preclude $s\rightarrow0$,} but this special case is illuminating because it
leads to many exact results. In particular, Eq. \eqref{double delta} leads to
the equilibrium distribution, $\Psi_{0}(g)$, being given by
\begin{equation}
\Psi_{0}(g)=\dfrac{1}{2}\sqrt{\dfrac{1}{2\pi m^{2}}}\cdot\left[\exp\left(
-\frac{(g+a)^{2}}{2m^{2}}\right)+\exp\left(  -\frac{(g-a)^{2}}{2m^{2}%
}\right)  \right]  .
\end{equation}
This result has a straightforward interpretation. After selection in offspring, only 
individuals with two genotypic values survive (and hence are present in the population), namely those with the trait values $a$ and $-a$. These two values become the trait values that are present in adults. Equilibrium corresponds to both trait values 
becoming equally weighted, and when mutation acts during the production of the next generation, the resulting peaks in the trait distribution derive variation only from mutations (with variance $m^{2}$).

\section{Changing environment}

\label{p1s3} We now focus on the adaptation of the population under
persistently changing environmental conditions. We consider the case where any
feature of the fitness function (such as a peak), that lies at a given trait
value in one generation, say $g_{0}$, will lie at the trait value $g_{0}+v$ in
the next generation, and $g_{0}+2v$ in the generation after that, ... . Thus the environment
`shifts' at a constant rate of $v$.
We call $v$ the \textit{velocity} of environmental change. Such change
corresponds to a time dependent fitness function, namely one that at time $t$
has the form $W(g-vt)$, where the \textit{shape} of the fitness function
remains constant, but the location of any feature increases its $g$-value by
$v$, each generation.

\subsection{Unimodal fitness function}

\label{p1s31} Similar to the static environment case, we first examine a
Gaussian (unimodal) form of the changing fitness function, $W(g-vt)$.

To simplify the analysis, we introduce the distribution, $\Phi$, in a `moving
frame' defined by
\begin{equation}
\Phi(g-vt,t)=\Psi(g,t).
\end{equation}
On substituting this form into Eq. \eqref{dyn}
and making the substitutions $g\rightarrow g+v(t+1)$, $h\rightarrow
h+vt$ we obtain%
\begin{equation}
\Phi(g,t+1)=\frac{\displaystyle\int_{-\infty}^{\infty}f(g-h+v)W(h)\Phi(h,t)dg}{\displaystyle\int_{-\infty}^{\infty
}W(h)\Phi(h,t)dh}.\label{Phi eq}%
\end{equation}
Compared with a static environment, we see that $v$ enters the problem purely
via the $v$ shift in the argument of the mutation distribution.

In Sections \ref{subsection211}-\ref{subsection215}  of the Supplementary Material we provide a
proof that with $W(g)$ having the Gaussian form of Eq. \eqref{Gauss w(g)}, and for
a Gaussian form of the initial distribution with arbitrary mean and arbitrary variance,
the `moving frame' distribution $\Phi(g,t)$ converges to a time independent distribution, and
so only depends on $g$. This corresponds to the  distribution $\Psi(g,t)$
converging over time to a \textit{rigidly moving distribution}, which is a
time-dependent distribution that depends on $g$ and $t$, but only in the
combination $g-vt$. We write this rigidly moving distribution as $\Phi_{\text{rigid}}(g-vt)$. Thus as
$t$ gets large%
\begin{equation}
\Psi(g,t)\rightarrow\Phi_{\text{rigid}}(g-vt).
\end{equation}
We find (see Sections \ref{subsection215}  of the Supplementary Material) that $\Phi_{\text{rigid}}(g-vt)$ has the form
\begin{equation}
\Phi_{\text{rigid}}(g-vt)=\sqrt{\dfrac{1}{2\pi\sigma_{0}^{2}}}\exp\left(  -\dfrac{\left[
(g-vt)+v\left(  1+\frac{s^{2}}{\sigma_{0}^{2}}\right)  \right]  ^{2}}%
{2\sigma_{0}^{2}}\right)   \label{psiv(g-vt)}
\end{equation}
where $\sigma_{0}^{2}$ is given in Eq. \eqref{sigma0 sq}. 
Thus at long time $t$, where the distribution of the trait, $\Psi(g,t)$, has converged to $\Phi_{\text{rigid}}(g-vt)$, the mean trait value at this time is $\mu_{v}(t)=\displaystyle\int_{-\infty}^{\infty}g\Psi(g,t)dg
=\displaystyle\int_{-\infty}^{\infty}g\Phi_{\text{rigid}}(g-vt)dg$ and using Eq. (\ref{psiv(g-vt)}) we obtain
\begin{equation}
\mu_{v}(t)=vt-v\left(  1+\frac{s^{2}}{\sigma_{0}^{2}}\right)
\end{equation}
while the variance of the trait value at this time equals the $v=0$ variance of Eq. \eqref{sigma0 sq}. The
mean fitness in this case is given by
\begin{equation}%
\begin{split}
\overline{W}_{v}= &  \displaystyle\int_{-\infty}^{\infty}W(g-vt)\Phi_{\text{rigid}}(g-vt)dg=\displaystyle\int_{-\infty}^{\infty}W(g)\Phi_{\text{rigid}}(g)dg\\
= &  \left(  1+\dfrac{\sigma_{0}^{2}}{s^{2}}\right)  ^{-1/2}\exp{\left(
-\dfrac{v^{2}(s^{2}+\sigma_{0}^{2})}{2\sigma_{0}^{4}}\right)  }\equiv
\overline{w}_{0}\exp{\left(  -\dfrac{v^{2}(s^{2}+\sigma_{0}^{2})}{2\sigma
_{0}^{4}}\right).  }%
\end{split}
\end{equation}

\subsection{Bimodal fitness function}

\label{p1s32} We have combined numerical and theoretical analyses to
investigate the behavior of the population distribution under a bimodal
fitness function, when the environment changes at a rate $v$. We restrict
considerations to a parameter regime such that when $v=0$ the equilibrium
distribution, $\Psi_{0}(g)$, is bimodal.

Again, in terms of the function $\Phi(g-vt,t)=\Psi(g,t)$ we find Eq.
(\ref{Phi eq}). The presence of $v$ in this equation has significant effects
on $\Phi(g,t)$ and hence $\Psi(g,t)$. In particular, numerical results reveal
that when the environmental change rate $v$ is relatively low the population
distribution maintains its bimodal structure, that is present in equilibrium
at $v=0$. However, at a sufficiently large value of $v$, one of the peaks
ultimately vanishes, and the distribution ceases to be bimodal. From Figure 2
in the main text, a transition, from bimodality to unimodality,
occurs in the rigidly moving solution, over a relatively narrow range of $v$.
The precise range of $v$ over which the transition occurs depends on parameter
values, with examples of the detailed behaviour of the transition given in
Section  \ref{subsection34}  of the Supplementary Material.

Finally, we theoretically investigate the dynamics under the bimodal fitness
function $W(g-vt)$ where $W(g)$ is the particular `double spike' form in Eq.
(\ref{double delta}). In Section \ref{subsection312}  of the Supplementary Material we prove
that under this fitness function, \textit{any} initial distribution of the trait will
ultimately reach a rigidly moving distribution $\Phi_{\text{rigid}}(g-vt)$. We also
provide exact expressions for the mean $\mu_{v}(t)$ and variance $\sigma
_{v}^{2}$ of the trait at time $t$ --- when its distribution has achieved a rigidly moving form.
These are given by
\begin{align}
\mu_{v}(t) &  =vt-v-a\tanh\left(  \frac{av}{m^{2}}\right)  \\
\sigma_{v}^{2} &  =m^{2}+a^{2}-a^{2}\tanh^{2}\left(  \frac{av}{m^{2}}\right)
\end{align}
respectively.

From earlier numerical findings, we observe that the rigidly moving
distribution of $g$ undergoes a crossover from bimodal to unimodal form, as
the environmental rate of change, $v$, is increased. We are able to determine
this threshold behaviour,  referred to as the \textit{critical environmental
velocity}, where a bimodal distribution becomes unimodal. In 
Section \ref{subsection314}  of the Supplementary Material we derive an exact result for the critical
velocity given by
\begin{equation}
v_{c}=\sqrt{a^{2}-m^{2}}-\dfrac{m^{2}}{a}\arcosh\left(  \dfrac{a}{m}\right)
.\label{vcsii}%
\end{equation}
Thus, when $v\leq v_{c}$ the rigidly moving distribution of $g$ has bimodal
form, while for $v>v_{c}$, it has a unimodal form.

Finally, as the environmental rate of change, $v$, is increased to $v_{c}$
from below, the right peak of the distribution of $g$ disappears. In
 Section \ref{subsection314}  of the Supplementary Material, we show that the position of this
peak, immediately prior to its disappearance, is given by
\begin{equation}
g_{c}-v_{c}t=\dfrac{m^{2}}{a}\arcosh\left(  \dfrac{a}{m}\right)
\end{equation}
When the environmental rate of change exceeds $v_{c}$, the right peak is
absent from the distribution.\newpage

\part[Detailed calculations]{Part II: Detailed calculations}

\label{part2}

\section{The model}

\subsection{Distribution of genotypic values}

\label{subsection12} The distribution of allelic effects of newly produced
individuals in generation $t=0,1,2,\cdots$ is given by
\begin{equation}
\phi(\mathbf{x},t)\equiv\phi(x_{1},x_{2},...,x_{L},t).
\end{equation}

Newly produced individuals first undergo selection, and with%
\begin{equation}
g=\sum_{i=1}^{L}x_{i}%
\end{equation}
the fitness of individuals with allelic effects $x_{1},x_{2}%
,\cdots,x_{L}$ in generation $t$ is
\begin{equation}
W(g,t)\equiv W\left(  \sum_{i=1}^{L}x_{i},t\right)  .
\end{equation}

Each allele in an individual is a copy, not necessarily perfect, of an allele
in its parent and alleles are taken to mutate independently of all other
alleles. The probability of a mutation per replication per allele is taken to
be $\mu$ at all loci. If a mutation occurs that alters the effect of the
allele at a particular locus, $r(x-x^{\prime})dx$ is the probability that the
offspring will inherit an allele with an effect lying in the infinitesimal
interval $(x,x+dx)$, where $x^{\prime}$ is the parental allelic effect and $r$ is the density of the distribution of mutation effects at the locus. The
distribution of allelic effects $\phi(\mathbf{x},t)$ obeys
\begin{equation}
\phi(\mathbf{x},t+1)=\dfrac{\displaystyle\bigintsss_{\mathbb{R}^{L}%
}M(\mathbf{x}-\mathbf{y})W\left(  \sum_{i=1}^{L}y_{i},t\right)  \phi
(\mathbf{y},t)d\mathbf{y}}{\displaystyle\bigintsss_{\mathbb{R}^{L}}W\left(
\sum_{i=1}^{L}y_{i},t\right)  \phi(\mathbf{y},t)d\mathbf{y}}\label{phi}%
\end{equation}
where
\begin{equation}
M(\mathbf{x})=\prod\limits_{i=1}^{L}\left[  \left(  1-\mu\right)
\delta(x_{i})+\mu r(x_{i})\right],             \label{Mdef}
\end{equation}
and $\delta(x_{i})$ is a Dirac delta function of argument $x_{i}$.

The distribution of genotypic values in generation $t$ is denoted by
$\Psi(g,t)$, which is defined by
\begin{equation}
\Psi(g,t)\overset{\text{def}}{\equiv}\displaystyle\bigintsss_{\mathbb{R}^{L}%
}\delta\left(  g-\sum_{i=1}^{L}x_{i}\right)  \phi(\mathbf{x},t)d\mathbf{x}.
\label{psi-def}%
\end{equation}

By introducing
\begin{equation}
1=\displaystyle\bigintsss_{-\infty}^{\infty}\delta\left(  h-\sum_{i=1}^{L}y_{i}\right)  dh,
\end{equation}
using the definition in Eq. \eqref{phi} yields \allowdisplaybreaks
\begin{equation}%
\begin{split}
\Psi(g,t+1)=  &  \displaystyle\bigintsss_{\mathbb{R}^{L}}\delta\left(
g-\sum_{i}x_{i}\right)  \phi(\mathbf{x},t+1)d\mathbf{x}\\
=  &  \displaystyle\bigintsss_{\mathbb{R}^{L}}\delta\left(  g-\sum_{i}%
x_{i}\right)  \cdot\dfrac{\displaystyle\bigintsss_{\mathbb{R}^{L}}
M(\mathbf{x}-\mathbf{y})W\left(  \sum_{i}y_{i},t\right)  \phi(\mathbf{y}%
,t)d\mathbf{y}}{\displaystyle\bigintsss_{\mathbb{R}^{L}} W\left(  \sum
_{i}y_{i},t\right)  \phi(\mathbf{y},t)d\mathbf{y}}d\mathbf{x}\\
=  &  \dfrac{\displaystyle\bigintsss_{\mathbb{R}^{2L}} \delta\left(
g-\sum_{i}x_{i}\right)  M(\mathbf{x}-\mathbf{y})W\left(  \sum_{i}%
y_{i},t\right)  \phi(\mathbf{y},t)d\mathbf{x}d\mathbf{y}}%
{\displaystyle\bigintsss_{\mathbb{R}^{L+1}}\delta\left(h-\sum_{i}%
y_{i}\right)  W\left(  \sum_{i}y_{i},t\right)  \phi(\mathbf{y},t)d\mathbf{y}%
dh}\\
=  &  \dfrac{\displaystyle\bigintsss_{\mathbb{R}^{2L}} \delta\left(
g-\sum_{i}x_{i}\right)  M(\mathbf{x}-\mathbf{y})W\left(  \sum_{i}%
y_{i},t\right)  \phi(\mathbf{y},t)d\mathbf{x}d\mathbf{y}}%
{\displaystyle\bigintsss_{-\infty}^{\infty} W(h,t)\Psi(h,t) dh}%
\end{split}                     \label{big}
\end{equation}

Here, we write the denominator as
\begin{equation}
\overline{W}(t)=\displaystyle\int_{-\infty}^{\infty} W(h,t)\Psi(h,t) dh.
\end{equation}

The numerator can be simplified as:
\begin{equation}%
\begin{split}
&  \displaystyle\bigintsss_{\mathbb{R}^{2L}} \delta\left(  g-\sum_{i=1}%
^{L}x_{i}\right)  M(\mathbf{x}-\mathbf{y})W\left(  \sum_{i=1}^{L}%
y_{i},t\right)  \phi(\mathbf{y},t)d\mathbf{x}\;d\mathbf{y}\\
=  &  \displaystyle\bigintsss_{\mathbb{R}^{2L+1}} \delta\left(  g-\sum
_{i=1}^{L}x_{i}\right)  \delta\left(  h-\sum_{i=1}^{L}y_{i}\right)
W\left(  \sum_{i=1}^{L}y_{i},t\right)  \phi(\mathbf{y},t)M(\mathbf{x}%
-\mathbf{y})d\mathbf{x}\;d\mathbf{y}\;dh\\
=  &  \displaystyle\bigintsss_{\mathbb{R}^{2L+1}} \delta\left(  g-\sum
_{i=1}^{L}(y_{i}+q_{i})\right)  \delta\left(  h-\sum_{i=1}^{L}%
y_{i}\right)  W\left(  \sum_{i=1}^{L}y_{i},t\right)  \phi(\mathbf{y}%
,t)M(\mathbf{q})d\mathbf{q}\;d\mathbf{y}\;dh\\
=  &  \displaystyle\bigintsss_{\mathbb{R}^{L+1}}\left[
\displaystyle\bigintsss_{\mathbb{R}^{L}}\delta\left(  g-\sum_{i=1}^{L}%
(y_{i}+q_{i})\right)  \delta\left(  h-\sum_{i=1}^{L}y_{i}\right)
W\left(  \sum_{i=1}^{L}y_{i},t\right)  \phi(\mathbf{y},t) d\mathbf{y}\right]
M(\mathbf{q})d\mathbf{q}\;dh\\
=  &  \displaystyle\bigintsss_{\mathbb{R}^{L+1}}\delta\left(  g-h-\sum_{i=1}^{L}q_{i}\right)  W(h,t)\Psi(h,t)M(\mathbf{q}%
)d\mathbf{q}\;dh\\
=  &  \displaystyle\bigintsss_{-\infty}^{\infty} f(g-h)W(h,t)\Psi(h,t)dh%
\end{split}
\end{equation}
where
\begin{equation}
f(g-h)=\displaystyle\bigintsss_{\mathbb{R}^{L}}\delta\left(
g-h-\sum_{i=1}^{L}x_{i}\right)  M(\mathbf{x})d\mathbf{x}
\end{equation}
($M(\mathbf{x})$ is defined in Eq. (\ref{Mdef})).

The final result is
\begin{equation}
\Psi(g,t+1)=\dfrac{\displaystyle\int_{-\infty}^{\infty} f(g-h)W(h,t)\Psi(h,t)dh}{\overline{W}(t)} \label{exact}%
\end{equation}

\subsection{Approximation for $f(g)$}

\label{subsection13} Using the representation
\begin{equation}
\delta(g)=\displaystyle\int_{-\infty}^{\infty}\exp(i\lambda g)\;\dfrac
{d\lambda}{2\pi},
\end{equation}
we have
\begin{equation}%
\begin{split}
f(g)= &  \displaystyle\bigintsss_{\mathbb{R}^{L}}\delta\left(  g-\sum
_{i=1}^{L}x_{i}\right)  M(\mathbf{x})d\mathbf{x}\\
= &  \displaystyle\bigintsss_{\mathbb{R}^{L+1}}e^{i\lambda\left(  g-\sum
_{i=1}^{L}x_{i}\right)  }\prod\limits_{i=1}^{L}\left[  \left(  1-\mu\right)
\delta(x_{i})+\mu r(x_{i})\right]  d\mathbf{x}\;\dfrac{d\lambda}{2\pi}\\
= &  \displaystyle\bigintsss_{\mathbb{R}^{L+1}}e^{i\lambda g}e^{-i\lambda
\sum_{i=1}^{L}x_{i}}\prod\limits_{i=1}^{L}\left[  \left(  1-\mu\right)
\delta(x_{i})+\mu r(x_{i})\right]  d\mathbf{x}\;\dfrac{d\lambda}{2\pi}\\
= &  \displaystyle\bigintsss_{\mathbb{R}^{L+1}}e^{i\lambda g}\prod
\limits_{i=1}^{L}e^{-i\lambda x_{i}}\left[  \left(  1-\mu\right)  \delta
(x_{i})+\mu r(x_{i})\right]  d\mathbf{x}\;\dfrac{d\lambda}{2\pi}\\
= &  \displaystyle\bigintsss_{-\infty}^{\infty}e^{i\lambda g}\left\{
\prod\limits_{i=1}^{L}\displaystyle\int_{-\infty}^{\infty}e^{-i\lambda x_{i}%
}\left[  \left(  1-\mu\right)  \delta(x_{i})+\mu r(x_{i})\right]
dx_{i}\right\}  \dfrac{d\lambda}{2\pi}\\
= &  \displaystyle\bigintsss_{-\infty}^{\infty}e^{i\lambda g}\left\{
\displaystyle\int_{-\infty}^{\infty}e^{-i\lambda x}\left[  \left(
1-\mu\right)  \delta(x)+\mu r(x)\right]  dx\right\}  ^{L}\dfrac{d\lambda}%
{2\pi}\\
= &  \displaystyle\bigintsss_{-\infty}^{\infty}e^{i\lambda g}\left[  \left(
1-\mu\right)  +\mu\displaystyle\int_{-\infty}^{\infty}e^{-i\lambda
x}r(x)dx\right]  ^{L}\dfrac{d\lambda}{2\pi}%
\end{split}
\end{equation}

There are two limiting cases of this expression.

\begin{enumerate}
\item $L\mu\gg1$

While not relevant to most asexual populations, this leads, equivalent to the central
limit theorem, to
\begin{equation}
f(g)=\sqrt{\frac{1}{2\pi m^{2}}}\exp\left(  -\frac{g^{2}}{2m^{2}}\right)
\label{f1}%
\end{equation}
in which
\begin{equation}
m^{2}=L\mu m_{1}^{2}\equiv Um_{1}^{2}%
\end{equation}
where $m_{1}^{2}$ is the variance of $r(x)$ and we have assumed the mean of
$r(x)$ is zero.

\item $L\mu\ll1$

In this regime, $f(g)$ primarily derives two contributions: (i) from no alleles
mutating, and (ii) from just one allele mutating. 

(i) From no alleles mutating,  $f(g)$ has the contribution $(1-L\mu)\delta(g)$

(ii) From one allele mutating, $f(g)$ has the contribution $L\mu r(g)$

Assuming the simplest form of $r(g)$, namely a Gaussian with mean zero and
variance $m_{1}^{2}$ we obtain
\begin{equation}
f(g)=(1-U)\delta(g)+U\sqrt{\frac{1}{2\pi m_{1}^{2}}}\exp\left(-\frac{g^{2}%
}{2m_{1}^{2}}\right).\label{f2}%
\end{equation}
\end{enumerate}
Neither of the two forms of $f(g)$ is likely to be fully accurate, with
reality lying somewhere in the middle. 

We have carried out analysis and simulations with $f(g)$ given by both Eqs.
\eqref{f1} and \eqref{f2}. and in each case found the \textit{qualitatively} same
phenomenon occurring: when there are two peaks in the equilbrium distribution
of $g$ for $v=0$, at a fast enough rate of environmental change, one peak is
absent from the rigidly moving trait distribution.

In the main text, we have used Eq. \eqref{f1} for ease of exposition.

\subsection{The stability of equilibrium under static environment}

\label{subsection14} In a static environment, we have
\begin{equation}
\Psi(g,t+1)=\dfrac{\displaystyle\int_{-\infty}^{\infty} f(g-h)W(h)\Psi(h,t)dh}{\displaystyle\int_{-\infty
}^{\infty}W(h)\Psi(h,t)dh}.
\end{equation}

In Section \ref{subsection216}, we theoretically demonstrate that for a Gaussian form
of the fitness function $W(g)$, any initial Gaussian distribution can approach an
equilibrium distribution, denoted as $\Psi_{0}(g)$, after many generations. 
In this section, we use a numerical approach, with various
fitness functions and initial distributions, to illustrate that, under the
iterative dynamics described above, the population distribution $\Psi(g,t)$
converges to the equilibrium distribution $\Psi_{0}(g)$ in a static environment, 
as long as $W(g)$ vanishes at large $|g|$.

(a) For $W(g)$ in Gaussian form, we denote
\begin{equation}
W(g)=\exp{\left( -\frac{g^{2}}{2s^{2}}\right) }\label{siw1}%
\end{equation}
and initial distribution
\begin{equation}
\Psi(g,0):=\dfrac{\exp{\left( -\frac{(g+b)^{2}}{p^{2}}\right) }+\exp{\left(
-\frac{(g-b)^{2}}{p^{2}}\right) }}{\displaystyle\int_{-\infty}^{\infty}\left[
\exp{\left( -\frac{(g+b)^{2}}{p^{2}}\right) }+\exp{\left( -\frac{(g-b)^{2}%
}{p^{2}}\right) }\right] dg}\label{sii1}%
\end{equation}
then the figure of equilibrium distribution $\Psi_{0}(g)$ is given below in
Figure \ref{f1si}. 
\begin{figure}[H]
\centering
\includegraphics[width=0.8\textwidth]{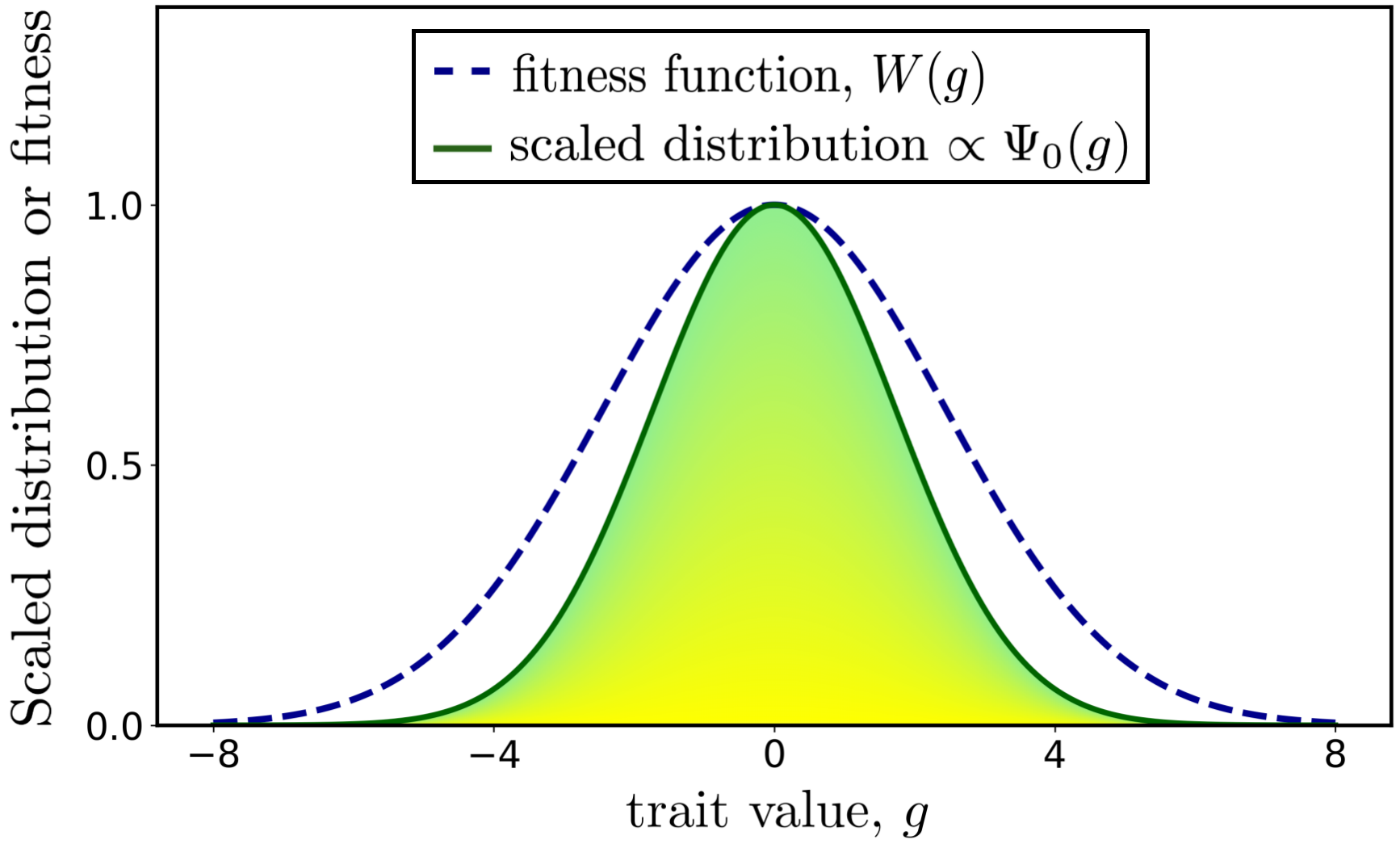}\caption{\textbf{Unimodal
fitness function and associated equilibrium distribution of genotypic trait
values in a static environment.} This figure presents a symmetric
\textit{unimodal} fitness function (dashed blue line) as a function of the
genotypic trait value, $g$. The fitness function exhibits a peak at $g=0$.
Additionally, the equilibrium genotypic distribution, scaled to a maximum
height of unity, is plotted (green line) against $g$. For this figure, we
utilized the mutational distribution specified in Eq. \eqref{fgsiiiii}, the
fitness function in Eq. \eqref{siw1} and we started the iteration with the initial distribution in Eq.
\eqref{sii1}. We used the parameter values $m=1$, $s=\sqrt{6}$, $b=2$ and 
$p=\sqrt{1.5}$.}%
\label{f1si}
\end{figure}

\begin{figure}[H]
\centering
\includegraphics[width=0.8\textwidth]{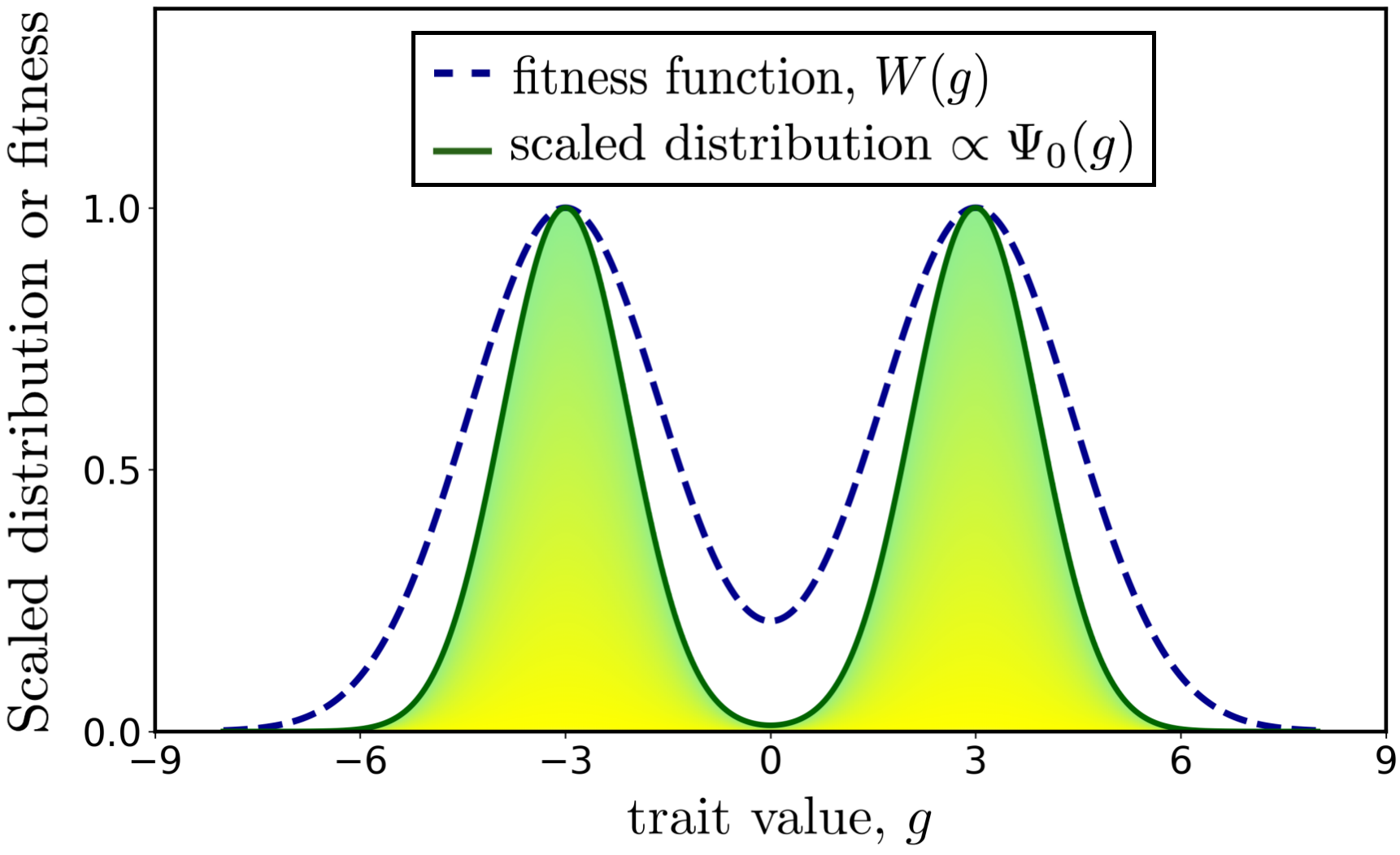}\caption{\textbf{Bimodal
fitness function and associated equilibrium distribution of genotypic trait
values in a static environment.} This figure displays a symmetric
\textit{bimodal} fitness function (dashed blue line) as a function of the
genotypic trait value, $g$. The fitness function features peaks near $g=-a$
and $g=a$. Additionally, the equilibrium genotypic distribution, normalized to
a maximum height of unity, is shown (green line) against $g$. In constructing
the figure, we employed the mutational distribution defined by Eq.
\eqref{fgsiiiii}, the fitness function in Eq. \eqref{siw2} and we started the iteration with the initial distribution in Eq. \eqref{sii2}. We used the parameter values $m=0.5$,
$s=\sqrt{2}$, $a=3$, $c=1$ and $d=0$.}%
\label{f2si}
\end{figure}

(b) For $W(g)$ in bimodal form, we denote
\begin{equation}
W(g):=\exp{\left( -\frac{(g+a)^{2}}{s^{2}}\right) }+\exp{\left( -\frac
{(g-a)^{2}}{s^{2}}\right) }\label{siw2}%
\end{equation}
and initial distribution
\begin{equation}
\Psi(g,0):=\sqrt{\frac{1}{2\pi c^{2}}}\cdot\exp{\left( -\frac{(g+d)^{2}%
}{2c^{2}}\right) }.\label{sii2}%
\end{equation}
The form of the equilibrium distribution, $\Psi_{0}(g)$, is given in Figure \ref{f2si}.

\begin{figure}[H]
\centering
\includegraphics[width=0.8\textwidth]{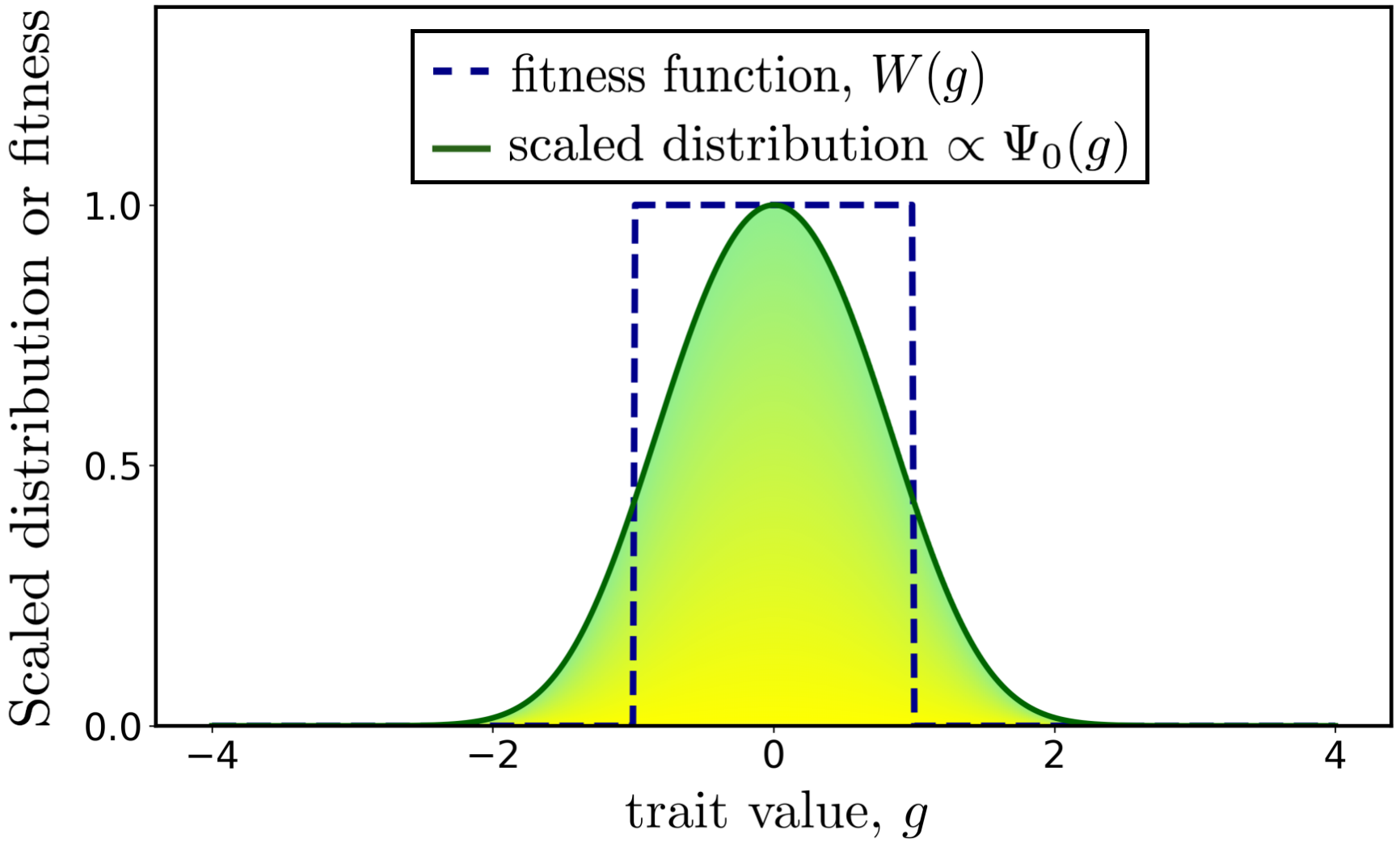}\caption{\textbf{Fitness
function $\bm{W(g)}$ in step-function form and associated equilibrium distribution
of genotypic trait values in a static environment.} This figure illustrates a
\textit{step-function form} fitness function (dashed blue line) plotted
against the genotypic trait value, $g$. The fitness function reaches a peak at
$g=0$. The equilibrium genotypic distribution, scaled to a maximum height of
unity, is also shown (green line) as a function of $g$. For this figure, we
employed the mutational distribution from Eq. \eqref{fgsiiiii}, the fitness
function from Eq. \eqref{siw3}, and and we started the iteration with the initial distribution
in Eq. \eqref{sii3}. We used the parameter values $m=0.5$, $a=1$, $c=1$ and $d=2$.}%
\label{f3si}
\end{figure}

(c) For $W(g)$ a step function, we used \begin{numcases}{W(g)=}
	1, & $|g|\le a$\\
	0, & $|g|>a$   \label{siw3}
\end{numcases}
and the initial distribution
\begin{equation}
\Psi(g,1):=\sqrt{\frac{1}{2\pi c^{2}}}\cdot\exp{\left( -\frac{(g+d)^{2}%
}{2c^{2}}\right).}\label{sii3}%
\end{equation}
The form of the equilibrium distribution, $\Psi_{0}(g)$, is given in Figure \ref{f3si}.

\newpage
\section{Unimodal fitness function}
In this Section, we consider the case where the fitness function is unimodal, with a Gaussian form. We take the initial distribution of the trait, $\Psi(g,0)$, to have a Gaussian form, 
but with an arbitrary mean and an arbitrary variance. In the first parts, \ref{subsection211} - \ref{subsection215}, we investigate
the dynamics of the distribution in the presence of environmental change that occurs
at a constant rate, $v$. In particular, we show how the distribution evolves to have a rigidly moving form, 
namely a Gaussian distribution with a time-independent variance that is independent of $v$.
In Section \ref{subsection216}, we extract results for the special $v=0$ case, which represents a static environment, and we determine the form of the equilibrium distribution. Then, in Sections \ref{subsection217} and \ref{subsection218}, we analyse two limiting cases where the `range' of the fitness function approaches $0$ and $\infty$, corresponding to very strong and very weak selection, respectively. 

Finally, we show that the variance of the distribution of the trait is not always independent of
$v$ at long times (as was found when the fitness function is Gaussian). In Section \ref{subsection219} we illustrate, with a numerical example, the case of a non-Gaussian unimodal fitness function where the long time variance of the distribution depends on $v$.

In the second part, Section \ref{subsection22}, we consider the case where the environment changes in a single jump. In Sections \ref{subsection221} and \ref{subsection222}, we show that, under the impact of such an environmental shift, the equilibrium distribution of the population will eventually revert to its equilibrium distribution in a static environment. In Section \ref{subsection223}, we generalize this result to multiple jumps and obtain the same conclusion.

To proceed, we assume a rigidly changing or `moving' fitness function of the form
\begin{equation}
W(g,t)=W(g-vt),
\end{equation}
where $W(g)$ is symmetric and unimodal and $v$ is a constant, that we call the
speed of the fitness optimum.

Equation \eqref{exact} then reduces to
\begin{equation}
\Psi(g,t+1)=\dfrac{\displaystyle\int_{-\infty}^{\infty} f(g-h)W(h-vt)\Psi(h,t)dh}{\displaystyle\int_{-\infty
}^{\infty}W(h-vt)\Psi(h,t)dh}\label{exactv}%
\end{equation}

\subsection{The fitness function is `moving' at constant velocity $v$}

\label{subsection21}

\subsubsection{Fitness function is Gaussian}

We base fitness on the function
\begin{equation}
    W(g)=\exp{\left(-\frac{g^{2}}{2s^{2}}\right)} \label{W(g) supp} 
\end{equation}
and shall use the mutation distribution
\begin{equation}
    f(g)=\sqrt{\frac{1}{2\pi m^{2}}}\cdot \exp{\left(-\frac{g^{2}}{2m^{2}}\right)}.\label{f(g)}
\end{equation}
Furthermore, we take the initial distribution, $\Psi(g,0)$, to be a \textit{Gaussian function of $g$, with an arbitrary mean and an
arbitrary variance.}

\label{subsection211} 
It directly follows from Eqs. (\ref{exactv}) - (\ref{f(g)}), along with the Gaussian form 
that has been adopted for $\Psi(g,0)$, that the distribution $\Psi(g,t)$ will retain a Gaussian form over time. 
In other words, with $f(g)$, $W(g)$ and $\Psi(g,0)$ all Gaussian functions, the
quantity $\displaystyle\int_{-\infty}^{\infty} f(g-h)W(h-vt)\Psi(h,0)dh$, which is proportional
to $\Psi(g,1)$, is also a Gaussian function of $g$. 

Using the notation $\mathbb{E}(g,t)$ and $\operatorname{Var}(g,t)$ to denote the mean and variance of the distribution
$\Psi(g,t)$, 
it further follows that
\begin{equation}
    \mathbb{E}(g,t+1)=\frac{vt\cdot\operatorname{Var}(g,t)+s^{2}\cdot \mathbb{E}(g,t)}{s^{2}
                           +\operatorname{Var}(g,t)}\label{mean}\\
\end{equation}
and
\begin{equation}
    \operatorname{Var}(g,t+1)=\frac{s^{2}\operatorname{Var}(g,t)}{s^{2}+\operatorname{Var}(g,t)}+m^{2}. \label{var}
\end{equation}

\subsubsection{Variance of the distribution}
\label{subsection212} We define
\begin{equation}
\sigma_{0}^{2}=\frac{m^{2}+m\sqrt{m^{2}+4s^{2}}}{2}%
\end{equation}
then from Eq. (\ref{var}) we obtain 
\begin{equation}
\frac{1}{\operatorname{Var}(g,t+1)-\sigma_{0}^{2}}=\frac{s^{2}+\sigma_{0}^{2}%
}{s^{2}+m^{2}-\sigma_{0}^{2}}\cdot\frac{1}{\operatorname{Var}(g,t)-\sigma
_{0}^{2}}+\frac{1}{s^{2}+m^{2}-\sigma_{0}^{2}}%
\end{equation}
and hence
\begin{equation}
\frac{1}{\operatorname{Var}(g,t+1)-\sigma_{0}^{2}}+\frac{1}{m\sqrt
{m^{2}+4s^{2}}}=\frac{s^{2}+\sigma_{0}^{2}}{s^{2}+m^{2}-\sigma_{0}^{2}}\left(
\frac{1}{\operatorname{Var}(g,t)-\sigma_{0}^{2}}+\frac{1}{m\sqrt{m^{2}+4s^{2}%
}}\right)
\end{equation}
This equation can be solved with the result
\begin{equation}
\frac{1}{\operatorname{Var}(g,t)-\sigma_{0}^{2}}+\frac{1}{m\sqrt{m^{2}+4s^{2}%
}}=\left( \frac{s^{2}+\sigma_{0}^{2}}{s^{2}%
+m^{2}-\sigma_{0}^{2}}\right) ^{t}\left[ \frac{1}{\operatorname{Var}(g,0)-\sigma_{0}^{2}}+\frac{1}%
{m\sqrt{m^{2}+4s^{2}}}\right].%
\end{equation}
and leads to
\begin{equation}
\operatorname{Var}(g,t)=\frac{1}{\left( \frac
{s^{2}+\sigma_{0}^{2}}{s^{2}+m^{2}-\sigma_{0}^{2}}\right) ^{t}\left[ \frac{1}{\operatorname{Var}%
(g,0)-\sigma_{0}^{2}}+\frac{1}{m\sqrt{m^{2}+4s^{2}}}\right] -\frac
{1}{m\sqrt{m^{2}+4s^{2}}}}+\sigma_{0}^{2}.%
\end{equation}
Since $\frac{s^{2}+\sigma_{0}^{2}}{s^{2}+m^{2}-\sigma_{0}^{2}} >1$
it follows that at large $t$
\begin{equation}
\operatorname{Var}(g,t)\rightarrow\sigma_{0}^{2}=\frac{m^{2}+m\sqrt{m^{2}+4s^{2}}%
}{2}.\label{var11111}%
\end{equation}
This result indicates that the variance asymptotically approaches a constant, corresponding to a rigidly
moving (Gaussian) form of the distribution.

\subsubsection{Mean of the distribution}
\label{subsection213} 
From Eq. \eqref{mean} at large $t$, where  $\operatorname{Var}(g,t)\approx\sigma_{0}^{2}$,
we have
\begin{equation}
\mathbb{E}(g,t+1)\approx\frac{v\sigma_{0}^{2}t+s^{2}\cdot
\mathbb{E}(g,t)}{s^{2}+\sigma_{0}^{2}}%
\end{equation}
We write this as 
\begin{equation}
\mathbb{E}(g,t+1)\approx p\mathbb{E}(g,t)+qt
\end{equation}
where  
\begin{equation}
    p=\frac{s^{2}}{s^{2}+\sigma_{0}^{2}}\quad\text{and}\quad q=\frac{v\sigma_{0}^{2}}{s^{2}+\sigma_{0}^{2}}
\end{equation}
and leads to 
\begin{equation}
\mathbb{E}(g,t+1)+\frac{q}{p-1}(t+1)+\frac{q}{(p-1)^{2}}\approx p\left[
\mathbb{E}(g,t)+\frac{q}{p-1}t+\frac{q}{(p-1)^{2}}\right].
\end{equation}
This equation can be solved with the result
\begin{equation}
\mathbb{E}(g,t)\approx\left[ \mathbb{E}(g,0)+\frac{q}{(p-1)^{2}%
}\right] p^{t}-\frac{q}{p-1}t-\frac{q}{(p-1)^{2}}.%
\end{equation}
Since $p<1$, it follows that as $t$ gets large $p^{t}\rightarrow0$ and we obtain
\begin{equation}
\mathbb{E}(g,t)\approx vt-\frac{v(s^{2}+\sigma_{0}^{2})}{\sigma_{0}^{2}}.%
\end{equation}

\subsubsection{Rigidly moving solution}

The above results make it plausible that no matter what the initial distribution is, 
after many generations, the population will adapt to the environmental change, by achieving 
a Gaussian distribution with a mean at time $t$ of 
$vt-\frac{v(s^{2}+\sigma_{0}^{2})}{\sigma_{0}^{2}}$
and a variance of $\sigma_{0}^{2}$. It can be verified that such a solution precisely 
satisfies Eq. (\ref{exactv}), when supplemented by Eqs. (\ref{W(g) supp}) and (\ref{f(g)}).  
We shall describe this distribution as a `rigidly moving solution'. It depends on $g$ and $t$, but only 
in the combination $g-vt$, thus its peak at time $t$ is $vt$ but its shape is unchanged with time.
We write the \textit{rigidly moving solution} as $\Phi_{\text{rigid}}(g-vt)$ and
\begin{equation}
\Phi_{\text{rigid}}(g-vt)= \sqrt{\dfrac{1}{2\pi\sigma_{0}^{2}}}\cdot\exp\left( -\dfrac{\left[
(g-vt)+\frac{v(s^{2}+\sigma_{0}^{2})}{\sigma_{0}^{2}}\right] ^{2}}{2\sigma
_{0}^{2}}\right). \label{rigid}
\end{equation}
At large $t$ we have
\begin{equation}
    \Psi(g,t) \rightarrow \Phi_{\text{rigid}}(g-vt).
\end{equation}
\label{subsection215} 
The distribution of the trait at time  $t$, when it has achieved the rigidly moving distribution, $\Phi_{\text{rigid}}(g-vt)$, has a mean
and variance, that we write as $\mu_{v}(t)$ and $\sigma_{v}^{2}$, respectively, which are given
by
\begin{equation}
\mu_{v}(t)=vt-v\left(  1+\frac{s^{2}}{\sigma_{0}^{2}}\right) \quad\text{ and
}\quad\sigma_{v}^{2}\equiv\sigma_{0}^{2}=\frac{m^{2}+m\sqrt{m^{2}+4s^{2}}}%
{2}.\label{TW results}%
\end{equation}
We note that the mean trait value in Eq. (S5.21)
has been found in the different context of  \textit{sexual population},
when fitness is Gaussian, under the approximation of a ``constant genetic
variance'' \cite{BurgerLynch1995}.

We can also obtain the mean fitness at large time $t$:
\begin{equation}
    \begin{split}
        \overline{W}_{v}&=\displaystyle\int_{-\infty}^{\infty}W(g-vt)\Phi_{\text{rigid}}(g-vt)dg=\displaystyle\int_{-\infty}^{\infty}W(g)\Phi_{\text{rigid}}(g)dg\\
    & =\displaystyle\int_{-\infty}^{\infty}\exp{\left(  \ -\frac{g^{2}}{2s^{2}}\right)  }\cdot\sqrt{\dfrac{1}{2\pi\sigma_{0}^{2}}}\cdot\exp\left(-\dfrac{\left[  g+\frac{v(s^{2}+\sigma_{0}^{2})}{\sigma_{0}^{2}}\right]  ^{2}}{2\sigma_{0}^{2}}\right)\\
    &=\left(1+\dfrac{\sigma_{0}^{2}}{s^{2}}\right)^{-1/2}\times\exp{\left(\ -\dfrac{v^{2}(s^{2}+\sigma_{0}^{2})}{2\sigma_{0}^{4}}\right).}  \label{mean W}
    \end{split}
\end{equation}

\subsubsection{Static environment}

\label{subsection216} Under static environmental conditions ($v=0$), we can obtain the equilibrium form of the 
distribution by taking $v \rightarrow 0$ in Eq. (\ref{rigid}), with the result
\begin{equation}
\Psi_{0}(g)=\sqrt{\frac{1}{2\pi\sigma_{0}^{2}}}\cdot\exp{\left( -\frac{g^{2}%
}{2\sigma_{0}^{2}}\right) }\label{unistaenvpsi}%
\end{equation}
where
\begin{equation}
\sigma_{0}^{2}=\frac{m^{2}+m\sqrt{m^{2}+4s^{2}}}{2}. 
\end{equation}
It can be verified that $\Psi_{0}(g)$ satisfies Eq. (\ref{exactv}) with $v=0$,
when supplemented by Eqs. (\ref{W(g) supp}) and (\ref{f(g)}).

We note, in particular, that the equilibrium distribution of the trait has a mean and a variance
that are given by
\begin{equation}
\mu_{0}=0\quad\text{ and }\quad\sigma_{0}^{2}=\dfrac{m^{2}+m\sqrt{m^{2}%
+4s^{2}}}{2},
\end{equation}
respectively.

We can also obtain the mean equilibrium fitness:
\begin{equation}
\overline{W}_{0}=  \displaystyle\int_{-\infty}^{\infty}W(g)\Psi_{0}(g)dg= \left( 1+\dfrac{\sigma_{0}^{2}}{s^{2}}\right) ^{-1/2}
\end{equation}
which coincides with the $v \rightarrow 0$ limit of Eq. (\ref{mean W}).

\subsubsection{Specific case 1: $s^{2}\rightarrow0$}

\label{subsection217} In this subsection, we consider results for the Gaussian fitness function 
in Eq. (\ref{W(g) supp}) in the limit $s^2\rightarrow 0$. 

Assuming $\operatorname{Var}(g,t) \ne 0$, when $s^{2}\rightarrow0$, Eqs. \eqref{mean} - \eqref{var} yield
\begin{numcases}{}
    \mathbb{E}(g,t+1)=vt\\               \label{Et+1}
    \operatorname{Var}(g,t+1)=m^{2}     
\end{numcases}
and the corresponding distribution in generation $t+1$ is
\begin{equation}
\Psi(g,t+1)=\frac{1}{\sqrt{2\pi m^{2}}}\exp{\left[ -\frac{\left(g-vt\right)^{2}}{2m^{2}}\right]}. \label{dist from delta}
\end{equation}
These results for generation $t+1$ are independent of the mean and variance of distribution in generation $t$.
To understand this, we first return to $s$ finite, and rescale both occurrences of the fitness in Eq. (\ref{exactv}) from 
$\exp(-g^2/(2s^2))$ to $(2\pi s^2)^{-1/2}\exp(-g^2/(2s^2))$. Then, when $s^2 \rightarrow 0$ the 
rescaled fitness function becomes a Dirac delta function: $(2\pi s^2)^{-1/2}\exp(-g^2/(2s^2)) \rightarrow \delta(g)$
and Eq. (\ref{exactv}) reduces to
\begin{equation}
\Psi(g,t+1)=\dfrac{\displaystyle\int_{-\infty}^{\infty} f(g-h)
\delta(h-vt)\Psi(h,t)dh}{\displaystyle\int_{-\infty
}^{\infty}\delta(h-vt)\Psi(h,t)dh}=\frac{f(g-vt)\Psi(vt,t)}{\Psi(vt,t)}=f(g-vt).
\end{equation}
This result coincides with the distribution in Eq. (\ref{dist from delta}), and explains the
origin of Eq. (\ref{dist from delta}).

The limit $s^2 \rightarrow 0$ corresponds to the strongest possible selection of a single trait value.

\subsubsection{Specific case 2: $s^{2}\rightarrow\infty$}

\label{subsection218} 

In this subsection, we consider results for the Gaussian fitness function 
in Eq. (\ref{W(g) supp}) in the limit $s^2\rightarrow \infty$. 

In the limit $s^{2}\rightarrow \infty$ the fitness function $W(g)$ reduces to a constant, 
corresponding to the absence of selection. From Eqs. \eqref{mean} - \eqref{var} for $s^{2}\rightarrow \infty$, we have
\begin{numcases}{}
    \mathbb{E}(g,t+1)=\mathbb{E}(g,t)\\
     \operatorname{Var}(g,t+1)=\operatorname{Var}(g,t)+m^{2}.
\end{numcases}
Thus the mean value of the trait value does not change from generation $t$ to
generation $t+1$, but the variance increases by $m^{2}$ each
generation. These results arise because in the absence of selection, only mutations act. 
However, mutations are unbiased, so have no effect on the mean, but mutations do add to 
the existing variance of the distribution, each generation.

\subsubsection{Non Gaussian fitness function}

\label{subsection219} 
\begin{figure}[H]
\centering
\includegraphics[width=1\textwidth]{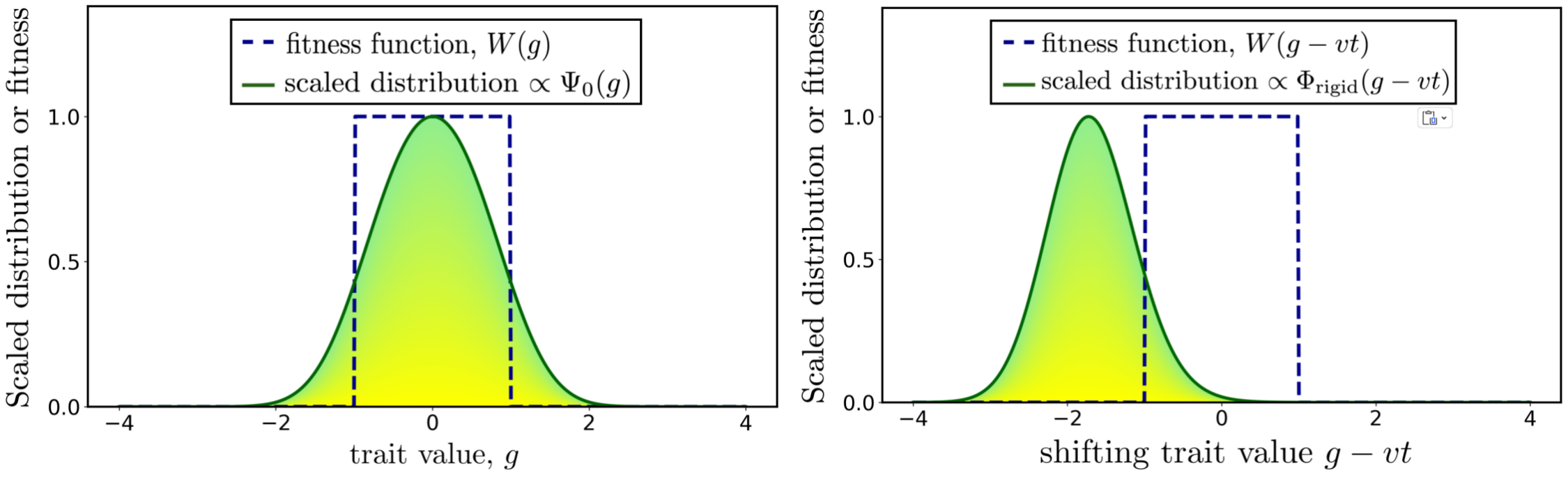}\caption{\textbf{Fitness
function $\bm{W(g)}$ in step-function form, and corresponding equilibrium
distribution of genotypic trait values in a static environment, alongside the
rigidly moving genotypic distribution in a uniformly changing environment.} In
the left panel we display a \textit{step-function form} of fitness function (dashed
blue line) plotted against the genotypic trait value, $g$. The equilibrium
genotypic distribution, scaled to a maximum height of unity, is also shown
(green line) as a function of $g$. For this plot, we used the mutational
distribution given in Eq. \eqref{fgsiiiii} with parameter value $m=0.5$ and set
$v=0$. In the right panel a unimodal fitness function, $W(g-vt)$, is shown (dashed
blue line) and plotted against the shifting genotypic trait value, $g-vt$, where $v$
represents the constant 'velocity' of environmental change. The rigidly moving
distribution, dependent on $g$ and $t$ only in the combination $g-vt$, is expressed 
as $\Phi_{\text{rigid}}(g-vt)$. A normalized version of $\Phi_{\text{rigid}}(g-vt)$, scaled to a maximum height of unity,
is also plotted (green line) against $g-vt$. For this figure, we used the
mutational distribution in Eq. \eqref{fgsiiiii} with parameter value $m=0.5$ and
$v=1$.}%
\label{f4si}
\end{figure}

In the previous subsection, we showed that at large $t$ the variance 
becomes independent of the environmental velocity, $v$ (see Eq. \eqref{var11111}).
However, this is not generally the case. For fitness functions that are not
Gaussian functions of $g$, we have numerical evidence that the variance of the
rigidly moving distribution will depend on $v$. Here we give one example with the `step function'
form of fitness function 
\begin{numcases}
{W(g)=}
	1, & $ |g|\leq 1$\\
	0, & $|g|>1$.
\end{numcases}
As can be seen from Figure \ref{f4si}, the equilibrium distribution in the left panel and
the rigidly moving distribution in the right panel have different shapes. The
distribution on the left has a different variance to the distribution on the right. 
This illustrates that the variance of the rigidly moving distribution is affected by
the environmental velocity, $v$.

\subsection{$W(g)$ has a single jump}

\label{subsection22} We assume that the fitness function $W(g)$ does not
change its form over time, which indicates that $v=0$. At a certain moment
$t$, $W(g)$ changes suddenly, with a magnitude of $e$:
\begin{equation}
W_{1}(g)=W(g+e)=\exp{\left[ -\frac{(g+e)^{2}}{2s^{2}}\right] }\label{w(g)jump}
\end{equation}

Set $\Psi(g,t)$ to be the distribution under fitness function $W(g)$. After
generation $t$, we change the fitness function to $W_{1}(g)$, then the dynamics of the distribution would be
\begin{equation}
\Psi(g,t+1)=\dfrac{{\displaystyle\int_{-\infty}^{\infty}}f(g-h%
)W_{1}(h)\Psi(h,t)dh}{{\displaystyle\int_{-\infty
}^{\infty}}W_{1}(h)\Psi(h,t)dh}.
\end{equation}

We shall use
\begin{equation}
    f(g)=\sqrt{\frac{1}{2\pi m^{2}}}\cdot \exp{\left(-\frac{g^{2}}{2m^{2}}\right)}.\label{f(g)jump}
\end{equation}
From Section \ref{subsection211}, the distribution at the jump time $t$, $\Psi(g,t)$, is a Gaussian function of $g$.
We shall use the notation $\mathbb{E}(g,t)$ and $\operatorname{Var}(g,t)$ to denote the mean and variance 
of the distribution $\Psi(g,t)$. We find from the above that
\begin{equation}
    \mathbb{E}(g,t+1)=\frac{s^{2}\cdot \mathbb{E}(g,t)-e\cdot\operatorname{Var}(g,t)}{s^{2}+\operatorname{Var}(g,t)}\label{mean1}
\end{equation}
and
\begin{equation}
    \operatorname{Var}(g,t+1)=\frac{s^{2}q^{2}}{s^{2}+q^{2}}+m^{2}=\frac{s^{2}\operatorname{Var}(g,t)}{s^{2}+\operatorname{Var}(g,t)}+m^{2}\label{var1}
\end{equation}

\subsubsection{Variance of the distribution}

\label{subsection221} Denote
\begin{equation}
\sigma_{0}^{2}:=\frac{m^{2}+m\sqrt{m^{2}+4s^{2}}}{2}%
\end{equation}

From the previous section, we have
\begin{equation}
\operatorname{Var}(g,t)=\frac{1}{\left[ \frac{1}{\operatorname{Var}%
(g,0)-\sigma_{0}^{2}}+\frac{1}{m\sqrt{m^{2}+4s^{2}}}\right] \left( \frac
{s^{2}+\sigma_{0}^{2}}{s^{2}+m^{2}-\sigma_{0}^{2}}\right) ^{t}-\frac
{1}{m\sqrt{m^{2}+4s^{2}}}}+\sigma_{0}^{2}%
\end{equation}
where
\begin{equation}
\frac{s^{2}+\sigma_{0}^{2}}{s^{2}+m^{2}-\sigma_{0}^{2}}=\frac{2s^{2}+m\left(
m+\sqrt{4s^{2}+m^{2}}\right) }{2s^{2}+m\left( m-\sqrt{4s^{2}+m^{2}}\right) }>1
\end{equation}

When $t\gg1$, we can see that
\begin{equation}
\operatorname{Var}(g,t)\approx\sigma_{0}^{2}=\frac{m^{2}+m\sqrt{m^{2}+4s^{2}}%
}{2}%
\end{equation}
which indicates that the variance will remain unchanged, leading to a rigidly
moving shape.

\subsubsection{Mean of the distribution}

\label{subsection222} From Eq. \eqref{mean}, we have
\begin{equation}
\mathbb{E}(g,t+1)=\frac{s^{2}\cdot\mathbb{E}(g,t)-e\cdot\operatorname{Var}%
(g,t)}{s^{2}+\operatorname{Var}(g,t)}%
\end{equation}

When $t\gg1$, $\operatorname{Var}(g,t)\approx\frac{m^{2}+m\sqrt{m^{2}+4s^{2}}%
}{2}:=\sigma_{0}^{2}$.
\begin{equation}
\Rightarrow\mathbb{E}(g,t+1)\approx\frac{s^{2}\cdot\mathbb{E}(g,t)-e\sigma
_{0}^{2}}{s^{2}+\sigma_{0}^{2}}%
\end{equation}

We denote  \begin{numcases}{}
		h=\frac{s^{2}}{s^{2}+\sigma_{0}^{2}}<1\\
		k=-\frac{e\sigma_{0}^{2}}{s^{2}+\sigma_{0}^{2}}	
	\end{numcases}
\begin{equation}
\iff\mathbb{E}(g,t+1)\approx h\mathbb{E}(g,t)+k
\end{equation}
\begin{equation}
\iff\mathbb{E}(g,t+1)+\frac{k}{h-1}\approx m\left[ \mathbb{E}(g,t)+\frac
{k}{h-1}\right]
\end{equation}

This is a geometric sequence, then we have
\begin{equation}
\mathbb{E}(g,t)\approx\left[ \mathbb{E}(g,0)+\frac{k}{h-1}\right]
h^{t}-\frac{k}{h-1}%
\end{equation}

When $t\gg1$, $h^{t}\rightarrow0$, we can obtain
\begin{equation}
\mathbb{E}(g,t)\approx-\frac{k}{h-1}=-e
\end{equation}

At a large generation $t$, the mean value of the distribution lies at
$\mathbb{E}(g,t)\approx-e$, which is also the fitness optimum. It indicates
that the population will finally adapt to the sudden change of environment.

\subsubsection{$W(g)$ has multiple jumps}
\label{subsection223}
After the population distribution achieves the equilibrium, we have
\begin{equation}
	\Psi_{d}(g)=\sqrt{\dfrac{1}{2\pi\sigma_{0}^{2}}}\cdot\exp{\left(-\dfrac{(g+e)^{2}}{2\sigma_{0}^{2}}\right)}
\end{equation}

According to \eqref{unistaenvpsi}, we have
\begin{equation}
	\Psi_{e}(g)=\Psi_{0}(g+e).
\end{equation}

This indicates that when the fitness function $W(g)$ undergoes a single jump—meaning the population experiences an environmental change in the form of a single extreme event—the shape of its equilibrium distribution remains consistent with that under a static environment. Moreover, the distance between the equilibrium distribution prior to the single jump and that following it is precisely the magnitude of the jump, denoted as $e$. In other words, the post-jump distribution can be viewed as a translation of the pre-jump distribution along the genotypic value axis.

From the results above, a straightforward generalization can be made: when environmental changes occur as several extreme events, resulting in the fitness function $W(g)$ exhibiting multiple jumps, we can denote these abrupt environmental shifts in chronological order as $e_{1}, e_{2}, \cdots, e_{n}$. After many generations of inheritance, the population will ultimately adapt to the final abrupt environmental change. Consequently, the shape of its equilibrium distribution will mirror that under a static environment, and the distance between the equilibrium distribution before the environmental changes and the final equilibrium distribution will equal the magnitude of the last jump, $e_{n}$. Thus, the final population distribution can be viewed as a translation of the initial distribution along the genotypic value axis.

\newpage

\section{Bimodal fitness function}

\label{subsection3} 
In this Section, we examine the case where the fitness function $W(g)$ is bimodal. Specifically, we assume that $W(g)$ takes the form described in Eq. \eqref{wgsiiiiii} and derive certain necessary and sufficient conditions for the bimodal structure of this function. To facilitate analytical tractability, we focus on a special case of the fitness function $W(g)$, which is represented by Eq. \eqref{twodeltawgsiiii}. In the first part, Section \ref{subsection31}, we investigate the dynamical and statistical properties of the ultimate state of the population distribution as it adapts to a continuously changing environment with a constant change rate $v$ per generation. In Sections \ref{subsection311} - \ref{subsection313}, we find that the population distribution ultimately reaches a form with invariant variance, evolving in a rigidly moving form. We also derive the explicit form, mean, and variance of this distribution. Further, in Section \ref{subsection314}, we explore a special case where the two peaks of the fitness function $W(g)$ have equal height and derive the explicit form of the rigidly moving population distribution. By setting the environmental change rate $v=0$, we obtain the equilibrium distribution in a static environment. We then observe that, when the environmental change rate $v$ is low, the population distribution maintains a bimodal shape; however, as $v$ exceeds a critical threshold $v_{c}$, this bimodal structure can no longer persist, and one peak disappears, signaling a reduction in biodiversity. We analytically derive the specific form of this critical threshold.

In the second part, Section \ref{subsection32}, we consider the scenario where the environment changes through a single jump. In Sections \ref{subsection321} - \ref{subsection325}, we show that, under such a change, the equilibrium distribution of the population will eventually revert to its equilibrium form in a static environment. In Section \ref{subsection326}, we generalize this result to multiple jumps and reach the same conclusion. 

Subsequently, in Section \ref{subsection33}, we provide a numerical example illustrating that, when the peaks of the fitness function $W(g)$ are of unequal heights, the equilibrium distribution of the population in a static environment may also exhibit a bimodal structure. Finally, in the main text, our numerical results reveal that the transition of the population distribution from bimodal to unimodal is very sharp as the environmental change rate $v$ increases. In Section \ref{subsection34}, by varying the variance $m^{2}$ of the mutation distribution, we present a series of numerical results showing that this sharp transition from bimodal to unimodal distribution is not a universal outcome.

First of all, we observe that in Eq. \eqref{dyn}, each instance of $W(g)$ can be scaled by an identical factor without altering the dynamics. In other words, we can equivalently apply the fitness function
\begin{equation}
W(g)=\exp\left(-\frac{(g+a)^{2}}{2s^{2}}\right) + \exp\left(-\frac{(g-a)^{2}}{2s^{2}}\right).
\label{wgsiiiiii}
\end{equation}

Notably, this function is unimodal when $s^{2}\geq a^{2}$, and becomes bimodal when $s^{2}<a^{2}$ \cite{eisenberger1964genesis}.

To achieve a solvable scenario, we consider the fitness function
\begin{equation}
W(g)=\frac{p}{\sqrt{2\pi s^{2}}}\exp\left(  -\frac{\left(  g+a\right)  ^{2}%
}{2s^{2}}\right)  +\frac{q}{\sqrt{2\pi s^{2}}}\exp\left(  -\frac{\left(
g-a\right)  ^{2}}{2s^{2}}\right)
\end{equation}
with
\begin{equation}
p+q=1
\end{equation}
in Eq. \eqref{dyn}. In the limit of $s\rightarrow0$ we have $W(g)$ reducing to%
\begin{equation}
W(g)=p\delta(g+a)+q\delta(g-a)\label{twodeltawgsiiii}
\end{equation}
where $\delta(x)$ denotes a Dirac delta function of argument $x$, so $W(g)$
consists of two arbitrarily sharp spikes, which indicates that the two peaks in the fitness function are arbitrarily narrow in width. Therefore, if $p=q=\dfrac{1}{2}$ then $W(g)$ is symmetric and has two equal height
`peaks' at $\pm a$.

\subsection{$W(g)$ is moving in a constant velocity $v$}

\label{subsection31} Denote
\begin{equation}
\Psi(g,t)=\Phi(g-vt,t)\label{Phi def}%
\end{equation}

Then we have
\begin{equation}
\Psi(g,t+1)=\dfrac{{\displaystyle\int_{-\infty}^{\infty}}f(g-h)W(h-vt)\Psi(h,t)dh}{{\displaystyle\int_{-\infty
}^{\infty}}W(h-vt)\Psi(h,t)dh}%
\end{equation}
\begin{equation}
\iff\Phi(g-v(t+1),t+1)=\dfrac{{\displaystyle\int_{-\infty}^{\infty}%
}f(g-h)W(h-vt)\Phi(h-vt,t)dh%
}{{\displaystyle\int_{-\infty}^{\infty}}W(h-vt)\Phi(g^{\prime
}-vt,t)dh}%
\end{equation}
\begin{equation}
\iff\Phi(g,t+1)=\dfrac{{\displaystyle\int_{-\infty}^{\infty}}f(g-g^{\prime
}+v)W(h)\Phi(h,t)dh}{{\displaystyle\int_{-\infty
}^{\infty}}W(h)\Phi(h,t)dh}\label{comoving}%
\end{equation}

Note that because of Eq. \eqref{Phi def} we have
\begin{equation}%
\begin{split}
\displaystyle\int_{-\infty}^{\infty}g\Psi(g,t)dg= & \displaystyle\int%
_{-\infty}^{\infty}g\Phi(g-vt,t)dg\\
= & \displaystyle\int_{-\infty}^{\infty}(g-vt)\Phi(g-vt,t)dg+\displaystyle\int%
_{-\infty}^{\infty}vt\Phi(g-vt,t)dg\\
= & \displaystyle\int_{-\infty}^{\infty}g\Phi(g,t)dg+vt
\end{split}
\end{equation}
and
\begin{equation}%
\begin{split}
\displaystyle\int_{-\infty}^{\infty}g^{2}\Psi(g,t)dg= & \displaystyle\int%
_{-\infty}^{\infty}g^{2}\Phi(g-vt,t)dg\\
= & \displaystyle\int_{-\infty}^{\infty}(g-vt+vt)^{2}\Phi(g-vt,t)dg\\
= & \displaystyle\int_{-\infty}^{\infty}g^{2}\Phi(g,t)dg+2vt\displaystyle\int%
_{-\infty}^{\infty}g\Phi(g,t)dg+v^{2}t^{2}%
\end{split}
\end{equation}
then we have
\begin{equation}
\displaystyle\int_{-\infty}^{\infty}g^{2}\Psi(g,t)dg-\left[ \displaystyle\int%
_{-\infty}^{\infty}g\Psi(g,t)dg\right] ^{2}=\displaystyle\int_{-\infty
}^{\infty}g^{2}\Phi(g,t)dg-\left[ \displaystyle\int_{-\infty}^{\infty}%
g\Phi(g,t)dg\right] ^{2}%
\end{equation}
so to get the \textit{true mean of the trait}, we have to add $vt$ to the mean
calculated from $\Phi(g,t)$. The variance of the trait can be calculated using
either $\Psi(g,t)$ or $\Phi(g,t)$.

\subsubsection{Recurrence relation}

\label{subsection311} Substituting the form of $W(g)$ in Eq. (\ref{twodeltawgsiiii}) into Eq.
\eqref{comoving} yields
\begin{equation}
\Phi(g,t+1)=\dfrac{f(g+v+a)p\Phi(-a,t)+f(g+v-a)q\Phi(a,t)}{p\Phi
(-a,t)+q\Phi(a,t)}.\label{general g}%
\end{equation}
We now set $g$ to be $-a$ and $a$ in this equation to obtain
\begin{numcases}{}
	\Phi(-a,t+1)=\dfrac{f(v)p\Phi(-a,t)+f(v-2a)q\Phi(a,t)}{p\Phi
(-a,t)+q\Phi(a,t)}\label{system}\\
	\Phi(a,t+1)=\dfrac{f(v+2a)p\Phi(-a,t)+f(v)q\Phi(a,t)}{p\Phi
(-a,t)+q\Phi(a,t)}
\end{numcases}

These equations are a dynamically complete description for $\Phi(-a,t)$ and
$\Phi(a,t)$. Let us define
\begin{equation}
\mathbf{X}(t)=\left(
\begin{array}
[c]{c}%
\Phi(-a,t)\\
\Phi(a,t)
\end{array}
\right) \label{X}%
\end{equation}
then the system in Eq. \eqref{system} can be written as
\begin{equation}
\mathbf{X}(t+1)=\frac{\mathbf{MDX}(t)}{\mathbf{F}^{T}\mathbf{DX}
(t)}\label{matrix eq}%
\end{equation}
where
\begin{equation}
\mathbf{M}=\left(
\begin{array}
[c]{cc}%
f(v), & f(v-2a)\\
f(v+2a), & f(v)
\end{array}
\right)  ,\qquad\mathbf{D}=\left(
\begin{array}
[c]{cc}%
p, & 0\\
0, & q
\end{array}
\right)  ,\qquad\mathbf{F}=\left(
\begin{array}
[c]{c}%
1\\
1
\end{array}
\right)  .
\end{equation}
For $t=1,2,\cdots$, the solution is
\begin{equation}
\mathbf{X}(t)=\frac{\left(  \mathbf{MD}\right)  ^{t}\mathbf{X}(0)}
{\mathbf{F}^{T}\mathbf{D}\left(  \mathbf{MD}\right)  ^{t-1}\mathbf{X}(0)}
.\label{solXt}%
\end{equation}

We can calculate the eigenvalues of $\mathbf{MD}$ \begin{numcases}{}
	\lambda_{1}=\frac{1}{2\sqrt{2\pi m^{2}}} \exp\left(-\frac{v^{2}}{2m^{2}}\right)\left[1+\sqrt{(p-q)^{2}+4pq\exp\left(-\frac{4a^{2}}{m^{2}}\right)}\right]\\
	\lambda_{2}=\frac{1}{2\sqrt{2\pi m^{2}}} \exp\left(-\frac{v^{2}}{2m^{2}}\right)\left[1-\sqrt{(p-q)^{2}+4pq\exp\left(-\frac{4a^{2}}{m^{2}}\right)}\right]
\end{numcases}

We can easily see that $0<\lambda_{2}<\lambda_{1}<1$.

Let $\boldsymbol{\alpha_{1}}^{(1)}$ and $\boldsymbol{\alpha_{1}}^{(2)}$ be the
corresponding right eigenvectors, $\boldsymbol{\beta}^{(1)T}$ and
$\boldsymbol{\beta} ^{(2)T}$ be the corresponding left eigenvectors, with the
properties \begin{numcases}{}
	\mathbf{MD}\boldsymbol{\alpha_{1}}^{(1)}=\lambda_{1}\boldsymbol{\alpha_{1}}^{(1)}\\
	\mathbf{MD}\boldsymbol{\alpha_{1}}^{(2)}=\lambda_{2}\boldsymbol{\alpha_{1}}^{(2)}\\
	\boldsymbol{\beta}^{(1)T}\mathbf{MD}=\lambda_{1}\boldsymbol{\beta}^{(1)T}\\
	\boldsymbol{\beta}^{(2)T}\mathbf{MD}=\lambda_{2}\boldsymbol{\beta}^{(2)T}\\
	\boldsymbol{\beta}^{(1)T}\boldsymbol{\alpha_{1}}^{(2)}=\boldsymbol{\beta}^{(2)T}\boldsymbol{\alpha_{1}}^{(1)}=0
\end{numcases}

Then we have \begin{numcases}{}
	\mathbf{MD}-\lambda_{1}\mathbf{I}=c_{1}\boldsymbol{\alpha_{1}}^{(2)}\boldsymbol{\beta}^{(2)T}\\
	\mathbf{MD}-\lambda_{2}\mathbf{I}=c_{2}\boldsymbol{\alpha_{1}}^{(1)}\boldsymbol{\beta}^{(1)T}
\end{numcases}
\begin{numcases}{\Rightarrow}	
	(\mathbf{MD}-\lambda_{1}\mathbf{I})\boldsymbol{\alpha_{1}}^{(2)}=(\lambda_{2}-\lambda_{1})\boldsymbol{\alpha_{1}}^{(2)}=c_{1}\boldsymbol{\alpha_{1}}^{(2)}\boldsymbol{\beta}^{(2)T}\boldsymbol{\alpha_{1}}^{(2)}\\
	(\mathbf{MD}-\lambda_{2}\mathbf{I})\boldsymbol{\alpha_{1}}^{(1)}=(\lambda_{1}-\lambda_{2})\boldsymbol{\alpha_{1}}^{(1)}=c_{2}\boldsymbol{\alpha_{1}}^{(1)}\boldsymbol{\beta}^{(1)T}\boldsymbol{\alpha_{1}}^{(1)}\\
	(\lambda_{2}-\lambda_{1})\mathbf{I}=c_{1}\boldsymbol{\alpha_{1}}^{(2)}\boldsymbol{\beta}^{(2)T}-c_{2}\boldsymbol{\alpha_{1}}^{(1)}\boldsymbol{\beta}^{(1)T}
\end{numcases}
\begin{numcases}{\Rightarrow}	
	\lambda_{2}-\lambda_{1}=c_{1}\boldsymbol{\beta}^{(2)T}\boldsymbol{\alpha_{1}}^{(2)}\\
	\lambda_{1}-\lambda_{2}=c_{2}\boldsymbol{\beta}^{(1)T}\boldsymbol{\alpha_{1}}^{(1)}
\end{numcases}
\begin{numcases}{\Rightarrow}	
	c_{1}=\frac{\lambda_{2}-\lambda_{1}}{\boldsymbol{\beta}^{(2)T}\boldsymbol{\alpha_{1}}^{(2)}}\\
	c_{2}=\frac{\lambda_{1}-\lambda_{2}}{\boldsymbol{\beta}^{(1)T}\boldsymbol{\alpha_{1}}^{(1)}}
\end{numcases}
\begin{equation}
\Rightarrow\frac{\boldsymbol{\alpha_{1}}^{(1)}\boldsymbol{\beta}^{(1)T}%
}{\boldsymbol{\beta}^{(1)T}\boldsymbol{\alpha_{1}}^{(1)}}+\frac
{\boldsymbol{\alpha_{1}}^{(2)}\boldsymbol{\beta}^{(2)T}}{\boldsymbol{\beta
}^{(2)T}\boldsymbol{\alpha_{1}}^{(2)}}=\mathbf{I}%
\end{equation}

Set \begin{numcases}{}
	\boldsymbol{\alpha}^{(1)}=\dfrac{\boldsymbol{\alpha_{1}}^{(1)}}{\boldsymbol{\beta}^{(1)T}\boldsymbol{\alpha_{1}}^{(1)}}\\
	\boldsymbol{\alpha}^{(2)}=\dfrac{\boldsymbol{\alpha_{1}}^{(1)}}{\boldsymbol{\beta}^{(2)T}\boldsymbol{\alpha_{1}}^{(2)}}
\end{numcases}
We can see that $\boldsymbol{\alpha}^{(1)}$ and $\boldsymbol{\alpha}^{(2)}$
are still the corresponding right eigenvectors to the eigenvalues of
$\mathbf{MD}$, $\lambda_{1}$ and $\lambda_{2}$.

Then $\boldsymbol{\alpha}^{(1)}$, $\boldsymbol{\alpha}^{(2)}$,
$\boldsymbol{\beta}^{(1)T}$ and $\boldsymbol{\beta} ^{(2)T}$ will obey
\begin{numcases}{}
	\boldsymbol{\alpha}^{(1)}\boldsymbol{\beta}^{(1)T}+\boldsymbol{\alpha}^{(2)}\boldsymbol{\beta}^{(2)T}=\mathbf{I}\\
	\boldsymbol{\beta}^{(i)T}\boldsymbol{\alpha}^{(j)}=\delta_{i,j}
\end{numcases}
\begin{equation}
\Rightarrow\mathbf{MD}=\boldsymbol{\alpha}^{(1)}\boldsymbol{\beta}%
^{(1)T}\mathbf{MD}+\boldsymbol{\alpha}^{(2)}\boldsymbol{\beta}^{(2)T}%
\mathbf{MD}=\lambda_{1}\boldsymbol{\alpha}^{(1)}\boldsymbol{\beta}%
^{(1)T}+\lambda_{2}\boldsymbol{\alpha}^{(2)}\boldsymbol{\beta}^{(2)T}%
\label{MD}%
\end{equation}

Substitute Eq. \eqref{MD} into Eq. \eqref{solXt}, we have
\begin{equation}%
\begin{split}
\mathbf{X}(t)= & \frac{\left[ \lambda_{1}^{t}\boldsymbol{\alpha}%
^{(1)}\boldsymbol{\beta}^{(1)T}+\lambda_{2}^{t}\boldsymbol{\alpha}%
^{(2)}\boldsymbol{\beta}^{(2)T}\right]  \mathbf{X}(0)}{\mathbf{F}%
^{T}\mathbf{D}\left[ \lambda_{1}^{t-1}\boldsymbol{\alpha}^{(1)}%
\boldsymbol{\beta}^{(1)T}+\lambda_{2}^{t-1}\boldsymbol{\alpha}^{(2)}%
\boldsymbol{\beta}^{(2)T}\right] \mathbf{X}(0)}\\
= & \frac{\left[ \lambda_{1}\boldsymbol{\alpha}^{(1)}\boldsymbol{\beta}%
^{(1)T}+\lambda_{2}\cdot\left( \frac{\lambda_{2}}{\lambda_{1}}\right)
^{t-1}\boldsymbol{\alpha}^{(2)}\boldsymbol{\beta}^{(2)T}\right]
\mathbf{X}(0)}{\mathbf{F}^{T}\mathbf{D}\left[ \boldsymbol{\alpha}%
^{(1)}\boldsymbol{\beta}^{(1)T}+\left( \frac{\lambda_{2}}{\lambda_{1}}\right)
^{t-1}\boldsymbol{\alpha}^{(2)}\boldsymbol{\beta}^{(2)T}\right] \mathbf{X}(0)}
.\label{Xtformula}%
\end{split}
\end{equation}

Since $\lambda_{1}$ is the larger (in magnitude) of the two eigenvalues, then
for $t^{\prime}\gg1$, we have
\begin{equation}
\mathbf{X}(t^{\prime})=\frac{\lambda_{1}\boldsymbol{\alpha}^{(1)}
\boldsymbol{\beta}^{(1)T}\mathbf{X}(0)}{\mathbf{F}^{T}\mathbf{D}%
\boldsymbol{\alpha}^{(1)}\boldsymbol{\beta}^{(1)T}\mathbf{X}(0)}=\frac
{\lambda_{1}\boldsymbol{\alpha}^{(1)}}{\mathbf{F} ^{T}\mathbf{D}
\boldsymbol{\alpha}^{(1)}} \equiv\left(
\begin{array}
[c]{c}%
X_{1}(t^{\prime})\\
X_{2}(t^{\prime})
\end{array}
\right)
\end{equation}
so we obtain a result independent of the initial state, and components are
essentially determined by a single eigenvector.

A measure of asymmetry of the distribution at generation $t^{\prime}$ is some
comparison of the two components of $\mathbf{X}(t^{\prime})$.

The actual distribution at generation $t^{\prime}$ follows from Eq.
\eqref{general g}:
\begin{equation}
\Phi(g,t^{\prime})=\dfrac{f(g+v+a)pX_{1}(t^{\prime})+f(g+v-a)qX_{2}(t^{\prime
})} {pX_{1}(t^{\prime})+qX_{2}(t^{\prime})}.
\end{equation}
Thus the relationship between $a$, $v$, the variance of $f(g)$, along with the
values of $p$ and $q$ will determine the bimodal (or not) nature of
$\Phi(g,t^{\prime})$.

Note that
\begin{equation}%
\begin{split}
{\int_{-\infty}^{\infty}g\Phi(g,t^{\prime})dg}= & \dfrac{(-v-a)pX_{1}%
(t^{\prime})+(-v+a)qX_{2}(t^{\prime})}{pX_{1}(t^{\prime})+qX_{2}(t^{\prime}%
)}\\
= & -v-a\cdot\dfrac{pX_{1}(t^{\prime})-qX_{2}(t^{\prime})}{pX_{1}(t^{\prime
})+qX_{2}(t^{\prime})}%
\end{split}
\end{equation}
and
\begin{equation}%
\begin{split}
\int_{-\infty}^{\infty}g^{2}\Phi(g,t^{\prime})dg= & \dfrac{[m^{2}%
+(-v-a)^{2}]pX_{1}(t^{\prime})+[m^{2}+(-v+a)^{2}]qX_{2}(t^{\prime})}%
{pX_{1}(t^{\prime})+qX_{2}(t^{\prime})}\\
= & m^{2}+v^{2}+a^{2}+2av\cdot\dfrac{pX_{1}(t^{\prime})-qX_{2}(t^{\prime}%
)}{pX_{1}(t^{\prime})+qX_{2}(t^{\prime})}%
\end{split}
\end{equation}
we have
\begin{equation}%
\begin{split}
& \int_{-\infty}^{\infty}g^{2}\Phi(g,t^{\prime})dg-\left[ \int_{-\infty
}^{\infty}g\Phi(g,t^{\prime})dg\right] ^{2}\\
= & m^{2}+a^{2}-a^{2}\cdot\left[ \dfrac{pX_{1}(t^{\prime})-qX_{2}(t^{\prime}%
)}{pX_{1}(t^{\prime})+qX_{2}(t^{\prime})}\right] ^{2}%
\end{split}
\end{equation}

\subsubsection{$X_{1}(t^{\prime})$ v.s. $X_{2}(t^{\prime})$ ($t^{\prime}\gg
1$)}

\label{subsection312} From above, for $t^{\prime}\gg1$, we have
\begin{equation}%
\begin{split}
\frac{X_{1}(t^{\prime})}{X_{2}(t^{\prime})}= & \dfrac{pf(v)X_{1}(t^{\prime
})+qf(v-2a)X_{2}(t^{\prime})}{pf(v+2a)X_{1}(t^{\prime})+qf(v)X_{2}(t^{\prime
})}\\
= & \dfrac{pf(v)\cdot\dfrac{X_{1}(t^{\prime})}{X_{2}(t^{\prime})}%
+qf(v-2a)}{pf(v+2a)\cdot\dfrac{X_{1}(t^{\prime})}{X_{2}(t^{\prime})}+qf(v)}%
\end{split}
\end{equation}
\begin{equation}
\iff pf(v+2a)\cdot\left[ \dfrac{X_{1}(t^{\prime})}{X_{2}(t^{\prime})}\right]
^{2}+[qf(v)-pf(v)]\cdot\dfrac{X_{1}(t^{\prime})}{X_{2}(t^{\prime})}-qf(v-2a)=0
\end{equation}
\begin{equation}
\dfrac{X_{1}(t^{\prime})}{X_{2}(t^{\prime})}=\frac{\exp\left( \frac
{2av+2a^{2}}{m^{2}} \right) }{2p}\left[ (2p-1)\pm\sqrt{\left( 1-2p\right)
^{2}+4p(1-p)\exp\left( -\frac{4a^{2}}{m^{2}}\right) }\right]
\end{equation}

Since
\begin{equation}
\sqrt{\left( 1-2p\right) ^{2}+4p(1-p)\exp\left( -\frac{4a^{2}}{m^{2}}\right)
}>|2p-1|
\end{equation}
and
\begin{equation}
X_{1}(t^{\prime})>0\;\;\&\;\;X_{2}(t^{\prime})>0
\end{equation}
we have
\begin{equation}
\dfrac{X_{1}(t^{\prime})}{X_{2}(t^{\prime})}=\frac{\exp\left( \frac
{2av+2a^{2}}{m^{2}} \right) }{2p}\left[ (2p-1)+\sqrt{\left( 1-2p\right)
^{2}+4p(1-p)\exp\left( -\frac{4a^{2}}{m^{2}}\right) }\right]
\end{equation}
which is independent of time $t^{\prime}$.

We have
\begin{equation}%
\begin{split}
\Phi(g,t^{\prime})= & \dfrac{f(g+v+a)pX_{1}(t^{\prime})+f(g+v-a)qX_{2}%
(t^{\prime})}{pX_{1}(t^{\prime})+qX_{2}(t^{\prime})}\\
= & \dfrac{f(g+v+a)p\cdot\dfrac{X_{1}(t^{\prime})}{X_{2}(t^{\prime}%
)}+f(g+v-a)q}{p\cdot\dfrac{X_{1}(t^{\prime})}{X_{2}(t^{\prime})}+q}>0
\end{split}
\end{equation}

Now we can calculate
\begin{equation}%
\begin{split}
{\int_{-\infty}^{\infty}g\Phi(g,t^{\prime})dg}= & v-a\cdot\dfrac
{pX_{1}(t^{\prime})-qX_{2}(t^{\prime})}{pX_{1}(t^{\prime})+qX_{2}(t^{\prime}%
)}\\
= & -v-a\cdot\dfrac{p\cdot\dfrac{X_{1}(t^{\prime})}{X_{2}(t^{\prime})}%
-q}{p\cdot\dfrac{X_{1}(t^{\prime})}{X_{2}(t^{\prime})}+q}%
\end{split}
\end{equation}
and
\begin{equation}%
\begin{split}
& \int_{-\infty}^{\infty}g^{2}\Phi(g,t^{\prime})dg-\left[ \int_{-\infty
}^{\infty}g\Phi(g,t^{\prime})dg\right] ^{2}\\
= & m^{2}+a^{2}-a^{2}\cdot\left[ \dfrac{pX_{1}(t^{\prime})-qX_{2}(t^{\prime}%
)}{pX_{1}(t^{\prime})+qX_{2}(t^{\prime})}\right] ^{2}\\
= & m^{2}+a^{2}-a^{2}\cdot\left[ \dfrac{p\cdot\dfrac{X_{1}(t^{\prime})}%
{X_{2}(t^{\prime})}-q}{p\cdot\dfrac{X_{1}(t^{\prime})}{X_{2}(t^{\prime})}%
+q}\right] ^{2}%
\end{split}
\end{equation}
which is independent of time $t^{\prime}$, this indicates that the
distribution at later time, $t^{\prime}$, approaches a \textit{rigidly moving
solution} which depends on $g$ and $t^{\prime}$, but only in the combination
$g-vt^{\prime}$, i.e., corresponding to a distribution that rigidly moves over
time. We represent this rigidly moving distribution as $\Phi_{\text{rigid}}(g-vt^{\prime
})$:
\begin{equation}
\Psi(g,t^{\prime})=\Phi(g-vt^{\prime},t^{\prime})=\Phi_{\text{rigid}}(g-vt^{\prime}),
\end{equation}
then
\begin{equation}
\Phi(g,t^{\prime})=\Phi_{\text{rigid}}(g)
\end{equation}
and \begin{numcases}{}
	X_{1}(t^{\prime})=\Phi(-a,t^{\prime})=\Phi_{\text{rigid}}(-a)\\
	X_{2}(t^{\prime})=\Phi(a,t^{\prime})=\Phi_{\text{rigid}}(a)	
\end{numcases}

\subsubsection{$p\neq q$}

\label{subsection313} The parameters $a$, $m$, $p$ and $q$ are predetermined,
the exponential function of $v$ can be factored out.
\begin{equation}%
\begin{split}
\dfrac{\Phi_{\text{rigid}}(-a)}{\Phi_{\text{rigid}}(a)}= & \frac{\exp\left( \frac{2av+2a^{2}}{m^{2}}
\right) }{2p}\cdot\left[ (2p-1)+\sqrt{\left( 1-2p\right) ^{2}+4p(1-p)\exp
\left( -\frac{4a^{2}}{m^{2}}\right) }\right] \\
= & \exp\left( \frac{2av}{m^{2}} \right) \cdot\frac{\exp\left( \frac{2a^{2}%
}{m^{2}} \right) }{2p}\cdot\left[ (2p-1)+\sqrt{\left( 1-2p\right)
^{2}+4p(1-p)\exp\left( -\frac{4a^{2}}{m^{2}}\right) }\right]
\end{split}
\end{equation}

As $v$ increases, the right peak $a$ will inevitably disappear.

\subsubsection{Special case: $p=q=\dfrac{1}{2}$}

\label{subsection314} From the previous section, we have
\begin{equation}
\dfrac{\Phi_{\text{rigid}}(-a)}{\Phi_{\text{rigid}}(a)}=\exp\left( \dfrac{2av}{m^{2}}\right)
\end{equation}
and obtain
\begin{equation}
    \begin{split}
        \Phi_{\text{rigid}}(g)=&\dfrac{f(g+v+a)\exp\left(  \dfrac{2av}{m^{2}}\right)+f(g+v-a)}{\exp\left(\dfrac{2av}{m^{2}}\right)+1}\\
        =&\dfrac{e^{av/m^{2}}f(g+v+a)+e^{-av/m^{2}}f(g+v-a)}{2\cosh\left(\dfrac{av}{m^{2}}\right)}\\
        =&\dfrac{2\exp{\left(-\dfrac{a^{2}-2av}{2m^{2}}\right)}\cdot\exp{\left(-\dfrac{(g+v)^{2}}{2m^{2}}\right)}\cosh\left(\dfrac{ag}{m^{2}}\right)}{\sqrt{2\pi m^{2}}\left[\exp\left(\dfrac{2av}{m^{2}}\right)+1\right]}\label{deltaequilibrium}
    \end{split}
\end{equation}

In a static environment, we have
\begin{equation}
	\begin{split}
		\Psi_{0}(g)=&\sqrt{\dfrac{1}{2\pi m^{2}}}\cdot\left[\exp{\left(-\dfrac{g^{2}+a^{2}}{2m^{2}%
}\right) }\cdot\cosh\left( \dfrac{ag}{m^{2}}\right)\right]\\
		=&\dfrac{1}{2}\sqrt{\dfrac{1}{2\pi m^{2}}}\cdot\left[ \exp\left(-\frac{(g+a)^{2}}{2m^{2}}\right) + \exp\left(-\frac{(g-a)^{2}}{2m^{2}}\right) \right]\label{staenvpsi0}
	\end{split}
\end{equation}

\begin{figure}[H]
	\centering
	\includegraphics[width=0.5\linewidth]{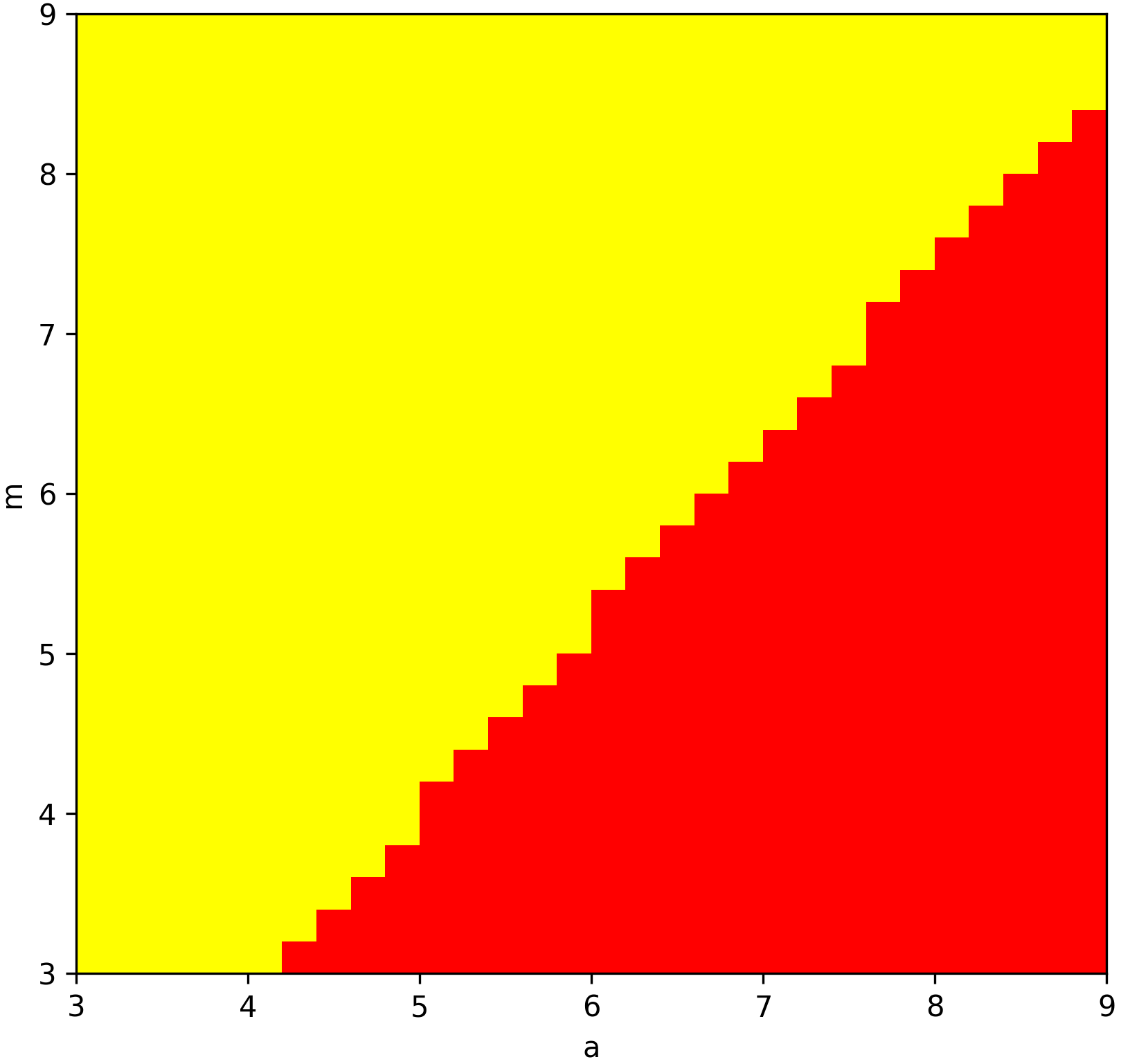}
    \caption{Bimodal–unimodal transition of the equilibrium distribution $\Psi_{0}(g)$ across parameter space. Red regions indicate bimodal distributions and yellow regions indicate unimodal distributions, based on numerical simulations for $a, m \in [3, 9]$ with fixed $s=\sqrt{6}$. The theoretical boundary $m=a$ reasonably matches the observed transition.}
    \label{heatmap}
\end{figure}

From above, this distribution is itself bimodal when $m^{2}<a^{2}$ and unimodal when $m^{2}\geq a^{2}$. To illustrate this we fixed $s = \sqrt{6}$ and performed a numerical parameter scan over $a, m\in [3, 9]$ in steps of $0.2$. For each parameter pair, we numerically constructed the equilibrium distribution and identified whether it was unimodal or bimodal. The results are summarized in Figure \ref{heatmap}. Parameter regions $(a, m)$ showing whether the equilibrium distribution $\Psi_0(g)$ is unimodal or bimodal. Red indicates a bimodal distribution, and yellow indicates unimodal. The theoretical boundary $m = a$ aligns well with the numerically observed transition.

We now examine the monotonicity of the function $\Phi_{\text{rigid}}(g)$, we denote
\begin{equation}
h(g):=\exp{\left[ -\dfrac{(g+v)^{2}}{2m^{2}}\right] }\cdot\cosh\left(
\dfrac{ag}{m^{2}}\right)
\end{equation}
we have
\begin{equation}
h^{\prime}(g)=\dfrac{\exp\left[ -\dfrac{(g+v)^{2}}{2m^{2}}\right] \cdot\left[
a\sinh\left( \dfrac{ag}{m^{2}} \right) -(g+v)\cosh\left( \dfrac{ag}{m^{2}%
}\right) \right] }{m^{2}}%
\end{equation}

For $H^{\prime}(g)=0$ we obtain
\begin{equation}
\tanh\left( \dfrac{ag}{m^{2}}\right) =\dfrac{g+v}{a}%
\end{equation}

Denote
\begin{equation}
l(g):=\tanh\left( \dfrac{ag}{m^{2}}\right) -\dfrac{g+v}{a}%
\end{equation}
we have \begin{numcases}{}
	\lim_{g\rightarrow-\infty}l(g)=+\infty\\
	\lim_{g\rightarrow+\infty}l(g)=-\infty\\
	l(g)\geq 0\iff h^{\prime}(g)\geq 0\iff\Phi_{\text{rigid}}(g)\uparrow\\
	l(g)<0\iff h^{\prime}(g)<0\iff\Phi_{\text{rigid}}(g)\downarrow
\end{numcases}
then we know that there must be a zero point $g_{0}$ for $l(g)$. Then we can
obtain
\begin{equation}
l^{\prime}(g)=\dfrac{a}{m^{2}}\sech^{2}\left( \dfrac{ag}{m^{2}}\right)
-\dfrac{1}{a}%
\end{equation}

(1) When $m^{2}\geq a^{2}$, we have
\begin{equation}
\dfrac{1}{a}\geq\dfrac{a}{m^{2}}\Rightarrow l^{\prime}(g)\leq0\Rightarrow
l(g)\downarrow
\end{equation}
then we know that there is only one zero point $g_{0}$ for $l(g)$.

(1.1) When $g\in(-\infty, g_{0})$, $l(g)\geq0$, then $h^{\prime}(g)\geq0$,
$h(g)\uparrow$;

(1.2) When $g\in(g_{0}, +\infty)$, $l(g)<0$, then $h^{\prime}(g)<0$,
$h(g)\downarrow$.

From (1.1) and (1.2), we know that $\Phi_{\text{rigid}}(g)$ has a unimodal form.\newline

(2) When $m^{2}<a^{2}$, by solving equation
\begin{equation}
l^{\prime}(g)=\dfrac{a}{m^{2}}\sech^{2}\left( \dfrac{ag}{m^{2}}\right)
-\dfrac{1}{a}=0
\end{equation}
we have \begin{numcases}{}
	g_{1}=-\dfrac{m^{2}}{a}\arcosh\left(\dfrac{a}{m}\right)\\
	g_{2}=\dfrac{m^{2}}{a}\arcosh\left(\dfrac{a}{m}\right)
\end{numcases}

(2.1) When $g\in(-\infty, g_{1})$, $l^{\prime}(g)\leq0$, $l(g)\downarrow$;

(2.2) When $g\in(g_{1}, g_{2})$, $l^{\prime}(g)>0$, $l(g)\uparrow$;

(2.3) When $g\in(g_{2}, +\infty)$, $l^{\prime}(g)\leq0$, $l(g)\downarrow$.

Only if $l(g_{1})<0$ and $l(g_{2})>0$, $l(g)$ has three zero points, which are
denoted as $g_{3}<g_{4}<g_{5}$. The condition is equivalent to
\begin{numcases}{\iff}
	l(g_{1})=\tanh\left(-\arcosh\left(\dfrac{a}{m}\right)\right)-\dfrac{-\dfrac{m^{2}}{a}\arcosh\left(\dfrac{a}{m}\right)+v}{a}<0\\
	l(g_{2})=\tanh\left(\arcosh\left(\dfrac{a}{m}\right)\right)-\dfrac{\dfrac{m^{2}}{a}\arcosh\left(\dfrac{a}{m}\right)+v}{a}>0
\end{numcases}
\begin{equation}
\iff\dfrac{m^{2}}{a}\arcosh\left( \dfrac{a}{m}\right) -a\tanh\left(
\arcosh\left( \dfrac{a}{m}\right) \right) <v<a\tanh\left( \arcosh\left(
\dfrac{a}{m}\right) \right) -\dfrac{m^{2}}{a}\arcosh\left( \dfrac{a}%
{m}\right)
\end{equation}
\begin{equation}
\iff\dfrac{m^{2}}{a}\arcosh\left( \dfrac{a}{m}\right) -\sqrt{a^{2}-m^{2}%
}<v<\sqrt{a^{2}-m^{2}}-\dfrac{m^{2}}{a}\arcosh\left( \dfrac{a}{m}\right)
\end{equation}

Denote
\begin{equation}
c(a):=\sqrt{a^{2}-m^{2}}-\dfrac{m^{2}}{a}\arcosh\left( \dfrac{a}{m}\right)
\end{equation}
then
\begin{equation}
\dot{c}(a)=\dfrac{\sqrt{a^{2}-m^{2}}}{a}+\dfrac{m^{2}\arcosh\left( \dfrac
{a}{m}\right) }{a^{2}}>0.
\end{equation}

We know $c(a)$ increases as $a$ increases, with $c(m)=0$, we have
\begin{equation}
\sqrt{a^{2}-m^{2}}-\dfrac{m^{2}}{a}\arcosh\left( \dfrac{a}{m}\right) >0
\end{equation}

Therefore, the above condition is equivalent to
\begin{equation}
0\leq v<\sqrt{a^{2}-m^{2}}-\dfrac{m^{2}}{a}\arcosh\left( \dfrac{a}{m}\right)
\end{equation}
we define the \textit{critical environmental velocity}
\begin{equation}
v_{c}=\sqrt{a^{2}-m^{2}}-\dfrac{m^{2}}{a}\arcosh\left( \dfrac{a}{m}\right) .\label{vcsi}
\end{equation}

(A) When $0\leq v<v_{c}$, we have the conclusion:

(2.4) When $g\in(-\infty, g_{3})$, $l(g)\geq0$, then $h^{\prime}(g)\geq0$,
$h(g)\uparrow$;

(2.5) When $g\in(g_{3}, g_{4})$, $l(g)<0$, then $h^{\prime}(g)<0$,
$h(g)\downarrow$;

(2.6) When $g\in(g_{4}, g_{5})$, $l(g)\geq0$, then $h^{\prime}(g)\geq0$,
$h(g)\uparrow$;

(2.7) When $g\in(g_{5}, +\infty)$, $l(g)<0$, then $h^{\prime}(g)<0$,
$h(g)\downarrow$.

From (2.4), (2.5), (2.6) and (2.7), we know that $\Phi_{\text{rigid}}(g)$ has a bimodal
form when $v<v_{c}$, and the peaks are located at $g=g_{3}$ (left peak) and
$g=g_{5}$ (right peak).\newline

For $v_{1}$, we denote the zero points of $l(g)$ as $g_{1,3}<g_{1,4}<g_{1,5}$;
for $v_{2}>v_{1}$, we denote the zero points of $l(g)$ as $g_{2,3}%
<g_{2,4}<g_{2,5}$. Then we have
\begin{equation}
l(g,v_{1})=\tanh\left( \dfrac{ag}{m^{2}}\right) -\dfrac{g+v_{1}}{a}%
>\tanh\left( \dfrac{ag}{m^{2}}\right) -\dfrac{g+v_{2}}{a}=l(g,v_{2}).
\end{equation}

(a) $l(g_{2,3},v_{1})>l(g_{2,3},v_{2})=0$, due to (2.1), we have
$g_{2,3}<g_{1,3}$;

(b) $l(g_{2,4},v_{1})>l(g_{2,4},v_{2})=0$, due to (2.2), we have
$g_{2,4}>g_{1,4}$;

(c) $l(g_{2,5},v_{1})>l(g_{2,5},v_{2})=0$, due to (2.3), we have
$g_{2,5}<g_{1,5}$.

We can see that as $v$ increases, the three zeros of $l(g)$ also shift
accordingly, with $g_{3}$ moving continuously to the left, and $g_{4}$,
$g_{5}$ moving closer to $g_{2}=\dfrac{m^{2}}{a}\arcosh\left( \dfrac{a}%
{m}\right) $. When these two points coincide with $g_{2}=\dfrac{m^{2}}%
{a}\arcosh\left( \dfrac{a}{m}\right) $, the population distribution
transitions from bimodal to unimodal. Furthermore, this means that the
\textit{original right peak at $g=g_{5}$ disappears}.\newline

(B) When $v=v_{c}$, $l(g)$ has two zero points $g_{6}<g_{7}$.

(2.8) When $g\in(-\infty, g_{6})$, $l(g)\geq0$, then $h^{\prime}(g)\geq0$,
$h(g)\uparrow$;

(2.9) When $g\in(g_{6}, +\infty)$, $l(g)\leq0$, then $h^{\prime}(g)\leq0$,
$h(g)\downarrow$;

From (2.8) and (2.9), we know that $\Phi_{\text{rigid}}(g)$ has a unimodal form when
$v=v_{c}$, and the peak is located at $g=g_{6}$.\newline

(C) When $v>v_{c}$, $l(g)$ has one zero point $g_{8}$.

(2.10) When $g\in(-\infty, g_{8})$, $l(g)>0$, then $h^{\prime}(g)>0$,
$h(g)\uparrow$;

(2.11) When $g\in(g_{8}, +\infty)$, $l(g)<0$, then $h^{\prime}(g)<0$,
$h(g)\downarrow$;

From (2.10) and (2.11), we know that $\Phi_{\text{rigid}}(g)$ has a unimodal form when
$v>v_{c}$, and the peak is located at $g=g_{8}$.\newline

Now, from Eq. \eqref{deltaequilibrium}, we denote
\begin{equation}
y(v):=\dfrac{\exp{\left[ -\dfrac{2(g-a)v+v^{2}}{2m^{2}}\right] }}{\exp{\left(
\dfrac{2av}{m^{2}}\right)+1}}>0
\end{equation}
then we have
\begin{equation}%
\begin{split}
\dfrac{y^{\prime}(v)}{y(v)}= & \dfrac{d}{dv}\ln y(v)\\
= & \dfrac{d}{dv}\left[ -\dfrac{2(g-a)v+v^{2}}{2m^{2}}-\ln\left( \exp{\left(
\dfrac{2av}{m^{2}}\right) +1}\right) \right] \\
= & \dfrac{-g+a-v}{m^{2}}-\dfrac{\dfrac{2a}{m^{2}}}{1+\exp{\left( -\dfrac
{2av}{m^{2}}\right) }}%
\end{split}
\end{equation}

Since $v\geq0$, we have
\begin{equation}
\dfrac{1}{1+\exp{\left( -\dfrac{2av}{m^{2}}\right) }}\in\left( \dfrac{1}%
{2},1\right)
\end{equation}
then we can obtain
\begin{equation}
\dfrac{y^{\prime}(v)}{y(v)}\leq\dfrac{-g+a-v}{m^{2}}-\dfrac{a}{m^{2}}=-\dfrac{g+v}{m^{2}}
\end{equation}
(3.1) For any fixed $g<0$ and $0<v<-g$, as $v$ increases, $y(v)$ increases,
then $\Phi_{\text{rigid}}(g)$ increases; for $v\geq-g$, as $v$ increases, $y(v)$
decreases, then $\Phi_{\text{rigid}}(g)$ decreases;

(3.2) For any fixed $g\geq0$, as $v$ increases, $y(v)$ decreases, then
$\Phi_{\text{rigid}}(g)$ decreases.\newline

What's more, we have
\begin{equation}
\Rightarrow{\int_{-\infty}^{\infty}g\Phi_{\text{rigid}}(g)dg=}-v-a\tanh\left( \frac
{av}{m^{2}}\right) .
\end{equation}
then the distribution $\Phi_{\text{rigid}}(g-vt)$ has a mean $\mu_{v}(t)$ and variance
$\sigma_{v}^{2}$ at time $t$, with the specific forms given by
\begin{equation}
\mu_{v}(t)=vt-v-a\tanh\left( \frac{av}{m^{2}}\right) \quad\text{ and }%
\quad\sigma_{v}^{2}=m^{2}+a^{2}-a^{2}\tanh^{2}\left( \frac{av}{m^{2}}\right)
\end{equation}
respectively.

This is independent of the sign of $a$, as it should be, and which we will
take as positive. We can write
\begin{equation}
{\int_{-\infty}^{\infty}g\Phi_{\text{rigid}}(g)dg}=-v- \operatorname*{sgn}(v)a\tanh\left(
\frac{a|v|}{m^{2}}\right) .
\end{equation}

So for $\dfrac{a|v|}{m^{2}}\gg1$ we have
\begin{equation}
{\int_{-\infty}^{\infty}g\Phi_{\text{rigid}}(g)dg}\simeq-v-\operatorname*{sgn}(v)a .
\end{equation}
and
\begin{equation}
\mu_{v}(t)=-vt-v-\operatorname*{sgn}(v)a\quad\text{ and }\quad\sigma_{v}%
^{2}=m^{2}\label{aaaaa}%
\end{equation}
This corresponds to the contribution to the mean arising from just one of the
peaks of $w$, according to the sign of $v$.

\subsubsection{Convergence rate}

\label{subsection315} From Eq. \eqref{Xtformula}, we can see that the smaller
$\dfrac{\lambda_{2}}{\lambda_{1}}$ is, the faster the sequence converges.
\begin{equation}%
\begin{split}
\dfrac{\lambda_{2}}{\lambda_{1}}= & \dfrac{1-\sqrt{(p-q)^{2}+4pq\exp\left(
-\frac{4a^{2}}{m^{2}}\right) }}{1+\sqrt{(p-q)^{2}+4pq\exp\left( -\frac{4a^{2}%
}{m^{2}}\right) }}\\
= & -1+\dfrac{2}{1+\sqrt{(p-q)^{2}+4pq\exp\left( -\frac{4a^{2}}{m^{2}}\right)
}}\in(0,1)
\end{split}
\end{equation}
we can see that the bigger $m^{2}$ is, the smaller $\dfrac{\lambda_{2}%
}{\lambda_{1}}$ is, and then the faster the sequence converges.

Denote
\begin{equation}
\dfrac{\lambda_{2}}{\lambda_{1}}:=k
\end{equation}
we have
\begin{equation}
\lambda_{2}=k\lambda_{1}%
\end{equation}
then
\begin{equation}
\dfrac{1}{\lambda_{1}-\lambda_{2}}=\dfrac{1}{(1-k)\lambda_{1}}=\sqrt
{\dfrac{2\pi m^{2}}{(p-q)^{2}+4pq\exp\left( -\frac{4a^{2}}{m^{2}}\right) }%
}\cdot\exp\left( \frac{v^{2}}{2m^{2}}\right)
\end{equation}
we can see that the smaller $\dfrac{\lambda_{2}}{\lambda_{1}}=k$ is, the
smaller $\dfrac{1}{\lambda_{1}-\lambda_{2}}$ is, and then the faster the
sequence converges. For constant $m^{2}$, the smaller $v$ is, the faster the
sequence converges.

\subsection{$W(g)$ has a single jump}

\label{subsection32}

\subsubsection{Recurrence relation}

\label{subsection321} For the fitness function
\begin{equation}
W(g)=p\delta(g+a)+q\delta(g-a)
\end{equation}
with $p+q=1$, we assume that the fitness function $W(g)$ does not change over
time, which indicates that $v=0$. At a certain moment $t$, the fitness function
suddenly changes, with a magnitude of $d$:
\begin{equation}
W_{1}(g)=p\delta(g+a+d)+q\delta(g-a+d).
\end{equation}

Substituting the above form of $w_{1}(g)$ into Eq. \eqref{comoving} yields
\begin{equation}
\Psi(g,t+1)=\dfrac{pf(g+a+d)\Psi(-a-d,t)+qf(g-a+d)\Psi(a-d,t)}{p\Psi
(-a-d,t)+q\Psi(a-d,t)} .\label{Phi1}%
\end{equation}
We now set $g$ to be $-a-d$ and $a-d$ in this equation to obtain
\begin{numcases}{}
	\Psi(-a-d,t+1)=\dfrac{pf(0)\Psi(-a-d,t)+qf(-2a)\Psi(a-d,t)}{p\Psi(-a-d,t)+q\Psi(a-d,t)}\\
	\Psi(a-d,t+1)=\dfrac{pf(2a)\Psi(-a-d,t)+qf(0)\Psi(a-d,t)}{p\Psi(-a-d,t)+q\Psi(a-d,t)}
\end{numcases}

Let us define
\begin{equation}
\mathbf{Y}(t)=\left(
\begin{array}
[c]{c}%
\Psi(-a-d,t)\\
\Psi(a-d,t)
\end{array}
\right)
\end{equation}
then the system can be written as
\begin{equation}
\mathbf{Y}(t+1)=\frac{\mathbf{NDY}(t)}{\mathbf{F}^{T}\mathbf{DY}(t)}%
\end{equation}
where
\begin{equation}
\mathbf{N}=\left(
\begin{array}
[c]{cc}%
f(0), & f(-2a)\\
f(2a), & f(0)
\end{array}
\right)  ,\qquad\mathbf{D}=\left(
\begin{array}
[c]{cc}%
p, & 0\\
0, & q
\end{array}
\right)  ,\qquad\mathbf{F}=\left(
\begin{array}
[c]{c}%
1\\
1
\end{array}
\right)  .
\end{equation}
For $t=1,2,\cdots$, the solution is
\begin{equation}
\mathbf{Y}(t)=\frac{\left( \mathbf{ND}\right) ^{t}\mathbf{Y}(0)}
{\mathbf{F}^{T}\mathbf{D}\left( \mathbf{ND}\right)  ^{t-1}\mathbf{Y}(0)}
.\label{solYt}%
\end{equation}

We can calculate the eigenvalues of $\mathbf{ND}$ \begin{numcases}{}
	\mu_{1}=\frac{1}{2\sqrt{2\pi m^{2}}}\left[1+\sqrt{(p-q)^{2}+4pq\exp\left(-\frac{4a^{2}}{m^{2}}\right)}\right]\\
	\mu_{2}=\frac{1}{2\sqrt{2\pi m^{2}}}\left[1-\sqrt{(p-q)^{2}+4pq\exp\left(-\frac{4a^{2}}{m^{2}}\right)}\right]
\end{numcases}

We can easily see that $\mu_{1}>\mu_{2}$.

Let $\boldsymbol{\gamma}^{(1)}$ and $\boldsymbol{\gamma}^{(2)}$ be the
corresponding right eigenvectors, $\boldsymbol{\eta}^{(1)T}$ and
$\boldsymbol{\eta}^{(2)T}$ be the corresponding left eigenvectors, with the
properties \begin{numcases}{}
	\boldsymbol{\gamma}^{(1)}\boldsymbol{\eta}^{(1)T}+\boldsymbol{\gamma}^{(2)}\boldsymbol{\eta}^{(2)T}=\mathbf{I}\\
	\boldsymbol{\eta}^{(i)T}\boldsymbol{\gamma}^{(j)}=\delta_{i,j}
\end{numcases}
\begin{equation}
\Rightarrow\mathbf{ND}=\boldsymbol{\gamma}^{(1)}\boldsymbol{\eta}%
^{(1)T}\mathbf{ND}+\boldsymbol{\gamma}^{(2)}\boldsymbol{\eta}^{(2)T}%
\mathbf{ND}=\mu_{1}\boldsymbol{\gamma}^{(1)}\boldsymbol{\eta}^{(1)T}+\mu
_{2}\boldsymbol{\gamma}^{(2)}\boldsymbol{\eta}^{(2)T}\label{ND}%
\end{equation}

Substitute Eq. \eqref{ND} into Eq. \eqref{solYt}, we have
\begin{equation}%
\begin{split}
\mathbf{Y}(t)= & \frac{\left[ \mu_{1}^{t}\boldsymbol{\gamma}^{(1)}%
\boldsymbol{\eta}^{(1)T}+\mu_{2}^{t}\boldsymbol{\gamma}^{(2)}\boldsymbol{\eta
}^{(2)T}\right] \mathbf{Y}(0)}{\mathbf{F}^{T}\mathbf{D}\left[ \mu_{1}^{t-1}
\boldsymbol{\gamma}^{(1)}\boldsymbol{\eta}^{(1)T}+\mu_{2}^{t-1}%
\boldsymbol{\gamma}^{(2)}\boldsymbol{\eta}^{(2)T}\right] \mathbf{Y}(0)}\\
= & \frac{\left[ \mu_{1}\boldsymbol{\gamma}^{(1)}\boldsymbol{\eta}^{(1)T}%
+\mu_{2}\cdot\left( \frac{\mu_{2}}{\mu_{1}}\right) ^{t-1}\boldsymbol{\gamma
}^{(2)}\boldsymbol{\eta}^{(2)T}\right]  \mathbf{Y}(0)}{\mathbf{F}%
^{T}\mathbf{D}\left[ \boldsymbol{\gamma}^{(1)}\boldsymbol{\eta}^{(1)T}+\left(
\frac{\mu_{2}}{\mu_{1}}\right) ^{t-1}\boldsymbol{\gamma}^{(2)}\boldsymbol{\eta
}^{(2)T}\right] \mathbf{Y}(0)}.\label{Ytformula}
\end{split}
\end{equation}

If $\mu_{1}$ is the larger (in magnitude) of the two eigenvalues then for
$t^{\prime\prime}\gg1$, we have
\begin{equation}
\mathbf{Y}(t^{\prime\prime})=\frac{\mu_{1}\boldsymbol{\gamma}^{(1)}
\boldsymbol{\eta}^{(1)T}\mathbf{Y}(0)}{\mathbf{F}^{T}\mathbf{D}%
\boldsymbol{\gamma}^{(1)}\boldsymbol{\eta}^{(1)T}\mathbf{Y}(0)}=\frac
{\lambda_{1}\boldsymbol{\gamma}^{(1)}}{\mathbf{F} ^{T}\mathbf{D}
\boldsymbol{\gamma}^{(1)}} \equiv\left(
\begin{array}
[c]{c}%
Y_{1}(t^{\prime\prime})\\
Y_{2}(t^{\prime\prime})
\end{array}
\right)
\end{equation}
so we obtain a result independent of the initial state, and components are
essentially determined by a single eigenvector.

A measure of asymmetry of the distribution at generation $t^{\prime\prime}$ is
some comparison of the two components of $\mathbf{Y}(t^{\prime\prime})$.

The actual distribution at generation $t^{\prime\prime}$ follows from Eq.
\eqref{Phi1}:
\begin{equation}
\Psi(g,t^{\prime\prime})=\dfrac{pf(g+a+d)Y_{1}(t^{\prime\prime}%
)+qf(g-a+d)Y_{2}(t^{\prime\prime})} {pY_{1}(t^{\prime\prime})+qY_{2}%
(t^{\prime\prime})}.
\end{equation}
Thus the relationship between $a$, $d$, the variance of $f(g)$, along with the
values of $p$ and $q$ will determine the bimodal (or not) nature of
$\Psi(g,t^{\prime\prime})$.

Note that
\begin{equation}%
\begin{split}
{\int_{-\infty}^{\infty}g\Psi(g,t^{\prime\prime})dg}= & \dfrac{(-a-d)pY_{1}%
(t^{\prime\prime})+(a-d)qY_{2}(t^{\prime\prime})}{pY_{1}(t^{\prime\prime
})+qY_{2}(t^{\prime\prime})}\\
= & -d-a\cdot\dfrac{pY_{1}(t^{\prime\prime})-qY_{2}(t^{\prime\prime})}%
{pY_{1}(t^{\prime\prime})+qY_{2}(t^{\prime\prime})}%
\end{split}
\end{equation}
and
\begin{equation}%
\begin{split}
\int_{-\infty}^{\infty}g^{2}\Psi(g,t^{\prime\prime})dg= & \dfrac
{[m^{2}+(-d-a)^{2}]pY_{1}(t^{\prime\prime})+[m^{2}+(-d+a)^{2}]qY_{2}%
(t^{\prime\prime})}{pY_{1}(t^{\prime\prime})+qY_{2}(t^{\prime\prime})}\\
= & m^{2}+d^{2}+a^{2}+2ad\cdot\dfrac{pY_{1}(t^{\prime\prime})-qY_{2}%
(t^{\prime\prime})}{pY_{1}(t^{\prime\prime})+qY_{2}(t^{\prime\prime})}%
\end{split}
\end{equation}
we have
\begin{equation}%
\begin{split}
& \int_{-\infty}^{\infty}g^{2}\Psi(g,t^{\prime\prime})dg-\left[ \int_{-\infty
}^{\infty}g\Psi(g,t^{\prime\prime})dg\right] ^{2}\\
= & m^{2}+a^{2}-a^{2}\cdot\left[ \dfrac{pY_{1}(t^{\prime\prime})-qY_{2}%
(t^{\prime\prime})}{pY_{1}(t^{\prime\prime})+qY_{2}(t^{\prime\prime})}\right]
^{2}%
\end{split}
\end{equation}

\subsubsection{$Y_{1}(t^{\prime\prime})$ v.s. $Y_{2}(t^{\prime\prime})$
($t^{\prime\prime}\gg1$)}

\label{subsection322} From above, we have
\begin{equation}%
\begin{split}
\frac{Y_{1}(t^{\prime\prime})}{Y_{2}(t^{\prime\prime})}= & \dfrac
{pf(0)Y_{1}(t^{\prime\prime})+qf(-2a)Y_{2}(t^{\prime\prime})}{pf(2a)Y_{1}%
(t^{\prime\prime})+qf(0)Y_{2}(t^{\prime\prime})}\\
= & \dfrac{pf(0)\cdot\dfrac{Y_{1}(t^{\prime\prime})}{Y_{2}(t^{\prime\prime}%
)}+qf(-2a)}{pf(2a)\cdot\dfrac{Y_{1}(t^{\prime\prime})}{Y_{2}(t^{\prime\prime
})}+qf(0)}%
\end{split}
\end{equation}
\begin{equation}
\iff pf(2a)\cdot\left[ \dfrac{Y_{1}(t^{\prime\prime})}{Y_{2}(t^{\prime\prime
})}\right] ^{2}+[qf(0)-pf(0)]\cdot\dfrac{Y_{1}(t^{\prime\prime})}%
{Y_{2}(t^{\prime\prime})}-qf(-2a)=0
\end{equation}
\begin{equation}
\dfrac{Y_{1}(t^{\prime\prime})}{Y_{2}(t^{\prime\prime})}=\frac{\exp\left(
\frac{2a^{2}}{m^{2}}\right) }{2p}\left[ (2p-1)\pm\sqrt{\left( 1-2p\right)
^{2}+4p(1-p)\exp\left( -\frac{4a^{2}}{m^{2}}\right) }\right]
\end{equation}

Since
\begin{equation}
\sqrt{\left( 1-2p\right) ^{2}+4p(1-p)\exp\left( -\frac{4a^{2}}{m^{2}}\right)
}>|2p-1|
\end{equation}
and
\begin{equation}
Y_{1}(t^{\prime\prime})>0\;\;\&\;\;Y_{2}(t^{\prime\prime})>0
\end{equation}
we have
\begin{equation}
\dfrac{Y_{1}(t^{\prime\prime})}{Y_{2}(t^{\prime\prime})}=\frac{\exp\left(
\frac{2a^{2}}{m^{2}} \right) }{2p}\left[ (2p-1)+\sqrt{\left( 1-2p\right)
^{2}+4p(1-p)\exp\left( -\frac{4a^{2}}{m^{2}}\right) }\right]
\end{equation}

When $p\neq q$, after a lot of generations, we can see that $\dfrac
{Y_{1}(t^{\prime\prime})}{Y_{2}(t^{\prime\prime})}$ is independent of the jump
magnitude $d$ and time $t$. Its value can be determined by the parameters $a$,
$m$, $p$ and $q$, which are set at the beginning.

Now we can calculate
\begin{equation}%
\begin{split}
{\int_{-\infty}^{\infty}g\Psi(g,t^{\prime\prime})dg}= & d-a\cdot\dfrac
{pY_{1}(t^{\prime\prime})-qY_{2}(t^{\prime\prime})}{pY_{1}(t^{\prime\prime
})+qY_{2}(t^{\prime\prime})}\\
= & -d-a\cdot\dfrac{p\cdot\dfrac{Y_{1}(t^{\prime\prime})}{Y_{2}(t^{\prime
\prime})}-q}{p\cdot\dfrac{Y_{1}(t^{\prime\prime})}{Y_{2}(t^{\prime\prime})}+q}%
\end{split}
\end{equation}
and
\begin{equation}%
\begin{split}
& \int_{-\infty}^{\infty}g^{2}\Psi(g,t^{\prime\prime})dg-\left[ \int_{-\infty
}^{\infty}g\Psi(g,t^{\prime\prime})dg\right] ^{2}\\
= & m^{2}+a^{2}-a^{2}\cdot\left[ \dfrac{pY_{1}(t^{\prime\prime})-qY_{2}%
(t^{\prime\prime})}{pY_{1}(t^{\prime\prime})+qY_{2}(t^{\prime\prime})}\right]
^{2}\\
= & m^{2}+a^{2}-a^{2}\cdot\left[ \dfrac{p\cdot\dfrac{Y_{1}(t^{\prime\prime}%
)}{Y_{2}(t^{\prime\prime})}-q}{p\cdot\dfrac{Y_{1}(t^{\prime\prime})}%
{Y_{2}(t^{\prime\prime})}+q}\right] ^{2},
\end{split}
\end{equation}
which are independent of time $t^{\prime\prime}$, this indicates that the
distribution at a later time, $t^{\prime\prime}$, maintains a constant mean
and variance, implying an equilibrium distribution $\Psi_{d}(g)$:
\begin{equation}
\Psi(g,t^{\prime\prime})=\Psi_{d}(g)
\end{equation}
and \begin{numcases}{}
	Y_{1}(t^{\prime\prime})=\Psi(-a-d,t^{\prime\prime})=\Psi_{d}(-a-d)\\
	Y_{2}(t^{\prime\prime})=\Psi(a-d,t^{\prime\prime})=\Psi_{d}(a-d)	
\end{numcases}
for large time $t^{\prime\prime}$.

\subsubsection{Special case: $p=q=\dfrac{1}{2}$}

\label{subsection323} From the previous section, we have
\begin{equation}
\dfrac{\Psi_{d}(-a-d)}{\Psi_{d}(a-d)}=1
\end{equation}
\begin{equation}
\Rightarrow\Psi_{d}(-a-d)=\Psi_{d}(a-d)
\end{equation}

We can obtain
\begin{equation}
	\begin{split}
		\Psi_{d}(g)=&\dfrac{f(g+a+d)+f(g-a+d)}{2}\\
		=&\dfrac{1}{2}\sqrt{\dfrac{1}{2\pi m^{2}}}\cdot\left[\exp{\left(-\dfrac{(g+a+d)^{2}}{2m^{2}}\right)}+\exp{\left(-\dfrac{(g-a+d)^{2}}{2m^{2}}\right)}\right]
	\end{split}
\end{equation}

According to \eqref{staenvpsi0}, we have
\begin{equation}
	\Psi_{d}(g)=\Psi_{0}(g+d)\label{eqqqqqqq}.
\end{equation}

After the jump, we have
\begin{equation}
{\int_{-\infty}^{\infty}g\Psi_{d}(g)dg=}-d.
\end{equation}

\subsubsection{Convergence rate}

\label{subsection325} From Eq. \eqref{Ytformula}, we can see that the smaller
$\dfrac{\mu_{2}}{\mu_{1}}$ is, the faster the sequence converges.
\begin{equation}%
\begin{split}
\dfrac{\mu_{2}}{\mu_{1}}= & \dfrac{1-\sqrt{(p-q)^{2}+4pq\exp\left(
-\frac{4a^{2}}{m^{2}}\right) }}{1+\sqrt{(p-q)^{2}+4pq\exp\left( -\frac{4a^{2}%
}{m^{2}}\right) }}\\
= & -1+\dfrac{2}{1+\sqrt{(p-q)^{2}+4pq\exp\left( -\frac{4a^{2}}{m^{2}}\right)
}}\in(0,1)
\end{split}
\end{equation}
we can see that the bigger $m^{2}$ is, the smaller $\dfrac{\mu_{2}%
}{\mu_{1}}$ is, and then the faster the sequence converges.

Denote
\begin{equation}
\dfrac{\mu_{2}}{\mu_{1}}:=k
\end{equation}
we have
\begin{equation}
\mu_{2}=k\mu_{1}%
\end{equation}
then
\begin{equation}
\dfrac{1}{\mu_{1}-\mu_{2}}=\dfrac{1}{(1-k)\mu_{1}}=\sqrt
{\dfrac{2\pi m^{2}}{(p-q)^{2}+4pq\exp\left( -\frac{4a^{2}}{m^{2}}\right)}}
\end{equation}
we can see that the smaller $\dfrac{\mu_{2}}{\mu_{1}}=k$ is, the
smaller $\dfrac{1}{\mu_{1}-\mu_{2}}$ is, and then the faster the
sequence converges.

\subsubsection{$W(g)$ has multiple jumps}
\label{subsection326}
From \eqref{eqqqqqqq}, we observe that when the fitness function $W(g)$ undergoes a single jump—indicating the population has experienced an environmental change as an extreme event—the shape of its equilibrium distribution remains identical to that in a static environment. Additionally, the distance between the equilibrium distribution before the single jump and that after it corresponds to the jump magnitude, $d$. This implies that the post-jump distribution can be interpreted as a translation of the pre-jump distribution along the genotypic value axis.

Building upon the above results, we can readily generalize: when environmental changes occur as multiple extreme events, such that the fitness function $W(g)$ experiences several jumps, we denote these abrupt changes in temporal sequence as $d_{1}, d_{2}, \cdots, d_{n}$. After numerous generations of inheritance, the population ultimately adapts to the most recent extreme environmental shift. Consequently, its equilibrium distribution shape aligns with that expected under a static environment, and the distance between the pre-change equilibrium distribution and the final equilibrium distribution is precisely the magnitude of the last jump, $d_{n}$. In essence, the final population distribution can be viewed as a translation of the initial distribution along the genotypic value axis.

\subsection{Non-equivalent fitness peaks}

\label{subsection33} For fitness function $W(g)$ with non-equivalent peaks
(peaks of different heights or different widths), we give one example of it
leading to a bimodal distribution of genotypic values at equilibrium.

We define fitness function
\begin{equation}
W(g):=\exp{\left( -\frac{(g+a)^{2}}{s^{2}}\right) }+k\exp{\left(
-\frac{(g-a)^{2}}{s^{2}}\right) }\label{bi Gauss}%
\end{equation}
which is illustrated in the following figure:
\begin{figure}[H]
\centering
\includegraphics[width=0.8\textwidth]{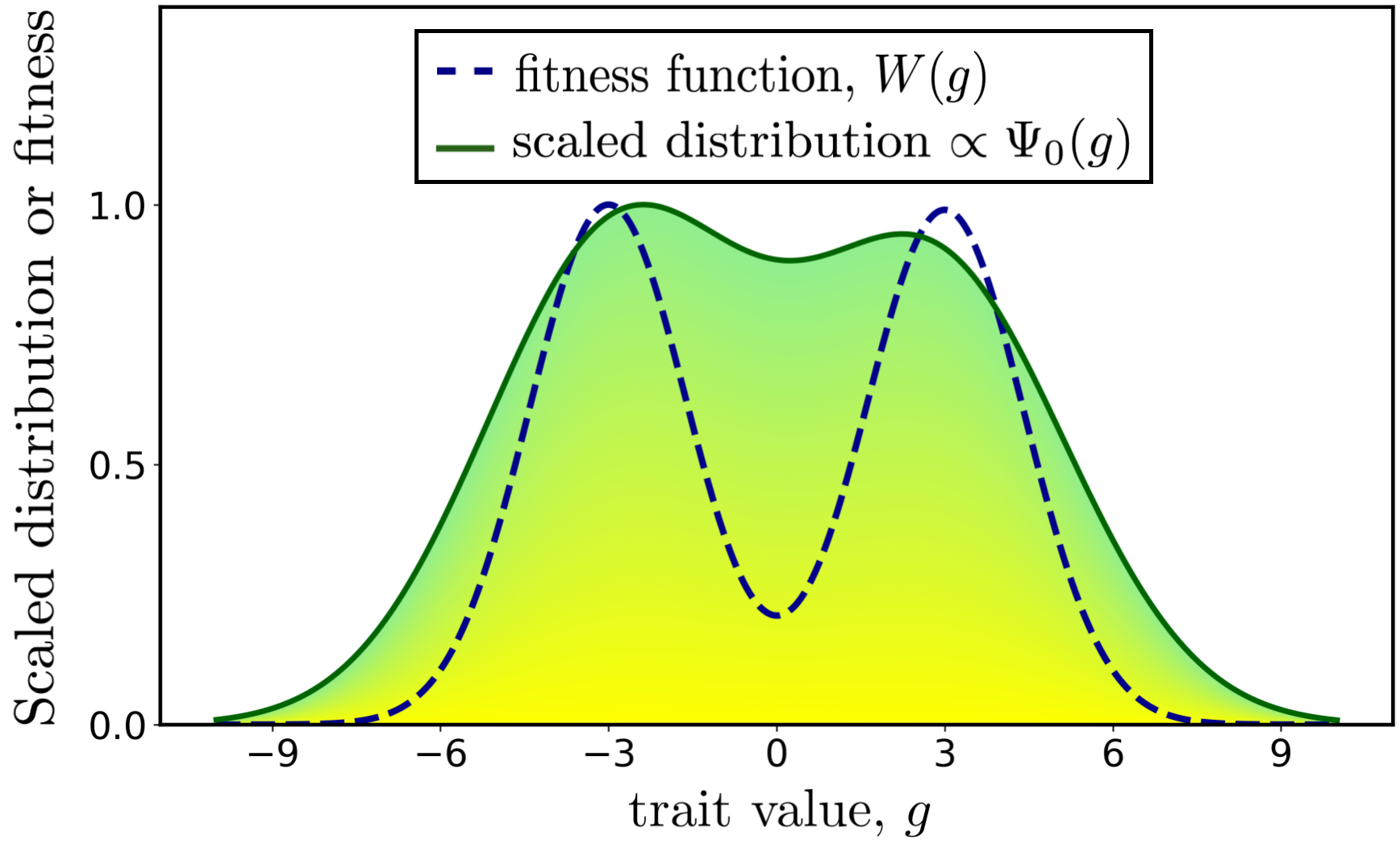}\caption{\textbf{Bimodal
fitness function and associated equilibrium distribution of genotypic trait
values in a static environment.} This figure presents a \textit{bimodal}
fitness function (dashed blue line) with peaks of varying heights, plotted
against the genotypic trait value, $g$. The fitness function exhibits peaks
near $g=-a$ and $g=a$. Additionally, the equilibrium genotypic distribution,
normalized to have a maximum height of unity, is displayed (green line) as a
function of $g$. For this figure, we used the mutational distribution from Eq.
\eqref{fgsiiiii} and the fitness function from Eq. \eqref{bi Gauss}, with
parameter values set to $m=2$, $s=\sqrt{2}$, $a=3$, and $k=0.99$.}%
\label{f5si}
\end{figure}

\subsection{The transition under different mutational distribution}

\label{subsection34} In the main text, for a bimodal fitness function in the
form of Eq. \eqref{bimodalfitnesssi}, we demonstrate through numerical simulations
that there exists a range of environmental velocities $v$ within which the
rigidly moving distribution remains bimodal. However, when the velocity $v$
exceeds this range, the rigidly moving distribution becomes unimodal. The
transition of the population distribution from bimodal to unimodal is, in
fact, continuous. This transition can sometimes be quite narrow, while at
other times, it may be more gradual. We illustrate this phenomenon by varying
the variance of the mutational distribution, $m^{2}$.

\begin{figure}[H]
    \centering
    \begin{subfigure}[b]{0.35\textwidth}
        \centering
        \includegraphics[width=\textwidth]{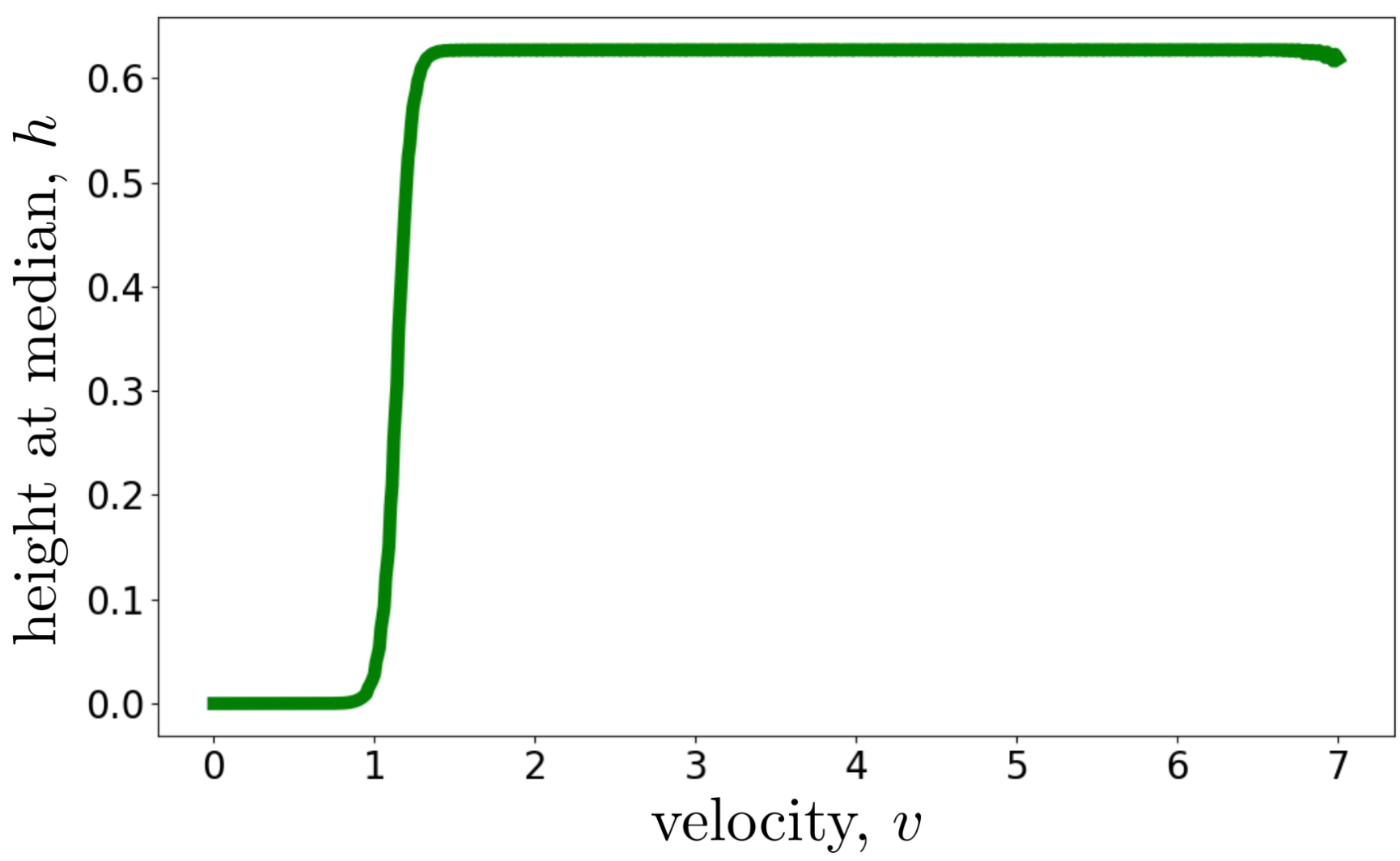}
        \caption{$m=0.5$}
    \end{subfigure}
    \hfill
    \begin{subfigure}[b]{0.35\textwidth}
        \centering
        \includegraphics[width=\textwidth]{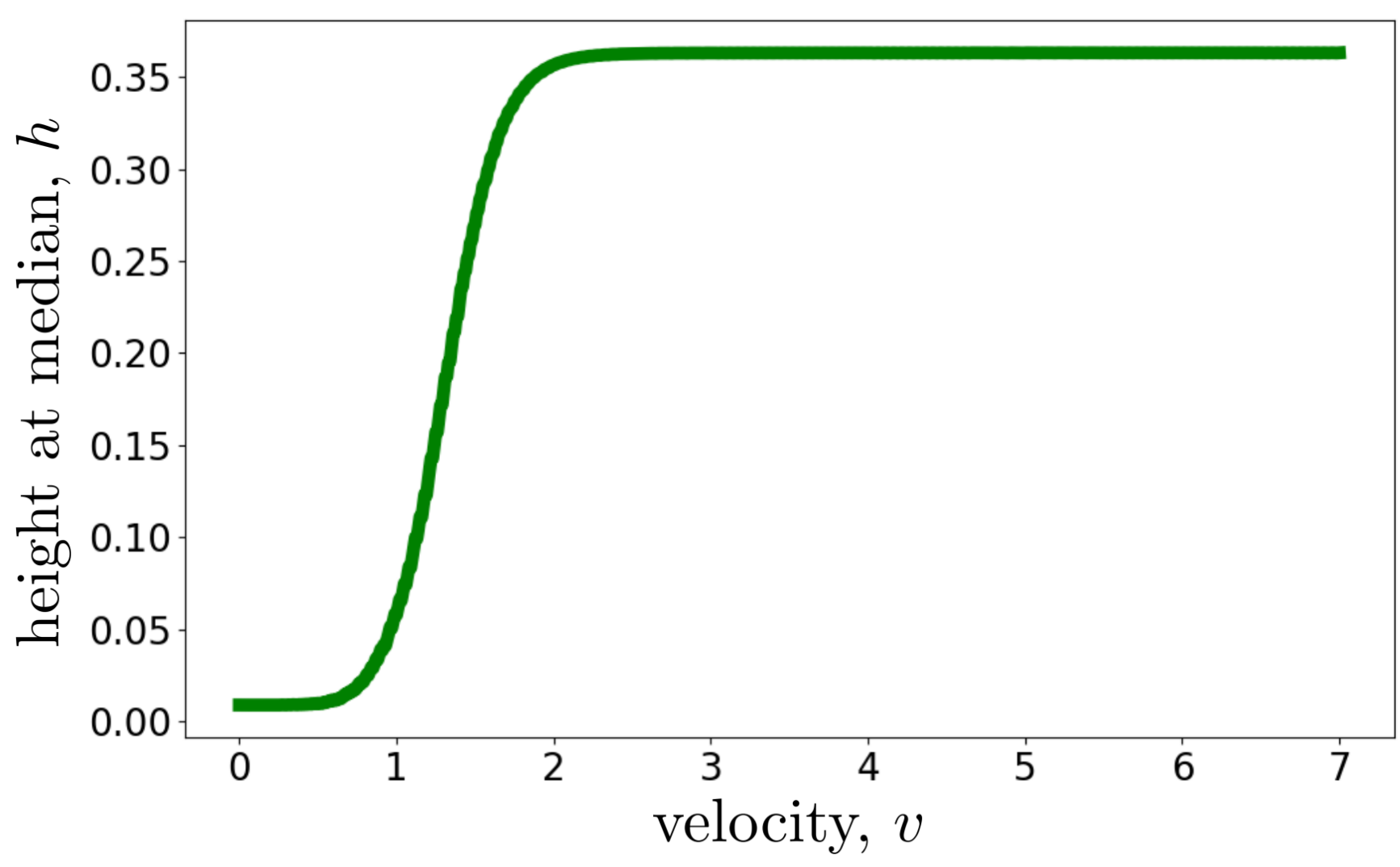}
        \caption{$m=1$}
    \end{subfigure}

    \vspace{0.5cm}

    \begin{subfigure}[b]{0.35\textwidth}
        \centering
        \includegraphics[width=\textwidth]{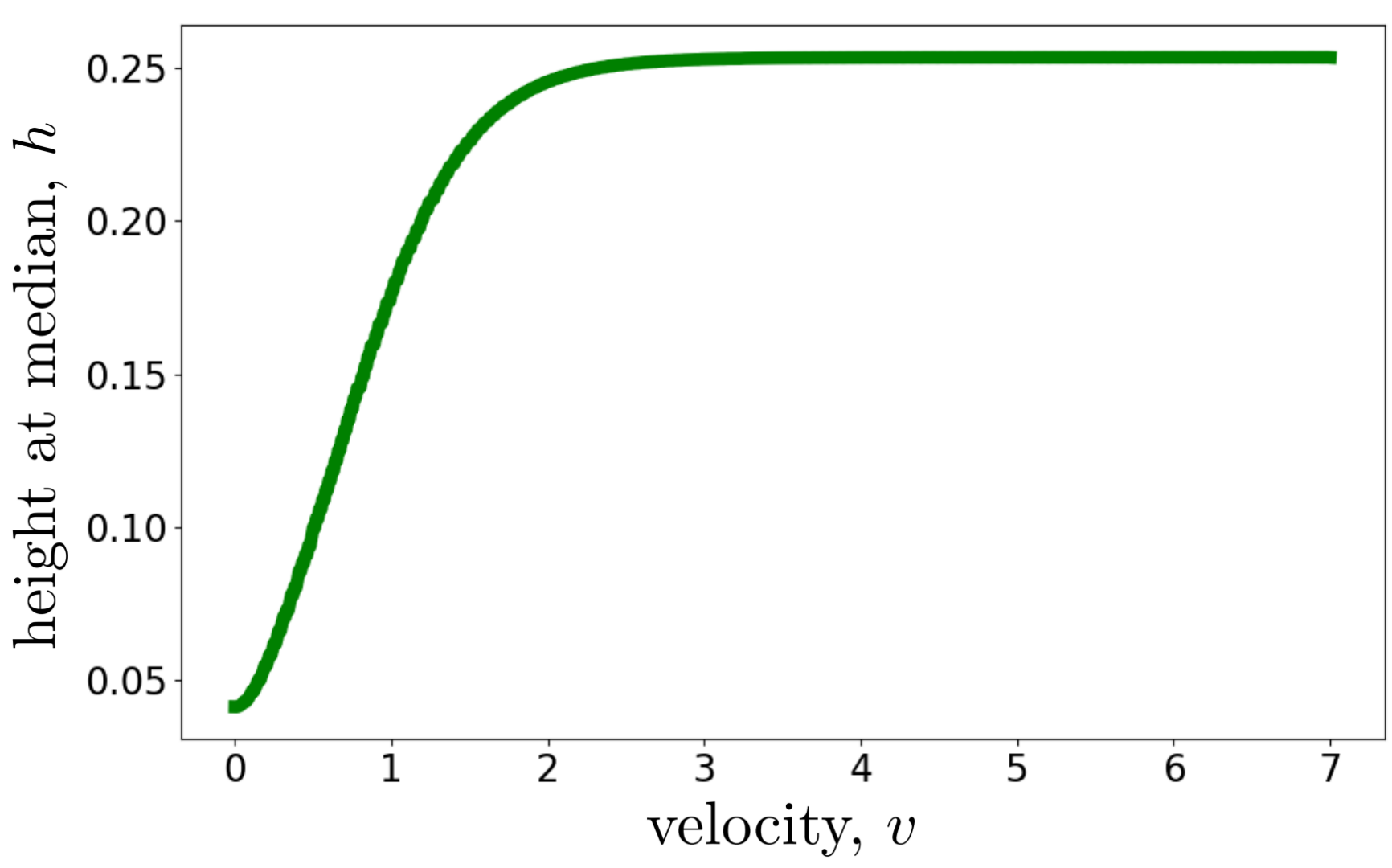}
        \caption{$m=1.5$}
    \end{subfigure}
    \hfill
    \begin{subfigure}[b]{0.35\textwidth}
        \centering
        \includegraphics[width=\textwidth]{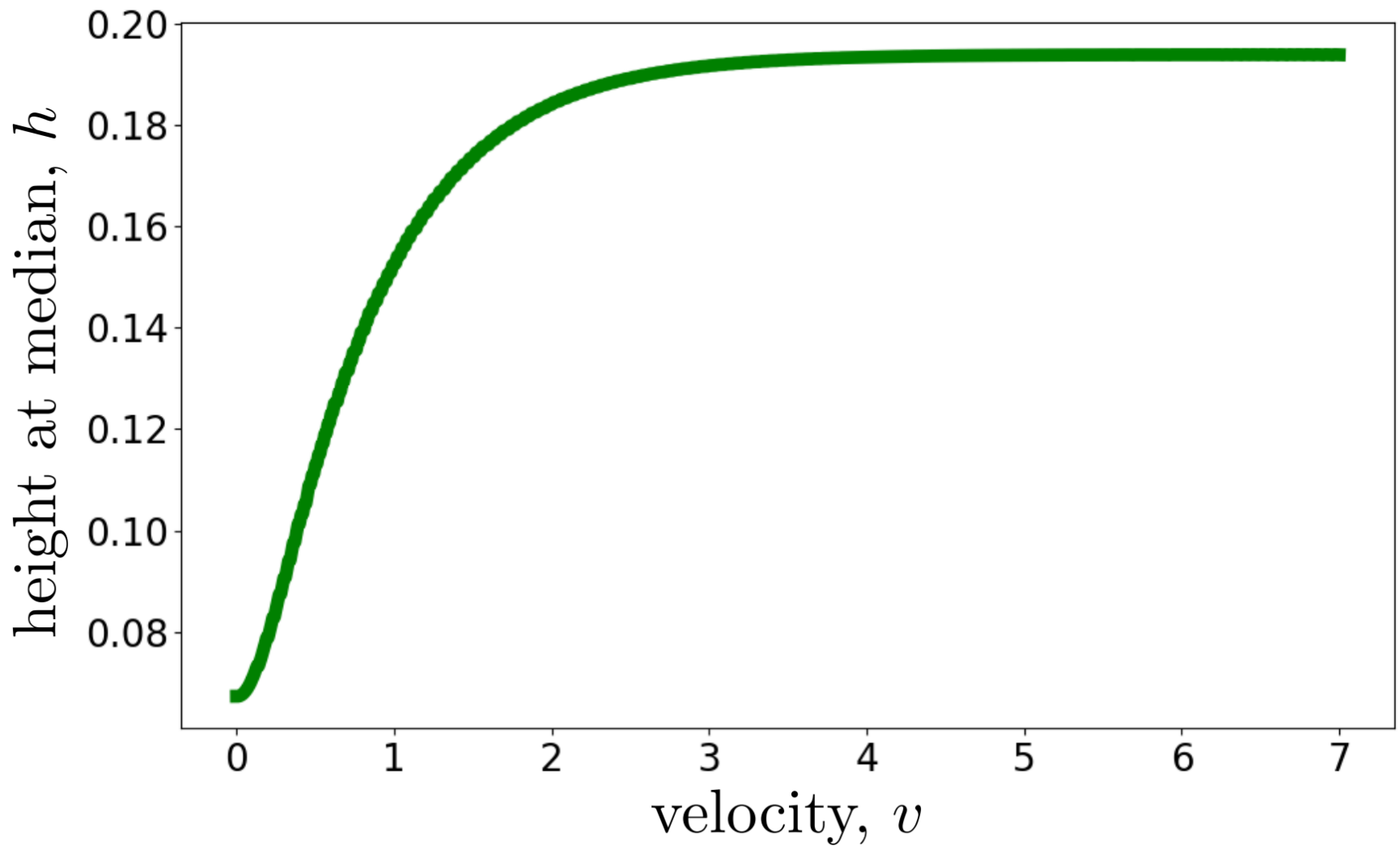}
        \caption{$m=2$}
    \end{subfigure}

    \vspace{0.5cm}

    \begin{subfigure}[b]{0.35\textwidth}
        \centering
        \includegraphics[width=\textwidth]{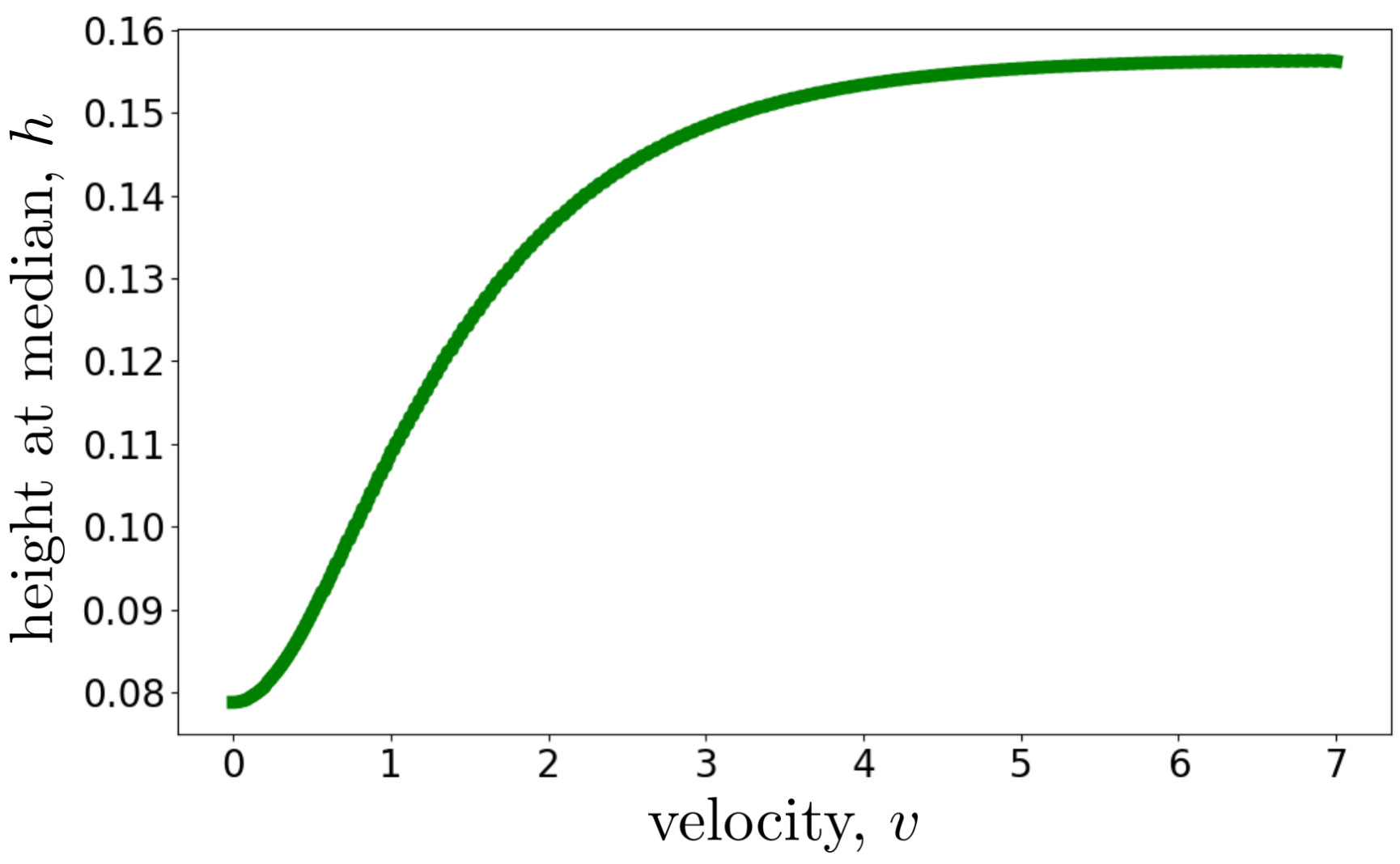}
        \caption{$m=2.5$}
    \end{subfigure}
    \hfill
    \begin{subfigure}[b]{0.35\textwidth}
        \centering
        \includegraphics[width=\textwidth]{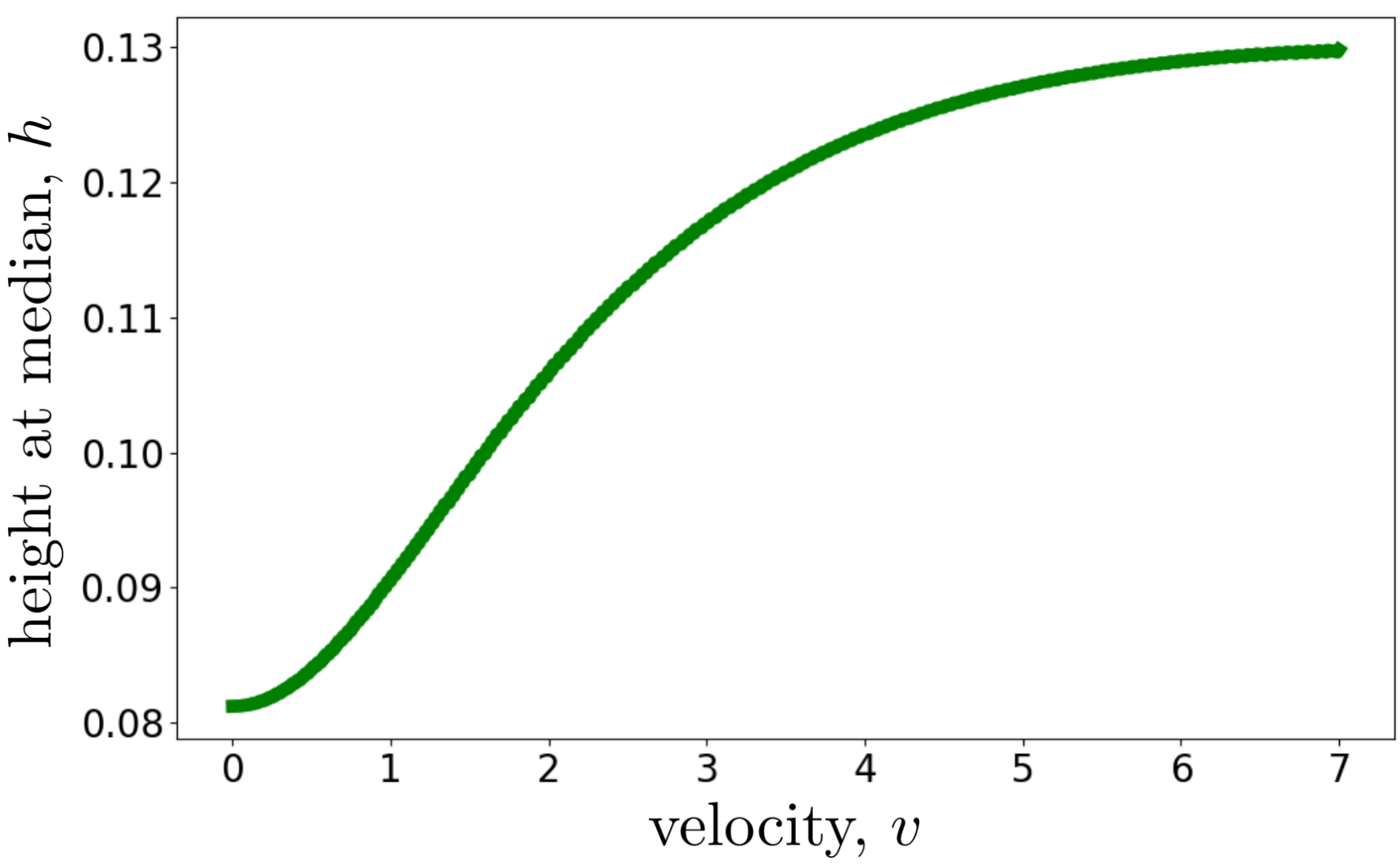}
        \caption{$m=3$}
    \end{subfigure}
    \caption{\textbf{Plot of the function describing the relationship between the height at the median of the distribution, $\bm{h}$, and the environmental velocity, $\bm{v}$, under varying variances $\bm{m^{2}}$ of the mutation distribution.} Specifically, $m^{2}$ takes values of $0.5^{2}$, $1^{2}$, $1.5^{2}$, $2^{2}$, $2.5^{2}$, and $3^{2}$.} 
    \label{f6si}
\end{figure}

We set the parameters of the fitness function to $s=\dfrac{\sqrt{2}}{2}$ and
$a=3$, and then conducted numerical simulations by setting the variance of the
mutational distribution to $m^{2}=0.5^{2}$, $1^{2}$, $1.5^{2}$, $2^{2}$, $2.5^{2}$, $3^{2}$. In Figure \ref{f6si}, we investigated the relationship
between the height at the median of the distribution $h$, and the
environmental velocity $v$. The height at the median can provide a rough
indication of whether the distribution is bimodal or unimodal. Specifically,
if the distribution is bimodal, the height at the median, $h$, should be
relatively low, as the median would lie near the valley between the peaks.
Conversely, if the distribution is unimodal, the height at the median, $h$,
should be relatively high, with the median located near the single peak.

\newpage
\bibliographystyle{ieeetr}
\bibliography{SIRef}